\documentclass[modern]{aastex63}
\usepackage[ruled,vlined,norelsize]{algorithm2e}
\usepackage[framemethod=tikz]{mdframed}
\usetikzlibrary{shadows}
\definecolor{captiongray}{HTML}{555555}
\mdfsetup{%
  innertopmargin=2ex,
  innerbottommargin=1.8ex,
  linecolor=captiongray,
  linewidth=0.5pt,
  roundcorner=5pt
}
\usepackage{enumitem}
\usepackage{booktabs}
\usepackage{xcolor}
\usepackage{amsmath}

\newcommand{\documentname}{\textsl{Article}}
\newcommand{\sectionname}{Section}
\renewcommand{\figurename}{Figure}
\newcommand{\equationname}{Equation}
\renewcommand{\tablename}{Table}

\newcommand{\bs}[1]{\boldsymbol{#1}}

\newcommand{\package}[1]{\textsl{#1}}
\newcommand{\acronym}[1]{{\small{#1}}}

\newcommand{\given}{\,|\,}

\newcommand{\dd}{\mathrm{d}}

\newcommand{\transp}{\ensuremath{^{\mathsf{T}}}}

\newcommand{\mean}[1]{\left\langle #1 \right\rangle}

\renewcommand{\vec}[1]{\ensuremath{\bs{#1}}}
\newcommand{\mat}[1]{\ensuremath{\mathbf{#1}}}

\newcommand{\Msun}{\ensuremath{\mathrm{M}_\odot}}

\newcommand{\kms}{\ensuremath{\mathrm{km}~\mathrm{s}^{-1}}}

\newcommand{\pc}{\ensuremath{\mathrm{pc}}}
\newcommand{\kpc}{\ensuremath{\mathrm{kpc}}}

\newcommand{\Kel}{\ensuremath{\mathrm{K}}}
\newcommand{\mas}{\ensuremath{\mathrm{mas}}}

\newcommand{\abunratio}[2]{\ensuremath{{[\mathrm{#1}/\mathrm{#2}]}}}

\newcommand{\logg}{\ensuremath{\log g}}
\newcommand{\Teff}{\ensuremath{T_{\textrm{eff}}}}

\definecolor{tabblue}{HTML}{4E79A7}
\definecolor{taborange}{HTML}{F28E2B}
\definecolor{tabgreen}{HTML}{59A14F}
\definecolor{tabred}{HTML}{E15759}
\definecolor{tabpurple}{HTML}{B07AA1}

\newcommand{\changes}[1]{{#1}}

\newcommand{\methodname}{\textsl{Orbital Torus Imaging}}

\newcommand{\gaia}{\textsl{Gaia}}
\newcommand{\dr}[1]{\acronym{DR}#1}
\newcommand{\apogee}{\acronym{APOGEE}}

\newcommand{\sdssiv}{\acronym{SDSS-IV}}

\newcommand{\nstars}{56,324}

\newcommand{\ofe}{\abunratio{O}{Fe}}
\newcommand{\mdisk}{\ensuremath{\mathrm{M}_\mathrm{disk}}}
\newcommand{\mratio}{\ensuremath{\mdisk / \mdisk^\star}}
\newcommand{\hz}{\ensuremath{h_{z, \odot}}}
\newcommand{\zsun}{\ensuremath{z_\odot}}
\newcommand{\vzsun}{\ensuremath{v_{z, \odot}}}

\renewcommand{\twocolumngrid}{}
\sloppypar
\renewcommand{\doi}[1]{{\footnotesize\href{https://doi.org/#1}{#1}}}

\shorttitle{Mapping Orbits with \methodname}
\shortauthors{price-whelan, hogg, johnston, ness, rix}

\begin{document}
\graphicspath{ {figures/} }
\DeclareGraphicsExtensions{.pdf,.eps,.png}

\title{\textbf{%
    \methodname: \\
    Using Element Abundances to Map Orbits and Mass in the Milky Way}}

\newcommand{\affcca}{Center for Computational Astrophysics, Flatiron Institute, 162 Fifth Ave, New York, NY 10010, USA}
\newcommand{\affccpp}{Center for Cosmology and Particle Physics, Department of Physics, New York University, 726 Broadway, New York, NY 10003, USA}
\newcommand{\affmpia}{Max-Planck-Institut f\"ur Astronomie, K\"onigstuhl 17, D-69117 Heidelberg, Germany}
\newcommand{\affcolumbia}{Department of Astronomy, Columbia University, New York, NY 10027, USA}
\newcommand{\affutah}{Department of Physics and Astronomy, University of Utah, 115 S. 1400 E., Salt Lake City, UT 84112, USA}
\newcommand{\affuw}{Department of Astronomy, University of Washington, Box 351580, Seattle, WA 98195, USA}
\newcommand{\affprinceton}{Department of Astrophysical Sciences, Princeton University, 4 Ivy Lane, Princeton, NJ~08544}
\newcommand{\affcarnegie}{The Observatories of the Carnegie Institution for Science, 813 Santa Barbara St., Pasadena, CA~91101}

\author[0000-0003-0872-7098]{Adrian~M.~Price-Whelan}
\affiliation{\affcca}

\author[0000-0003-2866-9403]{David~Wardell~Hogg}
\affiliation{\affcca}
\affiliation{\affmpia}
\affiliation{\affccpp}

\author[0000-0001-6244-6727]{Kathryn~V.~Johnston}
\affiliation{\affcca}
\affiliation{\affcolumbia}

\author[0000-0001-5082-6693]{Melissa~K.~Ness}
\affiliation{\affcolumbia}

\author[0000-0003-4996-9069]{Hans-Walter~Rix}
\affiliation{\affmpia}


\author[0000-0002-1691-8217]{Rachael L. Beaton}
\altaffiliation{Carnegie-Princeton Fellow}
\affiliation{\affprinceton}
\affiliation{\affcarnegie}

\author[0000-0002-8725-1069]{Joel~R.~Brownstein}
\affiliation{\affutah}

\author[0000-0002-1693-2721]{D.~A.~Garc\'ia-Hern\'andez}
\affiliation{Instituto de Astrof\'isica de Canarias (IAC), E-38205 La Laguna, Tenerife, Spain}
\affiliation{Universidad de La Laguna (ULL), Departamento de Astrofísica,
E-38206 La Laguna, Tenerife, Spain}

\author[0000-0001-5388-0994]{Sten Hasselquist}
\altaffiliation{NSF Astronomy and Astrophysics Postdoctoral Fellow}
\affiliation{\affutah}

\author[0000-0003-2969-2445]{Christian~R.~Hayes}
\affiliation{\affuw}

\author{Richard~R.~Lane}
\affiliation{Instituto de Astronom\'ia y Ciencias Planetarias de Atacama, Universidad de Atacama, Copayapu 485, Copiap\'o, Chile}

\author{Matthew~Shetrone}
\affiliation{University of California Observatories, UC Santa Cruz, 1156 High St., Santa Cruz, CA 95064}

\author{Jennifer~Sobeck}
\affiliation{Department of Astronomy, University of Washington, Seattle, WA 98195, USA}

\author[0000-0001-6761-9359]{Gail~Zasowski}
\affiliation{\affutah}

\begin{abstract}\noindent
  Many approaches to galaxy dynamics assume that the gravitational potential is
  simple and the distribution function is time-invariant.
  Under these assumptions there are traditional tools for inferring potential
  parameters given observations of stellar kinematics (e.g., Jeans models).
  However, spectroscopic surveys measure many stellar properties beyond
  kinematics.
  Here we present a new approach for dynamical inference, \methodname, which
  makes use of kinematic measurements and element abundances (or other invariant
  labels).
  We exploit the fact that, in steady state, stellar labels vary systematically
  with orbit characteristics (actions), yet must be invariant with respect to
  orbital phases (conjugate angles).
  The orbital foliation of phase space must therefore coincide with surfaces
  along which all moments of all stellar label distributions are constant.
  Both classical-statistics and Bayesian methods can be built on this;
  these methods will be more robust and require fewer assumptions than
  traditional tools because they require no knowledge of the (spatial) survey
  selection function and they do not involve second moments of velocity
  distributions.
  We perform a classical-statistics demonstration with red giant branch stars
  from the \apogee\ surveys:
  We model the vertical orbit structure in the Milky Way disk to constrain the
  local disk mass, scale height, and the disk--halo mass ratio (at fixed local
  circular velocity).
  We find that the disk mass can be constrained (na\"ively) at the few-percent
  level with \methodname\ using only eight element-abundance ratios,
  demonstrating the promise of combining stellar labels with dynamical
  invariants.
\end{abstract}


\section{Introduction}
\label{sec:intro}

An important goal of modern physics and astronomy is to understand the detailed
properties of dark matter on astrophysical scales as a way of informing
constraints on its fundamental nature \citep[see, e.g., recent reviews
by][]{Bullock:2017, Buckley:2018}.
Significant effort has therefore gone into developing and applying tools for
constraining the mass distributions (i.e., dark matter distributions) of Local
Group galaxies using only kinematic observations of tracers (i.e., stars;
\citealt{Jeans:1919b, Binney:2008} and references therein).
This challenge---using observations of tracer objects to constrain the
underlying force law---is conceptually similar to a much older problem faced by
physicists of the 17th century, who worked out that the gravitational force law
in the Solar System is proportional to the inverse square of the distance from
the Sun \citep{Newton:1687}.
That inference was based on observations that showed that orbits in the Solar
System are closed ellipses, with the Sun at one focus \citep{Kepler:1609}---that
is, they made use of observations of a few tracers (the planets) at many orbital
phases.

In our current efforts to map dark matter, we are instead in a regime where we
observe \emph{many} tracers (stars) but have little or no information about
orbital phase: We observe only a single snapshot (in time) of the stellar
orbits.
The closest we come to seeing orbits directly in the Milky Way is in the study
of stellar streams \citep[e.g.,][]{Johnston:1999, Helmi:1999, Eyre:2011,
Sanders:2013, Price-Whelan:2014, Bonaca:2018}, where ensembles of stars almost
directly encode (differential) phase information about the orbit of their
progenitor systems.
But stellar streams are relatively sparse in the Milky Way, and orbits in the
Galaxy have more degrees of freedom than orbits in the Solar System, meaning
that we need many more orbits to span the phase-space and obtain precise
constraints on the mass distribution.
One could argue then that our current task is much harder than Newton and Kepler
had it:
we seek to constrain the global, spatially-extended distribution of dark matter
around a galaxy---a time-evolving mass distribution with nontrivial shape and
radial distribution---with only a snapshot of the tracer kinematics (and often
only a subset of phase-space dimensions).

Given only a snapshot of the dynamics, the tools we use to constrain the
Galactic force field (i.e., the dark matter distribution) are therefore
typically statistical in nature and do not depend on knowing the orbits of
individual stars.
These methods generally rely on making strong assumptions about the distribution
function (\acronym{DF}) or mass model.
For example, in the case of Jeans modeling \citep[e.g.,][]{Jeans:1922,
Oort:1932, Bahcall:1984, Romanowsky:2003, Evans:2009, Watkins:2010, Walker:2011,
Zhai:2018, Buch:2019}, investigators make use of the Jeans equations, which
relate spatial derivatives of velocity second moments to derivatives of the
gravitational potential.
The equations are correct for any collisionless tracer population of stars, but
implementing Jeans models in practice requires assuming that the system is in
steady-state or equilibrium and requires specifying an explicit, parametrized
model for the underlying gravitational potential \citep[see, e.g.,][for a review
of such methods as applied to the problem of determining the local dark matter
density]{Read:2014}.
More general approaches here instead attempt to model the \acronym{DF}
explicitly \citep[e.g.,][]{BoHogg, McMillan:2013, Magorrian:2014, Binney:2014,
Bovy:2016, Magorrian:2019}, but these methods become computationally prohibitive
for large data sets or flexible model forms.
When applied in real-world contexts on stars in the Milky Way or dwarf
spheroidal galaxies, standard methods for inferring a mass distribution from
tracer kinematics therefore typically assume an equilibrium (or steady-state)
distribution function, (coordinate) separability of the \acronym{DF},
symmetries, simple parametrizations of the mass model (e.g., integrable models)
and \acronym{DF}, among others.


A contemporary ``challenge'' to applying these methods on modern data
(especially within the Milky Way) is that stellar kinematic data is now very
precise, and the phase-space is well sampled.
For example, data release (\dr{2}) from the \gaia\ mission
\citep{Gaia-Collaboration:2016, Gaia-Collaboration:2018} has provided full
phase-space kinematics (for a subset of stellar tracers) over a region several
kiloparsecs in size around the Sun.
These data have revealed evidence of dynamical disequilibrium throughout the
Galactic disk and halo \citep{Antoja:2018, Gaia-Collaboration:2018b,
Myeong:2018, Koppelman:2018, Eilers:2020}.
Standard methods for estimating dark matter properties using stellar kinematics
therefore rely on a list of assumptions that we now \emph{know} are strongly
violated in the real universe and Galaxy!
The fact that we still use and apply these methods is a reflection of the fact
that relaxing these assumptions make dynamical inferences\footnote{By
``dynamical inferences,'' we really mean statistical inferences of a mass model
(or parameters of a mass model) given noisy observations of stellar kinematics
(and other stellar labels).}
\emph{far} more challenging and computationally expensive.
For example, there are no straightforward methods for measuring the mass
distribution when the assumption of dynamical equilibrium is relaxed.

One reason to remain hopeful in our goal of precisely constraining the dark
matter is that, unlike Kepler and Newton, we have excellent working models of
stars and access to much more information per star than kinematics alone.
For example, spectroscopic surveys have measured the stellar parameters and
element abundances of hundreds of thousands, and soon to be millions, of stars
\citep[e.g.,][]{DR16, Martell:2017, Deng:2012} over large regions of the
Galactic disk and halo.
It is therefore promising to think that utilizing stellar ``labels'' (element
abundances, stellar ages, or other effectively-invariant stellar properties)
within dynamical inferences will provide additional information that may help
interpret or model the Galaxy \citep[see, e.g.,][for recent methods that begin
to move in this direction, within the context of equilibrium
models]{Sanders:2015, Das:2016, Binney:2016, Iorio:2020}.

In this \documentname, we are going to demonstrate that stellar surface
abundances can be used to illuminate the orbit structure in the Milky Way, and
are therefore extremely valuable for galaxy dynamics.
In our current methodology, like all other dynamical inferences, we will work
only under strong assumptions and we are therefore still in the business of
revealing the orbits indirectly.
However, this class of approaches, here referred to as \methodname\, is
qualitatively distinct from all previous methods for measuring the mass model.
There are reasonable regimes in which it will be more precise than any other
method, conditioned on the assumptions.

\section{Methodological Generalities}
\label{sec:generalities}

In a well-mixed, equilibrium population, stars are in a kinematic steady state.
As time goes on, stars move along their orbits: these orbits can be represented
by a set of dynamical invariants --- actions --- and a location along an orbit
can be represented by a set of phase --- conjugate angle --- variables.
These action--angle coordinates, ($\boldsymbol{J}, \boldsymbol{\theta}$), are
canonical coordinates (i.e., a transformation from, say, Cartesian position and
velocity) in which many dynamical computations are simpler or more efficient
\citep[see, e.g.,][]{Binney:2008}.
The actions $\boldsymbol{J}$ are integrals of motion like any other (for
example, energy or angular momentum), but are special in that they form a set of
momentum coordinates whose conjugate position coordinates are angle variables.
Action variables are defined as integrals of other canonical momentum
coordinates $\boldsymbol{p}$ over a closed loop in the conjugate position
coordinates $\boldsymbol{q}$,
\begin{equation}
  J_i = \oint p_i~\dd q_i \quad .
\end{equation}
As we will see later, if the motion in a given coordinate pair (e.g., $q_i,
p_i$) is separable from the other coordinates, the action integral simply
computes the area enclosed by the orbit in the two-dimensional phase-space
$(q_i, p_i)$.

For gravitational potentials with more than one degree of freedom, it is not
guaranteed that actions exist at all locations in phase space: generically,
regions of phase space can contain chaotic orbits and resonances.
However, when the resonant structure is weak (as is the case for many simple
models of Galactic potentials), most of phase-space is typically stable and
therefore admits transformation to action-space.
In action--angle coordinates, a steady state distribution function
(\acronym{DF}) $f$ is simply a function of the actions, $f(\boldsymbol{J})$.
In such situations, every location in angle-space is equally likely, or equally
probably populated in any snapshot of the kinematics of a tracers orbiting in
some mass distribution.

How, then, do stellar abundances relate to dynamics?
The ``chemical tagging'' insight \citep{Freeman:2002} notes that most stars also
effectively preserve their surface element abundances (and other stellar labels)
as they orbit, over many Galactic orbital timescales.
That is, in an integrable galaxy---a galaxy whose stellar orbits are associated
with three invariant actions---the element abundances and the dynamical actions
have something in common: They are invariant with time.
In these models, only the conjugate angle coordinates are time-dependent.
This means that, for a well-mixed population, the detailed element abundances
can only be a function of actions, and never a function of conjugate angles.
That is a remarkably informative constraint on the configuration of the Milky
Way in the space of positions (dimensionality 3), velocities (3), and
detailed abundances (10--30, depending on the spectroscopic survey).

Consider a collection of stars (localized, say, in phase space) for which we
have measured abundances for $D$ elements.
This collection of stars will in general have a diversity of element abundances:
Their abundances are drawn from some distribution in $D$-dimensional
element-abundance space.
In general, the abundance distribution will depend on position in phase-space:
In the Milky Way, there are observed radial and vertical abundance gradients in
the Galactic disk \citep[e.g.,][]{Hayden:2015, Queiroz:2020}.
However, any changes we observe in these element-abundance distributions with
respect to phase-space coordinates (i.e., the six-gradient with respect to the
three position and velocity components) must not project onto the directions of
increasing (or decreasing) conjugate angles in phase space.
All gradients with respect to phase-space coordinates of the element-abundance
distribution must be orthogonal to the directions of increase (or decrease) of
the conjugate angles, and lie in the subspace of the directions of increase (or
decrease) of the dynamical actions.
The trajectories of stars in the phase space (the dynamical tori) must therefore
lie along or describe level surfaces in the (ensemble mean) element-abundance
distribution!

In this \documentname, we demonstrate the utility of these gradients and
relations between element abundances and dynamical invariants in the context of
dynamical inferences in the Milky Way.
We consider stars in our Galaxy because here we can measure six-dimensional
kinematics and element-abundances for individual stars, enabling relatively
simple demonstrations of the concepts here.
However, we note that a generative model built on these concepts would, in
principle, be applicable in more general scenarios, such as for stars in Local
Group satellite galaxies where only a subset of phase-space coordinates are
measured.

\section{Data}
\label{sec:data}

In our toy demonstrations below, our main data source is a cross-match between
spectroscopic data from the \apogee\ surveys \citep{Majewski:2017} and
astrometric data from the \gaia\ mission \citep{Gaia-Collaboration:2016,
Gaia-Collaboration:2018}.

\apogee\ is a spectroscopic sub-survey and component of the Sloan Digital Sky
Survey IV (\sdssiv; \citealt{Eisenstein:2011, Blanton:2017}) whose main goal is
to map the chemical and dynamical properties of stars across the Milky Way disk.
The survey uses two nearly identical, high-resolution ($R \sim 22,500$;
\citealt{Wilson:2019}), infrared ($H$-band) spectrographs---one in the Northern
hemisphere at Apache Point Observatory (APO) using the SDSS 2.5m telescope
\citep{Gunn:2006}, and one in the Southern hemisphere at Las Campanas
Observatory (LCO) using the 2.5m du Pont telescope \citep{Bowen:1973}.
The primary survey targets are selected with simple color and magnitude cuts
(\citealt{Zasowski:2013, Zasowski:2017}, Santana et al. in prep., Beaton et al.
in prep.), but the sparse angular sky coverage and limited number of fibers per
field lead to a ``pencil-beam''-like sampling of the Milky Way stellar density.
\apogee\ spectra are reduced \citep{Nidever:2015} and then analyzed (i.e., to
measure stellar parameters and abundances) using the \apogee\ Stellar Parameters
and Chemical Abundance Pipeline (\acronym{ASPCAP}; \citealt{ASPCAP,
Holtzman:2018, Jonsson:2020}); here we use abundance measurements from the
standard \apogee\ pipeline.

Here we use a recent internal data product (which includes all data taken
through March 2020) from the \apogee\ surveys (post-\dr{16}) that includes
$\approx60\%$ more stars than the publicly-available \dr{16} catalogs
\citep{DR16, Jonsson:2020}, but was reduced and processed using the same
pipeline used to produce the \dr{16} release \citep[i.e., the pipeline described
in][]{Jonsson:2020}.
This \apogee\ catalog contains calibrated element abundance measurements for 18
elements, but these have a variety of physical origins and a range of
reliabilities and measurement precisions.
For demonstrations below, we therefore focus on a subset of eight, well-measured
(log) abundance ratios selected to have varied astrophysical origins:
\abunratio{Fe}{H}, \abunratio{C}{Fe}, \abunratio{N}{Fe}, \abunratio{O}{Fe},
\abunratio{Mg}{Fe}, \abunratio{Si}{Fe}, \abunratio{Mn}{Fe}, \abunratio{Ni}{Fe}.
In some cases, we focus on just a single element abundance, \abunratio{O}{Fe},
which is one of the most precisely and accurately determined element abundance
measured with the \dr{16} pipeline \citep{Jonsson:2020}.

\gaia\ is primarily an astrometric mission and survey
\citep{Gaia-Collaboration:2016} that obtains sky position, proper motion, and
parallax measurements for $>1$ billion stars, limited only by their apparent
magnitudes (\gaia\ $G \lesssim 20.7$).
Here we use parallax and proper motion measurements released in \gaia\ \dr{2}
\citep{Gaia-Collaboration:2018, Gaia-astrometric:2018}.

\begin{figure}[!tp]
  \begin{mdframed}
  \color{captiongray}
  \begin{center}
  \includegraphics[width=\textwidth]{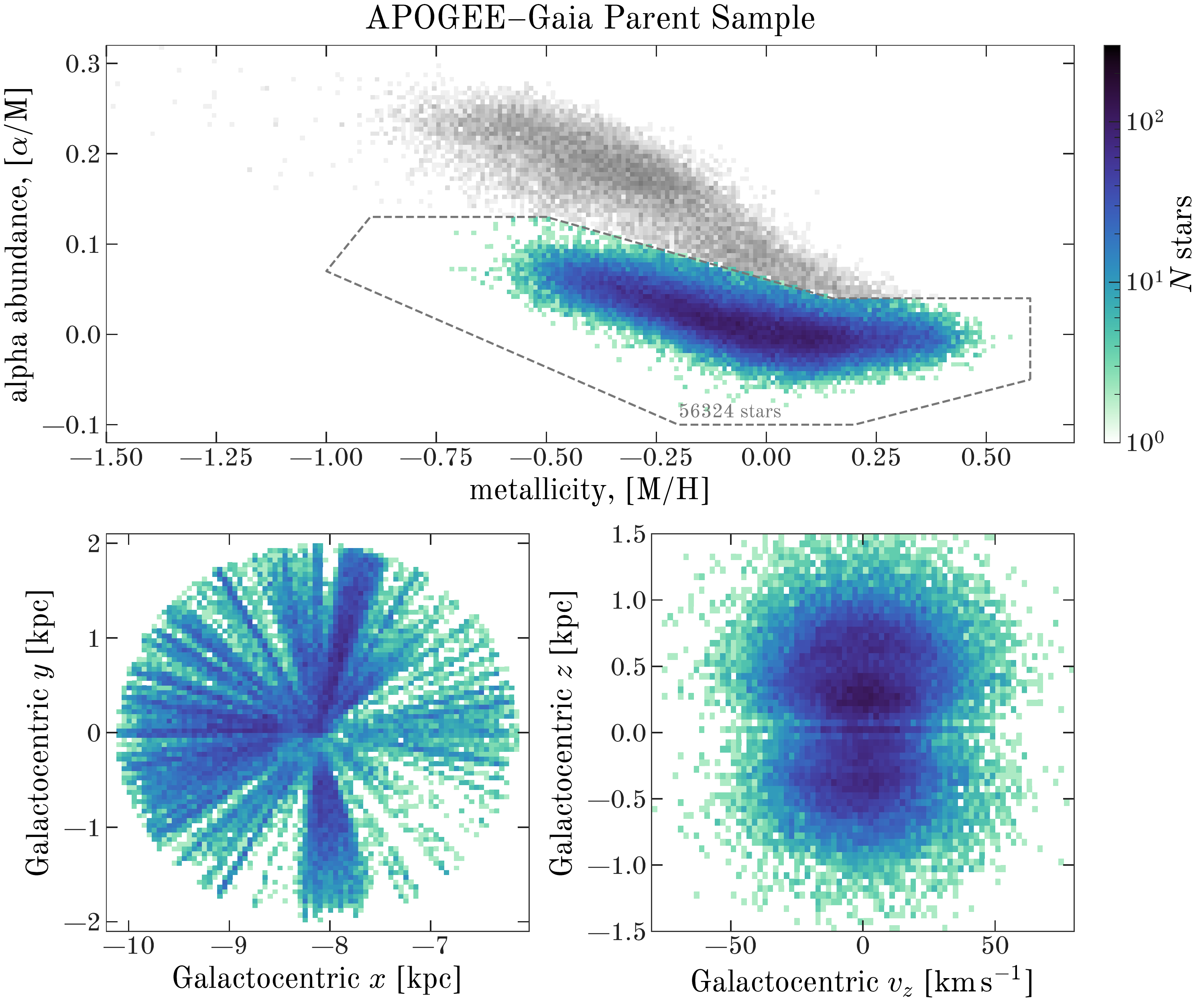}
  \end{center}
  \caption{%
    The data sample used in this project.
    Each panel shows a 2D histogram of the stars.
    \textsl{Top panel:}
    Bulk abundance ratios measured by \apogee\ and a selection boundary (dashed
    line) used to exclude the ``high-$\alpha$'' stars that are generally older
    and kinematically hotter.
    \textsl{Lower left panel:}
    Positions of the stars in the parent sample projected onto the Galactic
    plane, showing the spherical spatial cut and highly non-uniform (\apogee)
    spatial selection.
    Heliocentric distances to the stars are obtained by na\"ively inverting
    their \gaia\ parallax measurements.
    \textsl{Lower right panel:}
    The distributions of stars in the parent sample in vertical $z, v_z$ phase
    space.
    This panel shows that the sample is less populated at low $z$ (mainly
    because of the \apogee\ selection function), which in turn shows that the
    sample is less populated at certain values of vertical angle, $\theta_z$.
    The methods presented in this paper do not require that all angles are
    equally populated in the sample.
    \label{fig:mh-am-xy}}
  \end{mdframed}
\end{figure}

We cross-match the \apogee\ sample to \gaia\ \dr{2} using the \apogee-provided
\acronym{2MASS} \citep{Skrutskie:2006} identifiers, and the \gaia-provided
cross-match between \gaia\ \dr{2} and the final \acronym{2MASS} point source
catalog \citep{Gaia-crossmatch:2019}.
We then apply a number of quality cuts and other selections to limit the catalog
to chemically thin-disk, red giant branch (RGB) stars with well-measured stellar
parameters (\logg, \Teff) and abundances (the abundance ratios listed above),
high signal-to-noise parallax measurements, but excluding stars targeted in
stellar clusters and dwarf galaxies, as enumerated below.
In detail, our selections include:
\begin{itemize}
  \item \acronym{ASPCAP} quality flag (\texttt{ASPCAPFLAG}) must not contain
    \texttt{STAR\_BAD} or \texttt{STAR\_WARN},
  \item $3500~\Kel < \Teff < 6500~\Kel$,
  \item $1.5 < \logg < 3.4$,
  \item combined spectroscopic signal-to-noise ${\rm SNR} > 20$,
  \item \apogee\ targeting bit flags must not indicate that the source was part
    of a special program to observe stellar clusters, dwarf galaxies, M31 stars,
    stellar streams, or moving groups; this \emph{excludes} stars with the
    following bits enabled:
    \begin{itemize}
      \item \texttt{APOGEE\_TARGET1}: (9, 18, 24, 26)
      \item \texttt{APOGEE\_TARGET2}: (10, 18)
      \item \texttt{APOGEE2\_TARGET1}: (9, 18, 20, 21, 22, 23, 24, 26)
      \item \texttt{APOGEE2\_TARGET2}: (10)
      \item \texttt{APOGEE2\_TARGET3}: (5, 14, 15)
    \end{itemize}
  \item stars are part of the ``low-$\alpha$'' (or ``chemical thin disk'')
    population (see polygonal selection in the top panel of
    \figurename~\ref{fig:mh-am-xy}),
  \item \gaia\ parallax $\varpi > 0.5~\mas$,
  \item \gaia\ parallax signal-to-noise $\varpi / \sigma_\varpi > 8$.
\end{itemize}
The \apogee\ catalog contains a small number of duplicates (duplicated source
identifier \texttt{APOGEE\_ID}); to avoid duplication in our sample in these
cases, we keep only the entry with highest signal-to-noise.
We select the low-$\alpha$ sequence primarily because the high- and low-$\alpha$
sequences have different scale heights \citep[e.g.,][]{Bovy:2016}, and our toy
model is not currently flexible enough to account for this (however, it is
possible to include this complexity in our framework, as will be explored in
future extensions of this model).
The final parent sample contains \nstars\ RGB stars with high-quality \apogee\
and \gaia\ data.

For each star in the parent sample, we compute na\"ive distance estimates by
inverting the parallax, $d = 1/\varpi$.
While this is generally not a safe way of computing distance from parallaxes
(see, e.g., \citealt{Bailer-Jones:2015}), our sample stars are (by construction)
relatively nearby and have high signal-to-noise parallax measurements.
Using a parallax signal-to-noise selection has its own consequences, especially
in that it makes the sample selection function complex and non-intuitive.
However, our methodology should be relatively insensitive to selection effects,
and since the demonstrations that make use of these distance measurements below
are meant to be illustrative examples, we ignore these details in what follows.
This sample is visualized in bulk element abundance ratios and Galactocentric
positions in \figurename~\ref{fig:mh-am-xy}.

\section{Milky Way mass model and action--angle computations}
\label{sec:mw-model}

Our methodology and demonstrations below rely on computing actions and angles
for stars, which depend on the mass distribution of the Milky Way and the solar
position and motion with respect to a Galactocentric reference frame.

For the solar position, we use the recent, precise measurement of the
Sun--Galactic center distance from the \acronym{GRAVITY} collaboration, $r_\odot
= 8.122~\kpc$ \citep{Gravity:2018}, and initially adopt $z_\odot = 20.8~\pc$ as
the solar height above the Galactic midplane \citep{Bennett:2019}.
In Galactocentric Cartesian coordinates, we use a right-handed coordinate system
such that the Sun is at $\bs{x}_\odot = (-8.1219, 0, 0.0208)~\kpc$ and the
solar velocity is $\bs{v}_\odot = (12.9, 245.6, 7.78)~\kms$ \citep{Drimmel:2018,
Reid:2004, Gravity:2018}.

We represent the density distribution (or gravitational potential) of the
Milky Way using an idealized, four-component mass model consisting of a
spherical Hernquist bulge \citep{Hernquist:1990}, spherical Hernquist nucleus,
an axisymmetric Miyamoto-Nagai disk \citep{Miyamoto:1975}, and a spherical
Navarro-Frenk-White dark matter halo \citep{Navarro:1996}.
Most of the parameters of these components are fixed to their default values
from the \texttt{MilkyWayPotential} class implemented in the \package{gala}
Python package \citep[v1.2;][]{gala}:
briefly, the bulge parameters (mass and scale radius) and disk parameters (total
mass, scale height, and scale radius) are initially set to match the
\texttt{MWPotential2014} implemented in \package{galpy} \citep{Bovy:2015}, and
the dark matter halo parameters (virial mass and scale radius) are initially set
by fitting the enclosed mass profile of the mass model to a compilation of
recent enclosed mass measurements.\footnote{As described in
\url{https://gala.adrian.pw/en/latest/potential/define-milky-way-model.html}.}
However, the mass of the disk is then adjusted to match a circular velocity of
$v_{\rm circ}(R_\odot) = 229~\kms$ at the solar radius \citep{Eilers:2019}.
Our fiducial mass model therefore adopts the default \texttt{MilkyWayPotential}
parameters except for the disk mass, which is set to $\mdisk^\star = 6.526
\times 10^{10}~\Msun$.
Later in this \documentname, we vary the disk mass \mratio\ and disk scale
height at the solar position \hz, but we adjust the dark matter halo mass to
keep the circular velocity at the solar radius, $v_{\rm circ}(R_\odot)$,
constant.

In a given potential model, we compute actions, $\bs{J} = (J_R, J_\phi, J_z)$,
and angles, $\bs{\theta} = (\theta_R, \theta_\phi, \theta_z)$, for a star using
the ``St\"ackel Fudge'' \citep{Binney:2012, Sanders:2012} as implemented in
\package{galpy} \citep{Bovy:2015}.
We solve for the focal length, $\Delta$, of the locally-approximating St\"ackel
potential for each star's orbit by numerically computing the orbit of each star
for four orbital periods.\footnote{For computational efficiency, we actually
compute the locally-fitting St\"ackel potential focal length parameter $\Delta$
\citep{Sanders:2012} using \package{gala} \citep{gala}.}

We assess the accuracy of using the St\"ackel Fudge for orbits with large
vertical excursions from the Galactic disk by comparing actions computed with
the St\"ackel Fudge to actions computed with a more accurate action solver.
For a set of trial orbits (meant to span the range of vertical actions we see in
our sample of \apogee\ stars) over a range of disk masses (from
$\mratio=0.5$--$1.5$), we compute actions both with the St\"ackel Fudge and with
the ``O2GF'' method defined in \citet{Sanders:2014, Sanders:2016}.
The O2GF method works by numerically integrating an orbit for a given star and
solving for the generating function to transform from actions computed in a toy
potential model to the actions in the potential model of interest (as defined in
\citealt{Sanders:2014}, and implemented in \package{gala}).
This method has some tuning parameters related to the total orbital integration
time and time step, and the number of Fourier components to include in the
Fourier expansion used to represent the generating function, $N_{\rm max}$.
We use an Isochrone potential as our toy potential model, set the total
integration time for each star to 128 (radial) orbital periods, $T=128\,P_r$,
and set the time step to $\Delta t = P_r / 256$. For all stars, we set $N_{\rm
max} = 8$ (following \citealt{Sanders:2016}).
We find that in all cases, the values of the actions agree to within $<10\%,
<0.1\%, <0.5\%$ for $J_R, J_\phi, J_z$, respectively, with the largest
disagreements only affecting a small fraction of the stars in our sample
($<8\%$) with the largest excursions ($\gtrsim 3$ scale heights) from the
Galactic plane.

For computational efficiency, we therefore use the St\"ackel Fudge as our
primary method for computing actions.
To additionally speed up computation, we typically also parallelize the
computation of the actions (over stars) using the Python package
\package{schwimmbad} \citep{schwimmbad}.

\section{Motivation from Observed Element Abundance Gradients}
\label{sec:motivation}

\begin{figure}[!tp]
  \begin{mdframed}
  \color{captiongray}
  \begin{center}
  \includegraphics[width=\textwidth]{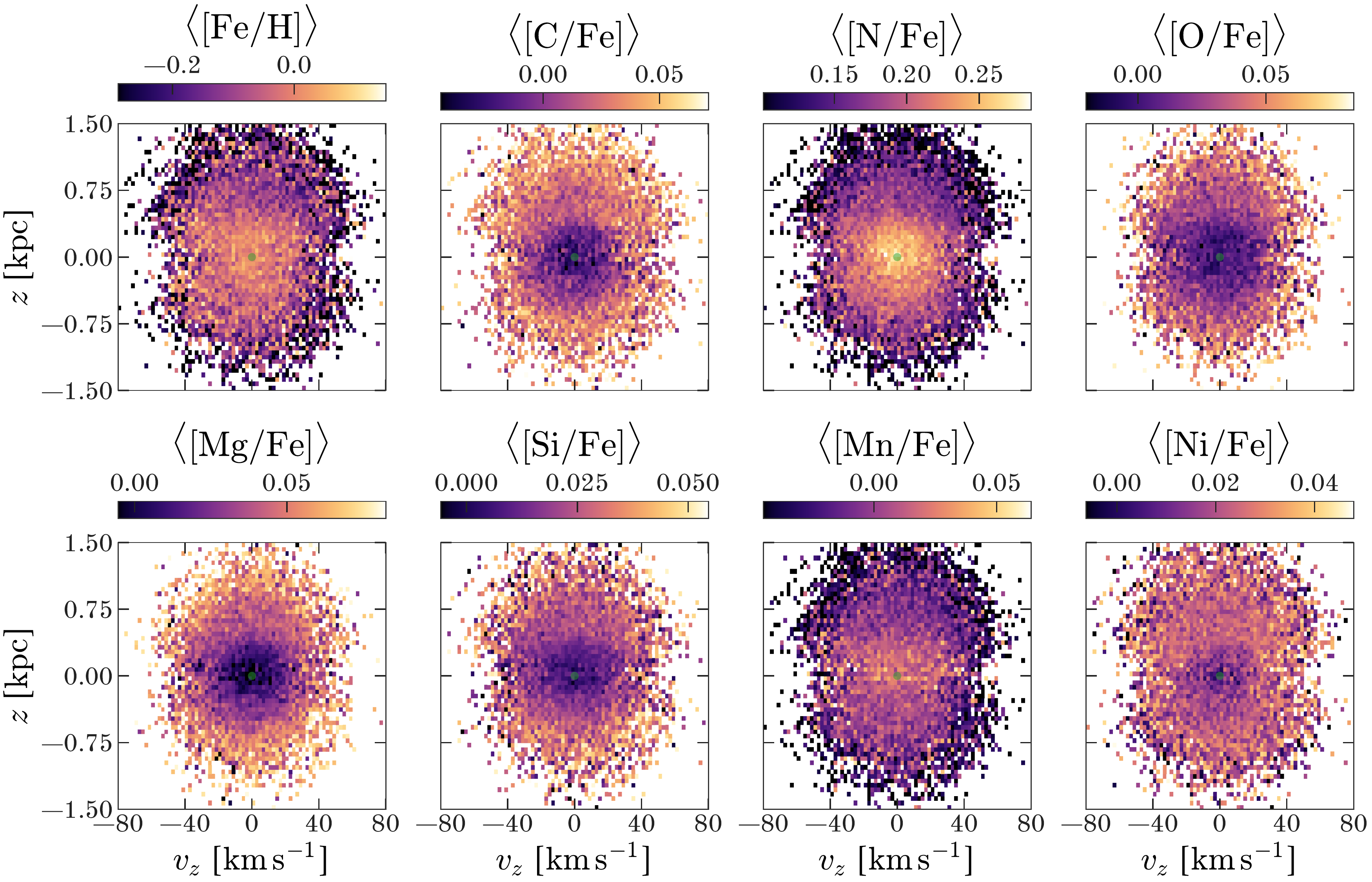}
  \end{center}
  \caption{%
    The means of various element abundance ratios as a function of Galactic
    vertical height $z$ and vertical velocity $v_z$.
    Averages are taken in $z, v_z$ boxels.
    The stars at lower $|z|$ and lower $|v_z|$ (that is, the stars with
    lower overall vertical action $J_z$) show higher overall metallicity
    on average, but lower $\alpha$-to-iron.
    These plots are somewhat affected by \apogee\ selection effects, in
    that different $z, v_z$ boxels are projections through different extents
    in Galactocentric radius (see \figurename~\ref{fig:mh-am-xy}).
    This explains some of the visible asymmetries.
  \label{fig:zvz-grid}
  }
  \end{mdframed}
\end{figure}

\begin{figure}[!tp]
  \begin{mdframed}
  \color{captiongray}
  \begin{center}
  \includegraphics[width=\textwidth]{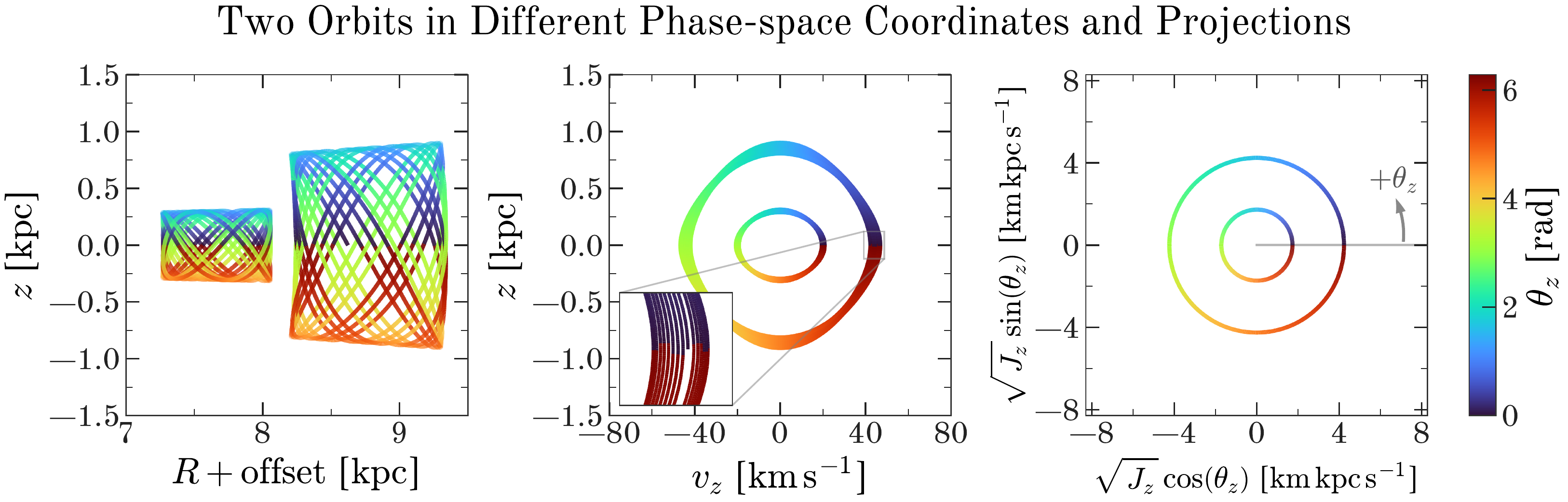}
  \end{center}
  \caption{%
    \textsl{Left panel:} Two orbits projected onto the plane of
    Galactic vertical height $z$ and Galactocentric cylindrical radius
    $R$. The orbits fill the surfaces of 3-tori in 6-d phase space.
    \textsl{Middle panel:} The same two orbits, but projected onto the plane
    of vertical height $z$ and vertical velocity $v_z$. In this projection,
    it becomes clearer that the orbital lines are
    colored by the angle $\theta_z$ that is conjugate to vertical action $J_z$.
    The inset shows that conjugate angle wraps non-trivially, because the action
    (by construction) wraps at constant angular velocity, whereas the vertical period
    is a (weak) function of the other orbital phases.
    Note that although this projection is close to ``face on'' for these two
    orbits, the fact that they fills the surfaces of 3-tori means that they
    project to finite-width bands in $z, v_z$ space.
    \textsl{Right panel:} The same two orbits, but now plotted in vertical-action,
    vertical-angle space. In this space, the two orbits trace perfect circles.
  \label{fig:zvz-demo}
  }
  \end{mdframed}
\end{figure}

A significant motivation for this work came from plots of elemental abundance
ratios of stars as a function of vertical height $z$ and vertical velocity $v_z$
in Galactocentric Cartesian coordinates.
As an example, \figurename~\ref{fig:zvz-grid} shows the mean abundance ratios of
stars in bins of their vertical phase-space coordinates, using data from the
\apogee\ and \gaia\ surveys (see \sectionname~\ref{sec:data}).
The stars shown in this figure are selected following the quality cuts defined
above (\sectionname~\ref{sec:data}) and are selected to lie in the low-alpha
sequence (\figurename~\ref{fig:mh-am-xy}).
In these plots, the eye is drawn to \emph{abundance gradients}:
The stars at small heights and small vertical velocities have different
abundance ratios, on average, than stars at large heights and large vertical
velocities.
But these positional and velocity gradients are related:
Stars at large absolute vertical velocities $v_z$ will, as they orbit, reach
large absolute vertical positions $z$ (far from the Galactic plane, that is),
and stars at large absolute vertical positions will, in the future, reach large
absolute vertical velocities.
That is, the stars will orbit in the Galaxy, which projects onto this $z$--$v_z$
plane as (to zeroth order) roughly elliptical trajectories.
For example, \figurename~\ref{fig:zvz-demo} shows two Galactic orbits computed
in a 3D model for the Milky Way (see \sectionname~\ref{sec:mw-model}) in
different projections of phase-space coordinates:
In the space of $z$--$v_z$ (center panel), a Galactic orbit will form a
close-to-elliptical band whose enclosed area scales with the vertical action,
$J_z$, whose thickness depends on the eccentricity of the orbit, and a given
position on its ``ellipse'' can nearly be mapped to a vertical angle,
$\theta_z$.

To very high precision, stars do not change their abundances as they orbit.
One consequence of this is a new method for inferring the orbit structure of the
Milky Way:
If two small neighborhoods in phase space lie on the same orbit---that is, they
correspond to the same dynamical actions but with different conjugate
angles---they must contain stars with the same distribution of element
abundances.
This prediction depends on many detailed assumptions; for example, that the
Galaxy is (approximately) phase mixed, and that the potential is (approximately)
time invariant and integrable.
Of course, the \emph{usefulness} of this prediction for inference depends on the
existence of element abundance gradients: if there are no
element-abundance-ratio gradients, there will be no information to exploit.

\begin{figure}[!tp]
  \begin{mdframed}
    \color{captiongray}
  \begin{center}
  \includegraphics[width=\textwidth]{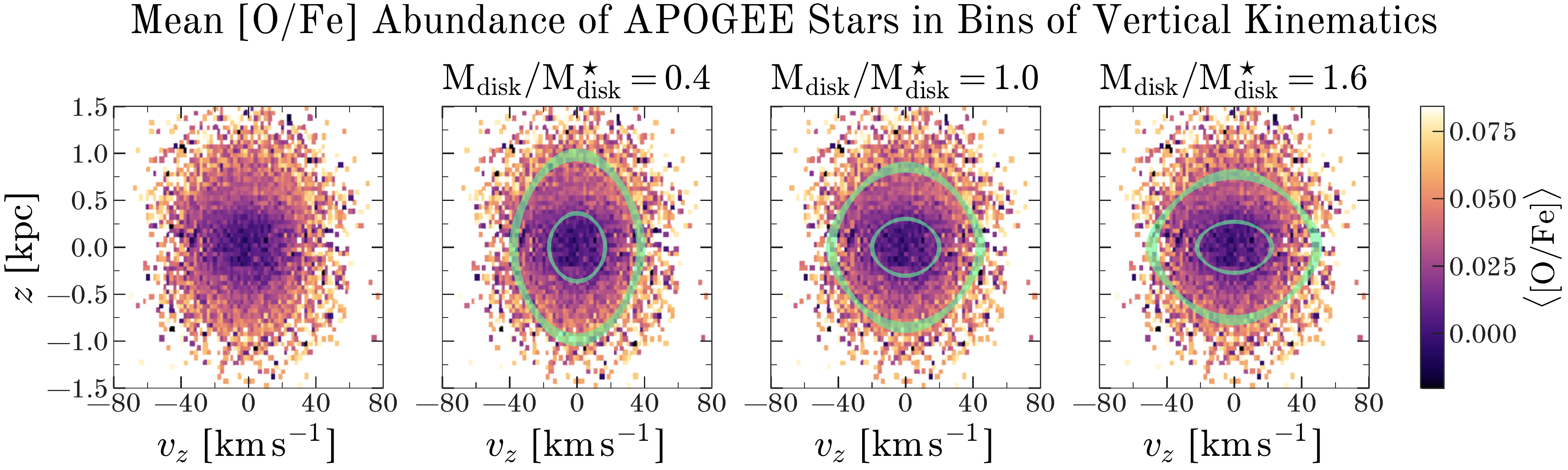}
  \end{center}
  \caption{%
    \textsl{Left panel:} A repeat of the $\ofe$ panel of
    \figurename~\ref{fig:zvz-grid}.
    \textsl{Other panels:} The same as the left panel, but with the two orbits
    from \figurename~\ref{fig:zvz-demo} over-plotted, for three different Milky
    Way potentials.
    These three potentials have the fiducial Milky Way disk mass (see text
    for details), or a disk less massive by a factor of 0.4 or more
    massive by a factor of 1.6, as noted in each panel title.
    All potentials are constrained to have a circular velocity at the
    Solar circle of $229\,\kms$.
    \emph{Which of the three panels appears most like the orbits are
    coincident with isopleths of the mean abundance?}
    This question is asked for illustrative purposes only: these plots distort
    the data by projection in phase space, and the methods we use for the
    inferences we perform (see \sectionname~\ref{sec:inferences}) do not rely
    on or use any projections.
  \label{fig:zvz-ofe}
  }
  \end{mdframed}
\end{figure}

For the sake of illustration and simplicity, we visualize and demonstrate these
concepts using the vertical kinematics of stars in our parent sample.
\figurename~\ref{fig:zvz-ofe} again shows the mean \abunratio{O}{Fe} abundance
ratios (in all panels), but now with two overlaid orbits (green overlaid bands)
computed in three different Milky Way models (with varied disk mass, as
indicated; see \sectionname~\ref{sec:mw-model}).
The two orbits were chosen for illustrative purposes (one with low $J_z$, one
with higher $J_z$), and are defined such that they have the same values of their
three actions $(J_R, J_\phi, J_z)$ in all mass models.
In the fiducial mass model ($\mratio = 1.0$), the two overlaid orbits nearly
follow mean abundance contours.
In the model in which the disk is made less massive (and the halo more massive
to keep the circular velocity constant; $\mratio = 0.4$), the orbits change
shape:
There is more positional extent to an orbit relative to its velocity extent.
If stars were traveling on these lower-disk-mass orbits, they would have to
obtain higher abundances when they are passing through the disk midplane,
and lower abundances when they are at their greatest absolute vertical heights,
which is absurd: Stars do not change their abundances as they orbit.
In our method below, we utilize this fact to infer the disk mass.

\section{Method, Assumptions, and Toy Applications}
\label{sec:inferences}

There are two families of approaches to \methodname.
In the first---which we will call \emph{classical} (in the sense of classical
statistics)---we make explicit the fact that the element abundance distribution
should not depend on the conjugate angles.
The best-fit or inferred mass-model parameters are those that lead to no
residual dependence of abundances (or mean abundance ratio, or any moment of the
abundance-ratio distribution) on angles.
In the second---which we will call \emph{generative}---we would construct a
predictive model for the high-dimensional abundance distribution as a function
of actions alone (and not angles).
The best-fit or inferred mass-model parameters are those that maximize the
combined probability (density), evaluated at the observed abundances.
The classical approaches are frequentist and the generative approaches produce
likelihoods and can be used in Bayesian inferences.

Both classical and generative implementations of \methodname\ depend on a
specific set of assumptions, enumerated below.
We emphasize that these assumptions are not necessarily correct, but that our
method is \emph{conditionally} correct, conditioning on these core assumptions:
\begin{description}
\item[integrable] Each stellar orbit is regular and has three
  dynamically-invariant actions and three conjugate angles. This assumption
  could in principle be relaxed in future implementations.
\item[phase mixed] \changes{At any position in action-space,} the stars in our
  sample are inherently uniformly distributed in angle variables; all angles are
  equally likely. This assumption will be violated in the data (as we discuss
  below; see \sectionname~\ref{sec:discussion}), but this is the fundamental
  assumption of the vast majority of inferences of the Milky Way mass
  distribution.
\item[properly selected] The selection function depends on position (or
  velocity) in the Milky Way, but not on element abundances at a given position.
  In detail, the \apogee\ selection function \emph{will} depend on abundances
  through gradients between stellar parameters and element abundance ratios, and
  implicit selections on stellar parameters (through, e.g., color and magnitude
  cuts). We attempt to mitigate these issues here by selecting a limited (in
  \logg) subset of red giant stars that are dominated by red clump stars (see
  \sectionname~\ref{sec:data}), and we discuss this further in
  \sectionname~\ref{sec:discussion}.
\item[measurable gradients] There exist gradients in the abundances with respect
  to the kinematics (i.e., actions) and those gradients are measurable at the
  precision of the individual stellar abundance measurements in our
  spectroscopic dataset.
\item[invariant abundances] Stellar surface element abundances are
  time-invariant. In detail, this assumption is violated by stellar and
  planetary evolution, as surface abundances can change, for example, during red
  giant branch evolution \citep[e.g.,][]{Iben:1965, Martig:2016} for solar-mass
  stars. However, we expect the magnitude of this to be unimportant and should
  not dominate our systematics as long as the timescales over which the
  abundances change are sufficiently different from the orbital timescales
  (hundreds of millions of years).
\end{description}

In this \documentname, we consider only classical-statistics approaches for
illustrative purposes.
However, we expect that generative approaches will produce at least slightly
more precise inferences, since they will be protected by the arguments and
proofs of Bayesian inference.
In addition to the core assumptions listed above, we also make additional
assumptions specific to our implementation and demonstrations.
In particular, here we assume that the 6D phase-space positions are sufficiently
accurate and precise such that we condition over these quantities directly (and
ignore their reported uncertainties).
For the parent sample used here, the median distance uncertainty (from inverting
the \gaia\ parallaxes) is $\sigma_d \approx 50~\pc$ and the median velocity
uncertainty is $\approx 2~\kms$.
We also here assume that stars in different parts of phase-space receive the
same quality of abundance measurements, an assumption that we know is weakly
violated because of known temperature \citep{Jonsson:2020} and \logg\ (Eilers et
al., in prep.) dependences on the \apogee\ abundance measurements, and stars
with different temperatures or surface gravities or luminosities can be seen at
different distances and $z$ heights.
However, our parent sample is selected to contain a relatively small range of
surface gravities along the red giant branch and predominantly consists of
red clump stars, so we expect these issues to be negligible.\footnote{We tried
repeating all subsequent analyses in this \documentname\ using a sample of
high-confidence Red Clump stars \citep[using the selection defined
in][]{Bovy:2014} and found no significant differences in our results.}
\changes{Finally, we additionally here assume that the Milky Way potential
model is time-independent.
This is not a strict or core assumption of \methodname: We only require that the
actions exist, and that the orbital phases of stars are mixed at any location in
action-space.
For example, this could still be satisfied in weakly time-dependent potentials
in which the actions remain adiabatically invariant.}

\subsection{Initial Demonstration of Fitting Procedure}

As an initial demonstration, we focus here on the vertical kinematics of stars
(but we do not assume separability) and we ask whether there is a choice of mass
model that leads to dynamical actions and conjugate angles such that the
abundance ratio distributions do not depend on the conjugate angle $\theta_z$.
To assess the dependence of a particular element abundance ratio
\abunratio{X}{Y} on the vertical angle, we define a ``mean abundance-ratio
deviation,'' $\Delta^{\abunratio{X}{Y}}$, for each $n$ star, which is the
difference between a given star's (logarithmic) abundance ratio and the mean of
the (logarithmic) abundance ratios of its kinematic neighbors (in action-space),
\begin{equation}
  \Delta^{\abunratio{X}{Y}}_n = \abunratio{X}{Y}_n -
    \mean{\abunratio{X}{Y}}_{\vec{J}}
    \quad . \label{eq:mean-abun-dev}
\end{equation}
That is, we use a given mass model to compute the three actions $\vec{J}$ for
each star in the parent sample and choose the closest $K$ neighbors in
$\vec{J}$-space (with the isotropic Euclidean metric distance) for each star to
compute the action-local mean abundance.
We actually perform a weighted mean in order to account for the fact that there
could be strong number-density gradients in action-space that could bias our
estimates of the mean abundance values (we discuss this in more detail in
Appendix~\ref{app:knn-weights}).
For a specific star, its neighbors are a function of the mass-model parameters
because all actions will change values in different mass models.
We choose to use $K=64$ based on experimentation: Smaller $K$ values more
accurately resolve the abundance gradients but have more shot noise, but larger
$K$ values smooth out the abundance gradients.
However, we find that our results are not very sensitive to this choice (within
a multiplication or division by a factor of a few).

Our goal then is to minimize (over mass model parameters) the dependence of the
mean abundance-ratio deviation distribution on the vertical angle $\theta_z$.
To quantify this dependence, we fit a smooth model of the form
\begin{equation}
  \Delta(\theta_z) = c_0 + a_1\,\cos  \theta_z + b_1\,\sin    \theta_z
                         + a_2\,\cos 2\theta_z + b_2\,\sin 2\theta_z \quad,
                         \label{eq:fouriermodel}
\end{equation}
where the five parameters $(c_0, a_1, a_2, b_1, b_2)$ are the coefficients of a
Fourier series expressed out to $m=2$.
The functional form captures our expectations for how the mean abundance-ratio
deviations will depend on vertical angle when our model is wrong:
if the orbit contours at a given action have the wrong shape, we expect there to
be an $m=2$ variation to the abundance-ratio deviations (e.g.,
\figurename~\ref{fig:zvz-ofe}).
If our assumed solar position relative to the midplane or solar velocity
relative to the local standard of rest are wrong, we expect there to be $\sin$
and $\cos$ (i.e., $m=1$) variations.
In the smooth model for the mean abundance deviations
(\equationname~\ref{eq:fouriermodel}), the parameter $a_1$ is sensitive to the
vertical component ($v_z$ component) of the local standard of rest or the Solar
motion.
Parameter $b_1$ is sensitive to the vertical ($z$) location of the Sun relative
to the disk midplane.
Parameter $a_2$ is sensitive to the local mass density concentrated in the disk,
or the disk-mass parameter \mratio.
Parameter $b_2$ should not exist and is included as a test of model assumptions;
in detail, it is sensitive to tilts in the coordinate system, and
non-phase-mixed structures.
As we note later, the ``best'' setting of the mass model parameters should
minimize a combination of these amplitudes.

We fit this model (\equationname~\ref{eq:fouriermodel}) to all of the
$N=\nstars$ individual abundance-ratio deviations and their uncertainties
$(\Delta_n, \sigma_{\Delta_n})_N$ (without binning) using least-squares fitting.
In detail, for a given mass model, we compute start by computing the actions and
angles for all $N$ stars in our sample.
We then use the three actions for each star to compute the mean abundance
deviations $\Delta_n$ and the associated uncertainty $\sigma_{\Delta_n}$ for
all stars (\equationname~\ref{eq:mean-abun-dev}).
We construct the design matrix $\mat{M}$ using the vertical angle values
$\theta_{z, n}$
\begin{equation}
  \mat{M} = \begin{pmatrix}
      1 & \cos{\theta_{z, 1}} & \sin{\theta_{z, 1}}
        & \cos{2\,\theta_{z, 1}} & \sin{2\,\theta_{z, 1}}\\
      \vdots & \vdots & \vdots & \vdots & \vdots \\
      1 & \cos{\theta_{z, N}} & \sin{\theta_{z, N}}
        & \cos{2\,\theta_{z, N}} & \sin{2\,\theta_{z, N}}
    \end{pmatrix}
\end{equation}
the ``data'' vector $\vec{y}$ using the mean abundance deviations
\begin{equation}
  \vec{y} = \begin{pmatrix} \Delta_1 & \cdots & \Delta_N \end{pmatrix}\transp
\end{equation}
and the covariance matrix $\mat{C}$ using the mean abundance deviation
uncertainties
\begin{equation}
  \mat{C} = \begin{pmatrix}
    \sigma_{\Delta_1}^2 & & \\
    & \ddots & \\
    & & \sigma_{\Delta_N}^2
    \end{pmatrix} \quad .
\end{equation}
We compute the best-fitting parameter vector $\hat{\vec{g}} = (c_0, a_1, a_2,
b_1, b_2)$ by solving the linear least-squares problem
\begin{equation}
  \hat{\vec{g}} = \left(\mat{M}\transp \, \mat{C}^{-1} \, \mat{M}\right)^{-1}
    \, \mat{M}\transp \, \mat{C}^{-1} \, \vec{y}
\end{equation}


\begin{figure}[!tp] 
  \begin{mdframed}
    \color{captiongray}
  \begin{center}
  \includegraphics[width=\textwidth]{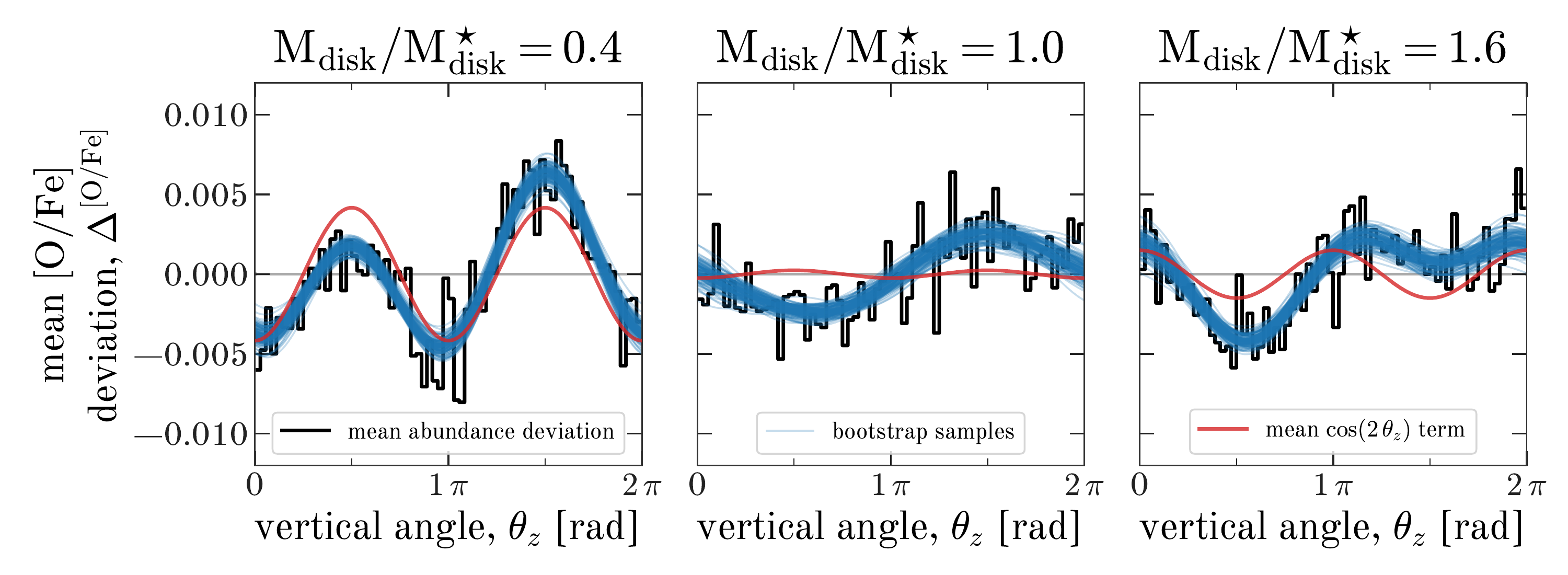}
  \end{center}
  \caption{%
    The mean $\ofe$ abundance deviation, $\Delta^\ofe$, as a
    function of vertical angle, for three different values of the mass of the
    disk (at fixed circular velocity at the Solar circle).
    The abundance deviation for each star in the sample is the difference
    between the abundance measured in each star and a mean of the $K=64$ nearest
    neighbors to that star in three- dimensional action space (for that mass
    model).
    In each panel, the mean abundance deviation is shown three ways.
    The black histogram shows the mean of the abundance deviation in small bins
    in vertical angle (binned for visualization purposes only: the binned data
    are never used in the analysis).
    The blue lines show continuous fits to the unbinned data, which are
    continuous linear combinations of sines and cosines (see
    \sectionname~\ref{sec:inferences}); there are 128 blue lines, one for each
    of 128 independent bootstrap trials.
    The red line shows the mean of the $\cos 2\,\theta_z$ terms across the 128
    bootstrap trials: The amplitude of this parameter is an indication of the
    goodness of fit of a particular choice of mass model parameters.
    A ``perfect'' fit would produce a red curve that is flat or has minimal
    amplitude.
    Comparing the three panels, the red curves show an amplitude of opposite
    sign in the highest-disk-mass panel relative to the other panels, suggesting
    that the best setting of the disk mass is in between the fiducial disk mass
    model and the higher-disk-mass model (as we find; see
    \sectionname~\ref{sec:inferences}).
  \label{fig:sinusoid-fits}
  }
  \end{mdframed}
\end{figure}

\NoCaptionOfAlgo
\begin{algorithm}[tp]
  \caption{\textbf{Procedure for computing the classical-statistics objective function}}\label{alg:objective}
  \SetAlgoLined
  \KwIn{Parameter vector $(\mdisk, \hz, \zsun, \vzsun)$ and bootstrap sample of parent stellar sample}
  \begin{enumerate}[itemsep=0mm]
    \item Compute the halo mass at fixed $v_{\rm circ}(R_\odot) = 229~\kms$
    \item Compute Galactocentric positions and velocities for each star
    \item Compute St\"ackel potential focal length parameter for each star
    \item Compute actions and angles for all stars (St\"ackel fudge)
    \item Compute action-local mean abundance deviation for each star (\equationname~\ref{eq:mean-abun-dev})
    \item Use linear least-squares to compute the optimal Fourier parameters (\equationname~\ref{eq:fouriermodel})
    \item Compute the objective function (\equationname~\ref{eq:objective}).
  \end{enumerate}
\end{algorithm}

\figurename~\ref{fig:sinusoid-fits} shows smooth fits of the above model
computed from bootstrapped resamplings of the data (blue curves) and, for
visualization only, we show binned means of the measured abundance-ratio
deviations (black histogram).
We emphasize that no binning is performed at any time in performing these fits.
The results shown in \figurename~\ref{fig:sinusoid-fits} are for abundance
deviations in $\ofe$, and just for three particular settings of the disk mass
parameters, \mratio (with all other mass model parameters fixed).
The data are bootstrapped prior to the construction of the abundance deviations,
because the abundance-deviation estimates depend on the data set in play, and
also the mass-model parameters.
Note that the sign of the $m=2$ term flips between the left and right panels of
\figurename~\ref{fig:sinusoid-fits}, indicating that the best-fit mass model
must be at an intermediate value of $\mratio$, close to but slightly larger than
the fiducial value.

\begin{figure}[!tp] 
  \begin{mdframed}
    \color{captiongray}
  \begin{center}
  \includegraphics[width=0.9\textwidth]{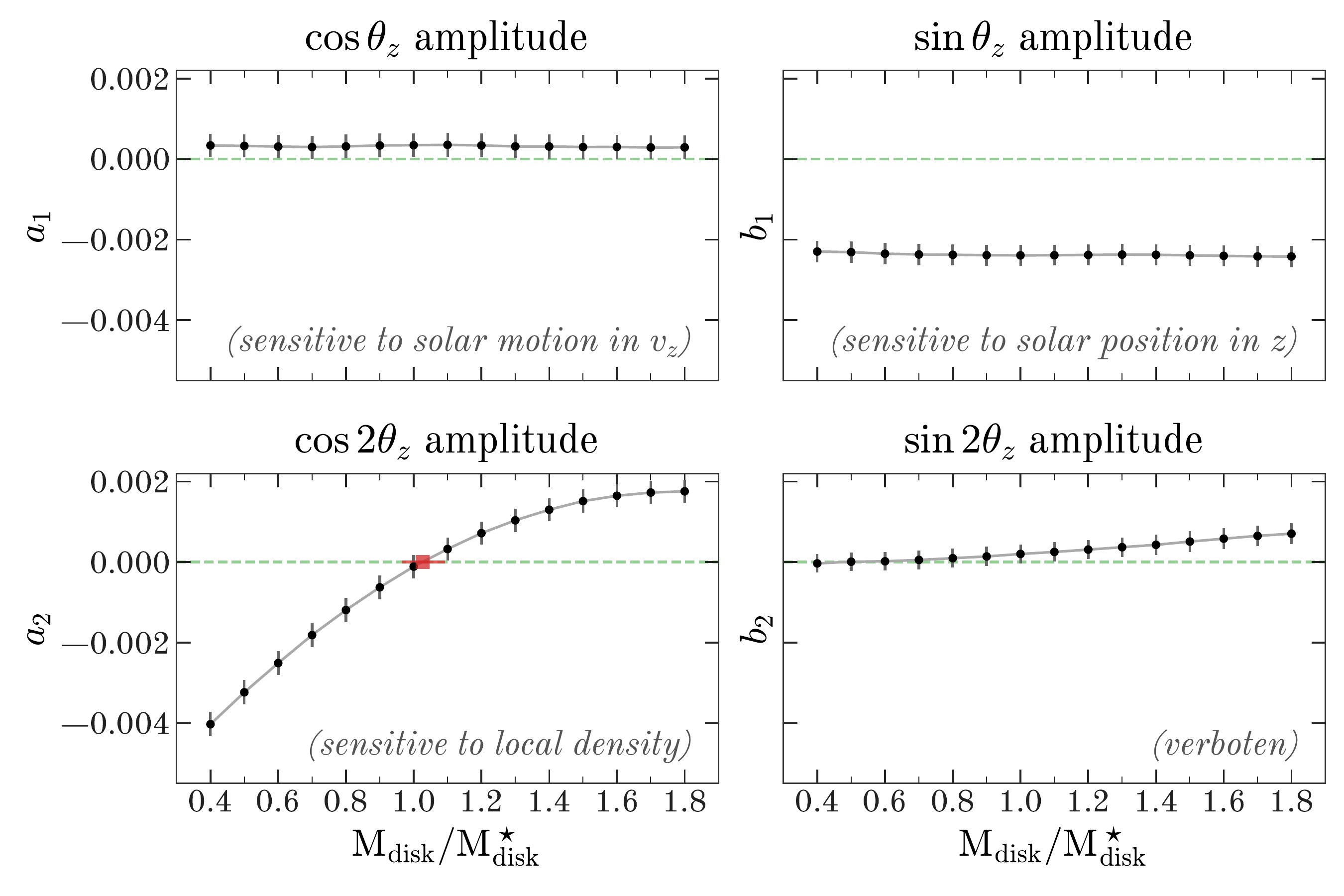}
  \end{center}
  \caption{%
    Parameters of the smooth sine-and-cosine fits to the dependence of \ofe\
    abundance deviation on vertical angle $\theta_z$ (the blue lines in
    \figurename~\ref{fig:sinusoid-fits}), as a function of the disk-mass
    parameter \mratio.
    The black markers show the values of the disk-mass parameter at which we
    performed the sine-and-cosine fits; the vertical error bars show the
    bootstrap uncertainties for each coefficient.
    The amplitude of the $\cos 2\theta_z$ term is the amplitude that is
    sensitive to the local density of the disk; it crosses zero when the model
    has the best-fit disk mass.
    The best-fit disk mass and its measurement uncertainty are shown as the
    square (red) marker and a horizontal error bar.
    The other terms shown have different dependencies on Milky Way parameters:
    The $\cos\theta_z$ would vary strongly if we varied the solar motion (the
    vertical component of the local standard of rest).
    The $\sin\theta_z$ term would vary strongly if we varied the location of the
    midplane of the disk.
    The $\sin 2\theta_z$ term cannot be non-zero; the fact that we find a
    non-zero value for this amplitude may suggest a weak violation of our model
    assumptions (see \sectionname~\ref{sec:inferences}).
    These figures show that we can precisely measure the disk mass.
  \label{fig:coeff-mdisk}
  }
  \end{mdframed}
\end{figure}

We repeat this fitting procedure for a grid of disk mass parameter values
$\mratio = 0.4$--$1.8$:
For each mass model (i.e., each setting of \mratio), we estimate the Fourier
coefficient parameters and uncertainties on the coefficients using 128 bootstrap
trials.
\figurename~\ref{fig:coeff-mdisk} shows the inferred coefficients for the
abundance ratio \ofe\ as a function of disk mass \mratio.
Conceptually, to turn this into a constraint on the disk mass, we then look for
the value of \mratio\ that minimizes the (absolute) value of $a_2$.
We obtain an estimate for the disk mass parameter \mratio\ and an associated
uncertainty by linearly interpolating the measurements and bootstrap error bars
onto the $a_2=0$ intercept.
Our best-fit disk mass and its uncertainty is shown as the square (red)
marker in \figurename~\ref{fig:coeff-mdisk} to emphasize that, though this is a
simple model and we only vary one parameter in this demonstration, the
measurement is encouragingly precise with a formal error bar of $\approx$7\% for
just a single element abundance.
We note that here we also apparently measure a finite value for the $\sin
2\theta_z$ term, which should not exist in the universe defined by our
assumptions (and is therefore labeled ``verboten'').
This implies that our assumptions seem to be lightly violated (the inferred
amplitude is only marginally significant), and we discuss this further in the
context of our assumptions below (\sectionname~\ref{sec:discussion}).

\subsection{Construction of a Mass Model Objective Function}
\label{sec:objective-func}

The demonstrations above show how we can compute best-fitting coefficients for
our model of the variations of the mean abundance deviations as a function of
vertical angle $\theta_z$.
To use these coefficients in a more general fitting procedure (i.e., to derive
constraints on the mass model parameters) we construct an objective function
that we can then optimize over the mass model parameters.
However, the upper panels of \figurename~\ref{fig:coeff-mdisk} show that, at all
values of the disk mass parameter \mratio, there is a finite $m=1$ amplitude for
both the $\cos$ and $\sin$ terms.
As noted before, these coefficients are sensitive to the solar motion and solar
position (in $z$), respectively.
We therefore consider four parameters in our objective function: the solar
position relative to the midplane \zsun, the solar $z$ velocity relative to the
local standard of rest \vzsun, the disk mass parameter \mratio, and the disk
scale height \hz\ \citep[the Miyamoto--Nagai scale height parameter, not an
exponential scale height;][]{Miyamoto:1975}.
For each setting of these parameters $\mratio, \hz, \zsun, \vzsun$, we compute
the Fourier coefficients (as described in the previous section) and minimize the
objective function
\begin{equation}
  f(a_1, b_1, a_2 \,;\, \mratio, \hz, \zsun, \vzsun) =
    a_1^2 + b_1^2 + a_2^2 \quad . \label{eq:objective}
\end{equation}
Note that here we ignore the constant term $c_0$ and the amplitude of the $\sin
2\theta_z$ term $b_2$\changes{, but we have verified that including these
coefficients in the objective function does not significantly (within a few per
cent) change the inferred parameters for any of the elements used here}.
\changes{This choice of objective function (\equationname~\ref{eq:objective}) is
somewhat arbitrary, but is sufficient for following the ``classical statistics''
approach we take in this article.
A Bayesian or likelihood-based formulation of the ideas described here could
instead construct a more physical model for $\Delta(\theta_z)$, and directly
fit for, or sample over (with priors), the mass model and Milky Way parameters;
we consider this out of scope for this work.}
Using the objective function above, we again perform 128 bootstrap trials per
element, and we perform the bootstrap resampling outside of the entire procedure
(so that each optimization is performed independently with a bootstrap sample).
We use Nelder-Mead optimization \citep{Gao:2012} to minimize our objective
function, as implemented in the \package{scipy} package \citep{Virtanen:2020}.

\begin{figure}[!tp]
  \begin{mdframed}
    \color{captiongray}
  \begin{center}
  \includegraphics[width=0.6\textwidth]{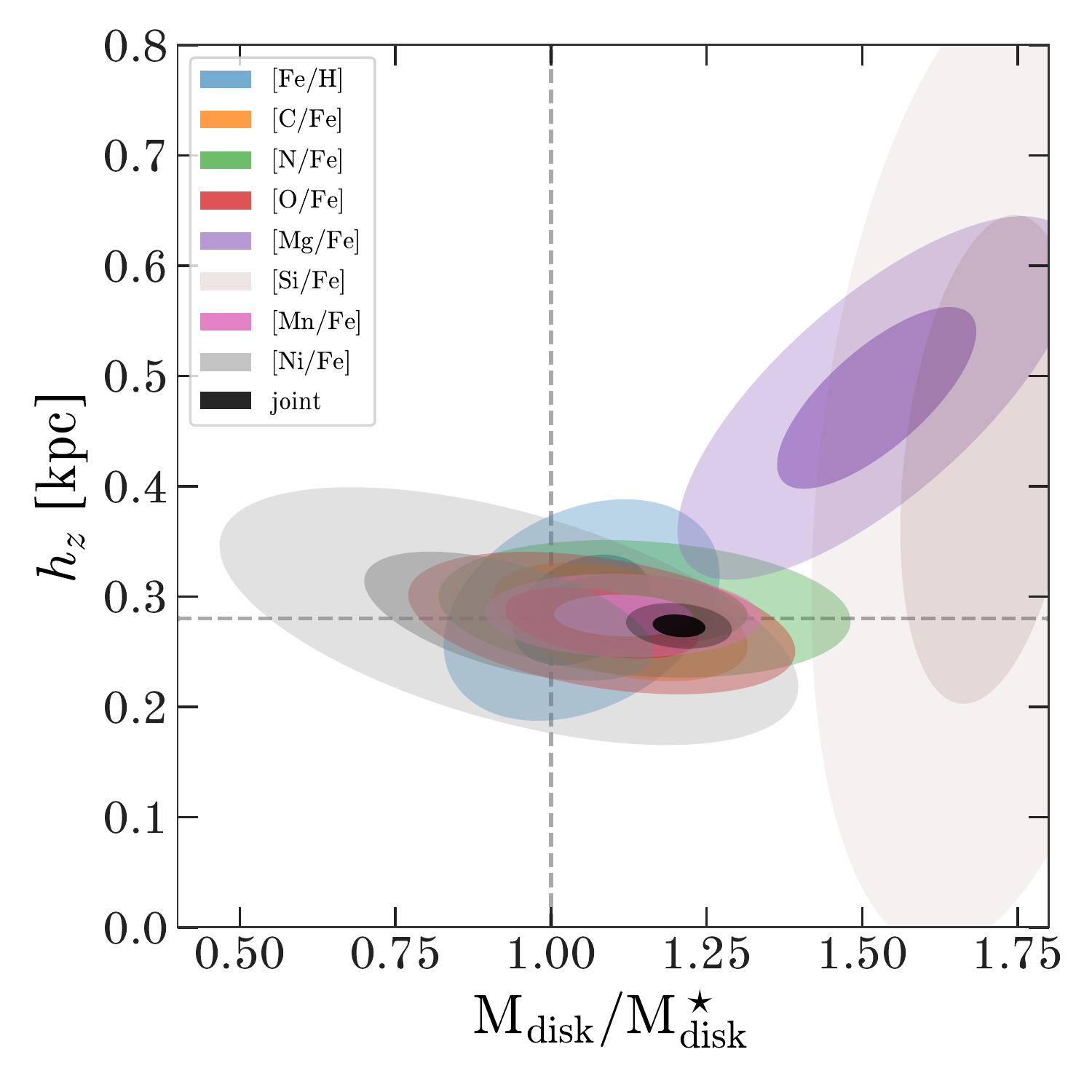}
  \end{center}
  \caption{%
    A summary of our joint constraints on the disk mass parameter \mratio\ and
    the scale height \hz\ for each element individually (colored ellipses) and
    the joint constraint (black).
    Each color shows one- and two-sigma error ellipses for each abundance ratio.
    The means and covariance matrices of each error ellipse are computed from
    optimizing the objective function (\equationname~\ref{eq:objective}) for 128
    bootstrap resamplings of the parent sample.
    Each individual element abundance ratio already provides fairly precise
    ($\lesssim10\%$) constraints on these parameters, but the joint constraint
    for these eight elements has a precision of $\approx$2.5\% for both of these
    parameters.
    The dashed vertical and horizontal lines indicate our fiducial values
    (\sectionname~\ref{sec:mw-model}).
  \label{fig:elem-ellipses-1}
  }
  \end{mdframed}
\end{figure}

\figurename~\ref{fig:elem-ellipses-1} shows a summary of our constraints on the
disk mass parameter and disk scale height for all of the elements shown in
\figurename~\ref{fig:zvz-grid} (colorful ellipses), and the joint constraint
from all of these elements combined (dark, black ellipse).
Here we show one- and two-sigma error ellipses (darker and lighter ellipses for
each color), with means and covariance matrices estimated from the optimization
results for the bootstrap resamplings of the data for each abundance ratio.

\begin{figure}[!tp]
  \begin{mdframed}
    \color{captiongray}
  \begin{center}
  \includegraphics[width=1\textwidth]{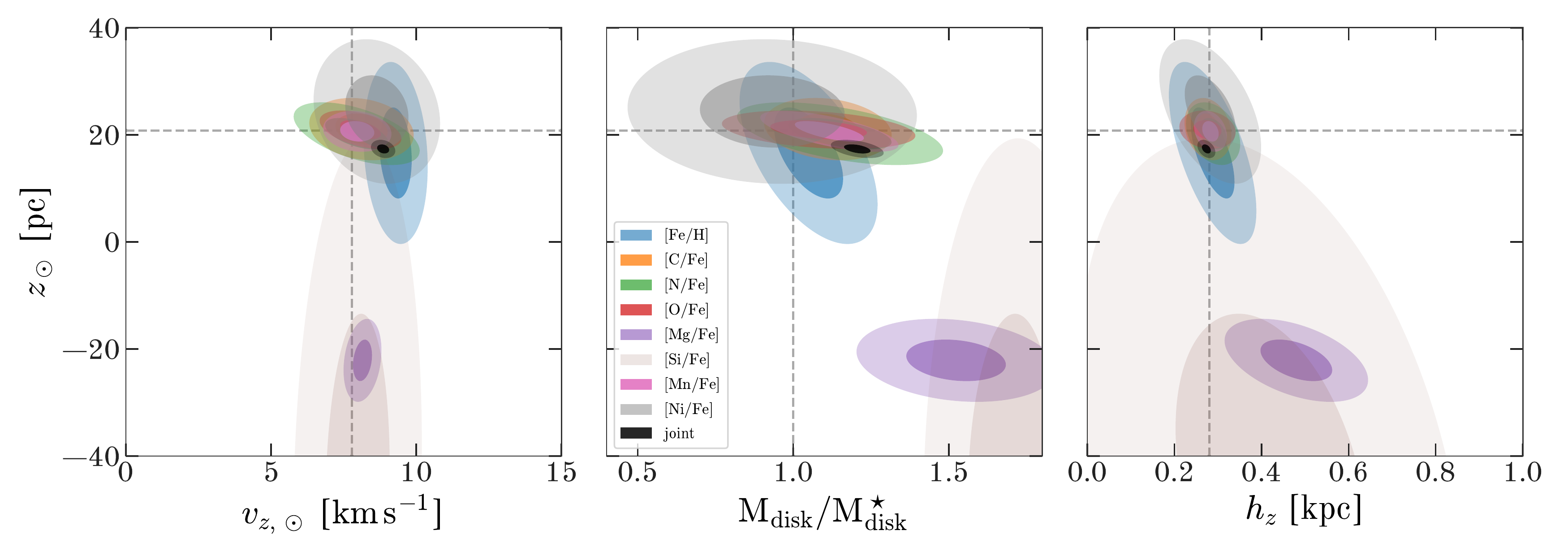}
  \end{center}
  \caption{%
  The same as \figurename~\ref{fig:elem-ellipses-1}, but showing other
  projections of our parameter space.
  Note that in the solar height above the midplane parameter \zsun, we find
  some significant disagreements between \abunratio{Mg}{Fe} and
  \abunratio{Si}{Fe} and the rest of the abundance ratios we consider.
  The dashed vertical and horizontal lines indicate our fiducial values
  (\sectionname~\ref{sec:mw-model}).
  \label{fig:elem-ellipses-2}
  }
  \end{mdframed}
\end{figure}

\figurename~\ref{fig:elem-ellipses-2} is similar to
\figurename~\ref{fig:elem-ellipses-1}, but shows our joint constraints on the
other projections of our parameter space:
This figure shows that a few of the element abundance ratios prefer a
significantly different solar position relative to the midplane:
While most elements are consistent with past measurements of the solar height of
$z_\odot \sim 20~\pc$ (\citealt{Bennett:2019} and \citealt{Bland-Hawthorn:2016}
and references therein), both \abunratio{Mg}{Fe} and \abunratio{Si}{Fe} suggest
that the sun is on the opposite side of the midplane!
While we do not have a simple explanation for this discrepancy, our constraints
from different element abundance ratios will effectively weight stars in
different parts of action space in ways that may amplify issues with our
assumptions.
In particular, there are known asymmetries in the vertical density and
kinematics of stars in the local disk, \changes{which affect stars with different
vertical actions with a different phase and amplitude.
In general, structures that are coherent in orbital phase (for example, the
known substructure in the local velocity distribution, e.g.,
\citealt{Hunt:2018}, and the vertical ``phase-space spiral'',
e.g., \citealt{Antoja:2018}).
}
If the abundance ratio gradients (with respect to vertical action) emphasize
stars with different vertical actions, we will in general find disagreements
between the results for different abundance ratios.

Our results are summarized in \tablename~\ref{tbl:results}, where we list the
joint constraints on our four parameters utilizing all eight abundance ratios
(center column), or excluding \abunratio{Mg}{Fe} and \abunratio{Si}{Fe} (right
column), which are clear outliers in their preferred solar position values.
We also include two derived quantities: the total disk mass $\mdisk$, and the
disk to halo mass ratio within the solar circle ${\rm M}_{\rm disk} / {\rm
M}_{\rm halo} (<8.1~\kpc)$, which we find is slightly larger than both  our
fiducial model $\approx 1.3$ \citep{gala} and the implied value for the
\texttt{MWPotential2014} \citep{Bovy:2015} of $\approx 1.7$.


\begin{table}[ht]
  \footnotesize
  \centering
  \begin{tabular}{c | c | c}
      \toprule
      Parameter & All eight abundance ratios & Excluding \abunratio{Mg}{Fe}, \abunratio{Si}{Fe}\\
      \midrule
      $\mratio$ & $1.21 \pm 0.04$ & $1.07\pm 0.05$\\
      $\hz$ & $0.27 \pm 0.01~\kpc$ & $0.28 \pm 0.01~\kpc$\\
      $\zsun$ & $16.9 \pm 0.8~\pc$ & $20.6 \pm 0.9~\pc$\\
      $\vzsun$ & $8.8 \pm 0.2~\kms$ & $8.4 \pm 0.3~\kms$\\
      \midrule
      $\mdisk$ &
        $7.89 \pm 0.26 \times 10^{10}~\Msun$ &
        $6.98 \pm 0.32 \times 10^{10}~\Msun$\\
      ${\rm M}_{\rm disk} / {\rm M}_{\rm halo} (<8.1~\kpc)$ &
        $2.11 \pm 0.21$ & $1.51 \pm 0.16$\\
      \bottomrule
  \end{tabular}
  \caption{A summary of our results from combining the constraints on the disk
  mass, scale height, solar position, and solar motion from eight independent
  element abundance ratios (center column). We also show joint results for all
  abundance ratios excluding \abunratio{Mg}{Fe} and \abunratio{Si}{Fe}, which
  are clear outliers in their preferred solar position parameter values.
  }
  \label{tbl:results}
\end{table}

\section{Discussion}
\label{sec:discussion}

We have defined and demonstrated a promising new method, \methodname, for using
measurements of both stellar kinematics and stellar surface abundances to infer
the underlying mass distribution of the Milky Way.
This method has promise in that it utilizes the high-dimensional stellar labels
that are measured by spectroscopic surveys to improve dynamical inferences.
Here we briefly review the method and its connection to other dynamical
inference methods, return to the assumptions that underpin \methodname\ and
discuss their applicability in the Milky Way, and discuss possible extensions or
reformulations of \methodname\ (e.g., in a Bayesian context).

\subsection{Fundamentals of \methodname}

The existence of element-abundance gradients in the Milky Way combined with the
assumption that stars do not (rapidly) change their surface abundances as
they orbit leads to the core concept of \methodname:
Stellar abundance ratios can depend on the three invariant actions, but they
cannot depend on the conjugate angles (or phases).
An implication of this is that gradients of stellar abundances with respect to
any 6-dimensional phase space coordinates will be locally tangent to the 3-tori
defined by the surfaces of constant actions at any position in phase space.
The 6-dimensional phase space is foliated by a complete set of orbital
3-tori, each of which is specified by the three actions.
Therefore, globally, each of these orbital tori will be a level 3-surface (in
the 6-dimensional phase space) of all moments or statistics of the
element-abundance distribution.

From an information theory perspective, this implies that any predictive model
for the element abundances of stars that depends on both actions and angles will
not predict the stellar abundances more precisely than a model that only depends
on the actions.
That is, in the 6-dimensional phase space, only the action coordinates provide
information about the element abundances.
In this context, we expect that the precision with which a mass model can be
constrained is better (i.e., the uncertainties on parameters are smaller) when
the abundance gradients are stronger.
We also expect that the precision will increase as the width or dispersion of
the local (in phase space) abundance distribution gets smaller.
This dispersion will have contributions from the intrinsic distribution of
abundances, which is related to both star formation and dynamical mixing, and
also from observational uncertainties.
An important consideration here is that inferences using \methodname\ will not
always continue to improve as the individual abundance measurements improve:
At some point, the finite scatter of the (kinematically-local) abundance
distribution will cause the improvements to saturate.
However, the precision will increase as the number of stars increases,
especially as the stars cover more of the range of possible conjugate angles.

In this \documentname, in our toy implementation, we have focused primarily on
fitting a model of the means of (logarithmic) abundance ratios of stars.
But \methodname\ is much more general:
Any statistical moments of any statistics of any invariant stellar labels would
also be allowed.
The method also does not assume that the element abundances are uniquely or even
specifically predicted by the actions.
This method only requires that there are gradients in some moments of the
abundance distribution.
Generically, the abundance ratio distributions at any point in action space are
expected to have (and are observed to have) substantial dispersion.

\subsection{Connection to Other Methods}


Like Jeans modeling, Schwarzschild modeling, or distribution function modeling,
\methodname\ provides a method for measuring the underlying acceleration field
traced by stars for which we only have access to instantaneous measurements of
phase-space coordinates.
However, unlike Jeans modeling, \methodname\ \emph{does not} require
separability or assumptions about symmetries of the acceleration field, nor does
it require accurately measuring second moments (or spatial derivatives of second
moments) of the velocity distribution.
We choose in this \documentname\ to focus on the vertical kinematics of stars,
but we do not assume that motion in $z$ is decoupled from the radial kinematics
(i.e., the orbits plotted in \figurename~\ref{fig:zvz-ofe} show finite-width
projections onto the vertical phase space).
We therefore expect \methodname\ to be more stable and less restrictive than
Jeans methods.

Another significant advantage of \methodname\ over other standard dynamical
inference methods is that it does not require knowledge of the survey selection
function, provided that the selection is made in position space (or phase space)
and does not act strongly on the element-abundance ratios that enter into the
inference.
This condition is sufficiently satisfied for the \apogee\ data used here, and
will generally be satisfied for other spectroscopic surveys of stars in the
Galaxy \citep[e.g., \acronym{GALAH};][]{Martell:2017, Buder:2018}.

In some ways, \methodname\ is related to the concept of ``extended distribution
functions'' \citep[EDFs; e.g.,][]{Sanders:2015,Das:2016}, in which the dynamical
distribution function in action space $p(\vec{J})$ (sometimes written as
$f(\vec{J})$) is extended into a distribution function in actions $\vec{J}$,
abundances $\vec{X}$, and (perhaps) other (near-) invariants to make a joint
distribution function $p(\vec{J},\vec{X})$.
So far, EDFs have been primarily used as tools to study the intrinsic properties
of the stellar tracer density distributions and to constrain models of diffusive
dynamical phenomena (e.g., radial migration).
That is, EDFs have been used in a mode where the Galactic mass model is fixed
and the tracer kinematics are used to constrain the intrinsic chemical and
kinematic structure of the tracer populations, but they have not yet been used
to infer the mass model for the Galaxy.
While is is possible, in principle, to use EDFs to simultaneously constrain the
mass model, most applications would require precise knowledge of the survey
selection functions (for any data used) and would therefore be challenging to
implement with existing survey data.

\methodname\ is also closely related to the concept of ``mono-abundance
populations'' \citep[MAPs; e.g.,][]{Bovy:2012, Bovy:2013, Bovy:2016,
Mackereth:2020}, which have been used to study the structure of the galaxy
divided into sub-populations that each have similar element abundances.
In these applications, analyses are typically carried out independently for each
MAP and therefore do not make use of any gradients in the element abundances
with respect to the kinematics.
Studies that make use of MAPs can therefore be seen as factorizing a conditional
distribution $p(\vec{J}\given\vec{X})$ out of the extended distribution function
$p(\vec{J},\vec{X})$ (although this is not precisely how they were formulated or
described).

In contrast EDFs or MAPs, \methodname\ can be seen as factorizing out (and
inferring) the conditional distribution $p(\vec{X}\given\vec{J})$ from the joint
distribution function $p(\vec{J}, \vec{X})$.
The great advantage of this factorization is that it does not require knowledge
of the survey selection function---it is \emph{conditioned} on the actions
$\vec{J}$, which are functions of the positions and velocities of
stars---provided that the selection acts in position- or action-space alone (and
not abundance-space).

The closest related dynamical methodology we are aware of is \textit{Orbital
Roulette} \citep{Beloborodov:2004}.
Like \methodname, \textit{Orbital Roulette} assumes that objects are not
observed at a special time and therefore that the distributions of phases for
all tracers in some observed population should be uniform.
This assumption is then turned into an estimator, which can be used to optimize
for the parameters of a mass model by making the distribution of phases approach
uniformity.
Our assumption is weaker: we only assume that the distribution of abundances is
invariant of phase.


\subsection{Returning to our Assumptions}

Like any dynamical inference method that seeks to measure properties of a mass
distribution from instantaneous measurements of tracer positions and velocities,
we have relied on a set of strong assumptions to formulate \methodname\
(\sectionname~\ref{sec:inferences}).
Already in our demonstrations above we see suggestions that these assumptions
are violated by the data in hand: For example, some elements prefer a finite
$\sin\,2\theta_z$ amplitude (as shown in the lower-right panel of
\figurename~\ref{fig:coeff-mdisk}), and the inferred solar position relative to
the Galactic midplane is significantly different for \abunratio{Mg}{Fe} and
\abunratio{Si}{Fe} as compared to the other elements (as shown in
\figurename~\ref{fig:elem-ellipses-2}).

For our particular toy application of \methodname\ to the \apogee\ sample
defined in this \documentname, and for more general applications using stars in
the Milky Way disk or halo, the ``phase mixed'' assumption is likely the most
strongly violated of our list of assumptions.
The Galactic disk is now known to show clear signatures of external
perturbations, intrinsic time-dependent phenomena, and significant phase-space
substructure \citep[e.g.,][]{Antoja:2018, Schonrich:2018, Hunt:2018,
Kamdar:2019, Monari:2019, Khanna:2019, Poggio:2020, Laporte:2020,
Coronado:2020}, all of which reflect the extent to which the stars in the disk
are not phase mixed.
In the case of vertical kinematics, the most striking illustration of this is
the recently-discovered phase-space spiral \citep{Antoja:2018}.
The spiral itself is only a $\approx$10\% perturbation to the local distribution
function \citep{Laporte:2019}, providing some limit to the amount with which
this will impact inferences with \methodname.
In principle, future implementations could make use of the spiral by explicitly
modeling the perturbation and its dependence on stellar kinematics and position
and stellar labels.
That is out of scope for this \documentname, but suggests possible
generalizations of \methodname\ that account for stellar populations that are
incompletely angle-mixed.
Beyond the Galactic disk in the stellar halo, as with Jeans modeling,
\methodname\ will be biased by the existence of unmixed substructures (streams,
shells, and merger remnants; e.g., \citealt{Grillmair:2016, Shipp:2018}).


Another key assumption that could lead to significant biases for certain stellar
samples is our assumption that the selection function only depends on
phase-space coordinates and not element abundances.
A different way of stating this assumption is that we require element abundances
to be measured and calibrated equivalently (in a statistical sense) at all
orbital phases.
In general, this assumption can be violated through a combination of subtle
effects: If a survey selection function implicitly depends on luminosity or
temperature, and there are systematic trends in the measured abundances with
surface gravity or temperature, the abundance distributions in any location of
phase-space will appear to be different from these effects alone.
For the \apogee\ sample used here, we have attempted to mitigate these issues by
selecting a small range of surface gravities (and therefore temperatures) along
the giant branch.
However, there are known systematic trends in \apogee\ between abundances and
stellar parameters (\citealt{Jonsson:2020, Wheeler:2020}, Eilers et al., in
prep.).
In principle, a joint dynamics and calibration model could be made that
simultaneously fits for the abundances as a function of conjugate angle and also
housekeeping data (like stellar surface gravity or spectrograph line-spread
function) that cause abundance systematics.

In our current implementation we have made use of transformations to
action-angle coordinates, which places implicit constraints on the mass models
we can consider.
In particular, any mass models must be integrable and not dominated by
resonances or chaotic regions, as we assume that the actions provide a
continuous foliation of the phase space.
At the precision with which we are operating, we do not think that this will be
a dominant issue in our analyses.
However, we are interested in the prospect of moving away from parametrizing the
mass model directly, as with a large enough sample of stars the abundance
gradients should be enough to image the orbital tori directly.

\changes{
As mentioned in \sectionname~\ref{sec:inferences}, dynamically, \methodname\
only strictly depends on the assumptions of phase-mixing and integrability
discussed above, which could be satisfied even in weakly time-dependent
potentials in which the actions remain adiabatically invariant.
However, there is substantial evidence for non-adiabatic time-dependence in the
dynamics of stars throughout the Galactic disk
\citep[e.g.,][]{Price-Whelan:2015, Antoja:2018, Xu:2020}.
In practice, time-dependence in the Milky Way could therefore cause the actions
to oscillate \citep[e.g.,][]{Penarrubia:2013, Buist:2015}, possibly with an
action-dependent amplitude \citep{Burger:2020}, which would lead to a change in
the mean abundance distribution as a function of actions.
Our implicit assumption that the potential is time-independent (through other
core assumptions) could therefore lead to biases in the inferred parameters of
the adopted time-independent models; This is a limitation of this work, and it
will be important to quantify the magnitude of the expected biases.
However, because our assumptions are slightly weaker, these biases may end up
being less severe than for other methods that strictly require time-independence
(e.g., Jeans modeling).
}

Much of the core methodology described here relies on the assumption that
stellar surface element abundances are invariants.
As mentioned above in our original list of assumptions, this is known to be
violated in detail due to gravitational settling, stellar evolution, or
planetary engulfment.
However, as long as the timescales over which surface abundances can change are
much shorter or much longer than the Galactic orbital timescales, the
distributions of element abundances at any orbital phase should not be affected
by these subtleties.
We therefore expect this to be inconsequential.

Finally, we have assumed all along that there are measurable gradients in
stellar abundances with respect to kinematics in the Galaxy or stellar
population of interest.
In the Milky Way, these are readily observed with existing spectroscopic
surveys, and our measurements of individual element abundances will only become
more numerous and precise in the coming years.

\subsection{Extensions of \methodname}

The implementation of \methodname\ used in this \documentname\ falls into the
category of classical (frequentist) statistics.
However, the concepts introduced here also lead naturally to probabilistic
generative model (Bayesian) versions of \methodname.
Bayesian implementations of this method at first look very different:
Instead of trying to minimize the dependence of abundance ratio statistics on
the conjugate angles (as we do here), Bayesian formulations would instead
involve building a flexible, forward model for the element-abundance
distribution as a function of the actions alone.
A non-intuitive aspect of this is that the angles would never explicitly appear
in the probabilistic model.
The Bayesian implementations would therefore look conceptually like extended
distribution functions \citep{Sanders:2015}, but while also varying the mass
models, or forward models of the distribution function
\citep[e.g.,][]{Magorrian:2014}, where element abundances or stellar labels are
treated as additional invariants.

In the implementation presented here, we use an abundance deviation that is
based on a nearest-neighbor interpolation in action space (see
\sectionname~\ref{sec:inferences} and \appendixname~\ref{app:knn-weights}).
That is a blunt tool; in principle, our results would improve if we instead
explicitly modeled the abundance distribution moments as functions of the
actions.
One option for this modeling would be to use a generative, physical model for
the star-formation history of the Galaxy (or sample), including the products of
nucleosynthesis and effects of chemical enrichment and stellar migration
\citep[similar to what is done in][]{Sanders:2015}.
Another option would be to use a flexible machine-learning method, such as a
Gaussian process.

We have parameterized the Milky Way mass model with
a very simple, few-component mass model.
Another option would be to instead parameterize the torus foliation of phase
space directly.
The representation of this foliation could be very general: Any foliation of
phase space with closed 3-tori is, in principle, allowed by dynamics.
A direct reconstruction of the foliation could then be interpreted in terms of
the force law, and hence the potential or mass model.
That would be a more data-driven approach---with far fewer assumptions---than
what has been executed here.

We used the means of
a set of hand-chosen abundance-ratio deviations
as our invariants for the inferences.
There are many other choices we could have made.
For example, we could have searched for maximally informative labels from among
the abundance ratios or functional combinations thereof. We could have used
moments other than the mean. And we could have used non-abundance labels, such
as stellar ages, angular momenta, or even binary-companion properties.
We could even have used the \emph{other actions $J_R, J_\phi$}
as invariant labels for the vertical-angle $\theta_z$ fits.
That is an interesting thought for future work, but the actions can only be used
as labels here if the survey selection function is precisely known and accounted
for. That condition obviates one of the principal advantages of
\methodname\ over other methods.

Lastly, here we have focused on the vertical kinematics of stars through the
vertical action $J_z$ and angle $\theta_z$, but in principle this
methodology---and any Bayesian extensions---would also work in all three actions
and angles.
In practice, the classical statistics approach would not work with the azimuthal
angle $\theta_\phi$ or action $J_\phi$ because our sample spans a small range of
$\theta_\phi$ (local to $\approx 2~\kpc$ around the Sun).
However, we could have equally used the radial action $J_R$ and angle $\theta_R$
here for demonstrations.




\section{Conclusions}
\label{sec:conclusions}

We have presented a new method---\methodname---for inferring the mass model (or
acceleration field) traced by a phase-mixed stellar distribution.
\methodname\ is novel in that it provides a way of using measurements of element
abundance ratios and other stellar labels to improve the precision of dynamical
inferences.
This method is also more statistically robust than traditional methods (for
example, Jeans modeling) in that it does not require measuring second moments of
the stellar velocity distribution or knowing the spatial (or phase-space) survey
selection function.
This method does, however, depend on the existence of element-abundance
gradients with respect to kinematics and our ability to measure these gradients:
The stars on different orbits (i.e., with different actions) must have---on
average or statistically---different compositions.
Fortunately these gradients are ubiquitous and measurable in the Milky Way (and
most galaxies; see \figurename~\ref{fig:zvz-grid}).

We have outlined and implemented a classical-statistics approach to \methodname\
and applied this to the vertical kinematics of a subsample of red giant branch
stars from the \apogee\ surveys.
The fundamental concept of this method is that stars do not change their
abundances appreciably over timescales comparable to Galactic orbital
timescales.
With that, and under the assumption that the stellar populations in the Galactic
disk are phase-mixed, this implies that the stellar abundance distribution at
any place in phase-space must be independent of phase or conjugate angles and
can only be a function of dynamical invariants, like orbital actions.
In our current implementation, we explicitly try to find a setting of the
gravitational potential parameters disk mass $\mdisk$ and scale height $\hz$
(and reference frame parameters, the solar position and velocity in $z$) that
minimize the dependence of the abundance distribution with vertical angle
$\theta_z$ (see \figurename~\ref{fig:sinusoid-fits}).
Turning this into an objective function (\sectionname~\ref{sec:objective-func}),
we then optimize over the four Milky Way parameters using eight independent
element abundance ratios (see \figurename~\ref{fig:elem-ellipses-1}).
Individually, the constraints are already encouragingly precise, but combining
the different abundance ratios provides joint constraints on the four parameters
that are precise to a few percent (see \tablename~\ref{tbl:results}).

Ongoing and near future spectroscopic surveys (e.g., SDSS-V, GALAH, DESI, 4MOST,
WEAVE) will provide samples of stars that are a factor of $\approx$10--100
times larger than that used in this work.
Combining these stellar data with kinematic measurements from upcoming data
releases from the \gaia\ mission will enable extremely precise constraints on
the Galactic mass distribution using \methodname.

\acknowledgments
It is a pleasure to thank
  Jo Bovy (Toronto),
  Anna-Christina Eilers (MIT),
  Jos\'e G. Fern\'andez-Trincado (U de Atacama),
  Suroor S. Gandhi (NYU),
  Matt Shetrone (UCO/Lick),
  David Spergel (Flatiron),
  Eugene Vasiliev (Cambridge),
  Adam Wheeler (Columbia),
  the Dynamics and Astronomical Data groups at the Flatiron Institute,
  and the Galaxy group at the MPIA
for valuable discussions and input.
We thank the anonymous referee for valuable, constructive feedback that
improved this article.
This research was conducted in part at the Aspen Center for Physics,
which is supported by National Science Foundation grant \acronym{PHY-1607611}.
K.V.J.'s contributions were enabled by NSF grant AST-1715582.
S.H. is supported by an NSF Astronomy and Astrophysics Postdoctoral Fellowship
under award AST-1801940.
DAGH acknowledges support from the State Research Agency (AEI) of the Spanish
Ministry of Science, Innovation and Universities (MCIU) and the European
Regional Development Fund (FEDER) under grant AYA2017-88254-P.

Funding for the Sloan Digital Sky
Survey IV has been provided by the
Alfred P. Sloan Foundation, the U.S.
Department of Energy Office of
Science, and the Participating
Institutions.

SDSS-IV acknowledges support and
resources from the Center for High
Performance Computing  at the
University of Utah. The SDSS
website is www.sdss.org.

SDSS-IV is managed by the
Astrophysical Research Consortium
for the Participating Institutions
of the SDSS Collaboration including
the Brazilian Participation Group,
the Carnegie Institution for Science,
Carnegie Mellon University, Center for
Astrophysics | Harvard \&
Smithsonian, the Chilean Participation
Group, the French Participation Group,
Instituto de Astrof\'isica de
Canarias, The Johns Hopkins
University, Kavli Institute for the
Physics and Mathematics of the
Universe (IPMU) / University of
Tokyo, the Korean Participation Group,
Lawrence Berkeley National Laboratory,
Leibniz Institut f\"ur Astrophysik
Potsdam (AIP),  Max-Planck-Institut
f\"ur Astronomie (MPIA Heidelberg),
Max-Planck-Institut f\"ur
Astrophysik (MPA Garching),
Max-Planck-Institut f\"ur
Extraterrestrische Physik (MPE),
National Astronomical Observatories of
China, New Mexico State University,
New York University, University of
Notre Dame, Observat\'ario
Nacional / MCTI, The Ohio State
University, Pennsylvania State
University, Shanghai
Astronomical Observatory, United
Kingdom Participation Group,
Universidad Nacional Aut\'onoma
de M\'exico, University of Arizona,
University of Colorado Boulder,
University of Oxford, University of
Portsmouth, University of Utah,
University of Virginia, University
of Washington, University of
Wisconsin, Vanderbilt University,
and Yale University.

This work has made use of data from the European Space Agency (\acronym{ESA})
mission \gaia\ (\url{https://www.cosmos.esa.int/gaia}), processed by the \gaia\
Data Processing and Analysis Consortium (\acronym{DPAC},
\url{https://www.cosmos.esa.int/web/gaia/dpac/consortium}). Funding for the
\acronym{DPAC}
has been provided by national institutions, in particular the institutions
participating in the \gaia\ Multilateral Agreement.


\software{
  \package{Astropy} \citep{astropy, astropy:2018},
  \package{gala} \citep{gala},
  \package{IPython} \citep{ipython},
  \package{matplotlib} \citep{Hunter:2007},
  \package{numpy} \citep{numpy:2020},
  \package{schwimmbad} \citep{schwimmbad},
  \package{scipy} \citep{Virtanen:2020}.
}

\appendix

\section{Re-weighting the $K$ nearest neighbors to account for steep gradients}
\label{app:knn-weights}

In standard $K$-nearest-neighbor regression, the estimate of the
(let's say scalar) label $y_\ast$ for a vector test point
$\vec{x}_\ast$ is the na\"ive mean $\mean{y}$ of the training labels $y_k$ for the $K$
nearest training-set neighbors $k$ in the vector space $\vec{x}$:
\begin{equation}
  y_\ast \leftarrow \mean{y} \equiv \frac{1}{K}\,\sum_{k=1}^K y_k
  \quad .
\end{equation}
Issues arise when the test point $\vec{x}_\ast$ is near the edge of
the training set or at a location of strong gradients in the density
in the training set.
In these cases, the na\"ive mean position $\mean{\vec{x}}$ of the $K$ neighbors
\begin{equation}
  \mean{\vec{x}} \equiv \frac{1}{K}\,\sum_{k=1}^K \vec{x}_k
\end{equation}
will be substantially displaced from the test point $\vec{x}_\ast$, and
the na\"ive mean of the labels $y_k$ will not be appropriate for the
test position.

We can correct for this problem by replacing the na\"ive mean of the labels
with a more sophisticated weighted mean.
We begin by defining scalar displacements $\xi_k$ of the neighbors away
from the test point
\begin{equation}
  \xi_k \equiv (\vec{x}_k - \vec{x}_\ast)\cdot(\mean{\vec{x}} - \vec{x}_\ast)
  \quad ,
\end{equation}
which are the individual neighbor vector displacements
projected onto the displacement between
the mean position and the test position.
Then we fit (by linear least squares) a linear model of the form
\begin{equation}
  y_k = a + b\,\xi_k + \mbox{noise}
  \quad ,
\end{equation}
where $a$ is an intercept and $b$ is a slope.
The intercept $a$ is the linear
fit interpolated to the position $\vec{x}_\ast$ of the test point.
In detail because this linear fit is just a two-parameter least-square fit,
it has a simple closed form:
\begin{align}
  y_\ast & \leftarrow \hat{a} = \frac{\sum_{k=1}^K w_k\,y_k}{\sum_{k=1}^K w_k}
  \\
  w_k & \equiv \xi_k\,\sum_{j=1}^K \xi_j - \sum_{j=1}^K \xi_j^2
  \quad ,
\end{align}
where $\hat{a}$ is the least-squares estimate of the intercept $a$, and that
estimate is itself a non-trivial weighted sum of the data with weights $w_k$.
One implementation note:
In the (rare) edge case that the displacement
$\mean{\vec{x}}-\vec{x}_\ast$ underflows the
floating-point representation, the weighted mean should be replaced with the
na\"ive mean $\mean{y}$.

This new estimate $y_\ast=\hat{a}$ is much more accurate than the na\"ive mean $\mean{y}$
in the presence of gradients in the training set.
It is, in some sense, the first-order correction of na\"ive $K$-nearest
neighbors. It is the first in a series of corrections to account for non-trivial
training-set distributions.

\bibliographystyle{aasjournal}
\bibliography{stationary-tags}

\begin{thebibliography}{}
\expandafter\ifx\csname natexlab\endcsname\relax\def\natexlab#1{#1}\fi
\providecommand{\url}[1]{\href{#1}{#1}}
\providecommand{\dodoi}[1]{doi:~\href{http://doi.org/#1}{\nolinkurl{#1}}}
\providecommand{\doeprint}[1]{\href{http://ascl.net/#1}{\nolinkurl{http://ascl.net/#1}}}
\providecommand{\doarXiv}[1]{\href{https://arxiv.org/abs/#1}{\nolinkurl{https://arxiv.org/abs/#1}}}

\bibitem[{{Ahumada} {et~al.}(2020){Ahumada}, {Allende Prieto}, {Almeida},
  {Anders}, {Anderson}, {Andrews}, {Anguiano}, {Arcodia}, {Armengaud},
  {Aubert}, {Avila}, {Avila-Reese}, {Badenes}, {Balland }, {Barger},
  {Barrera-Ballesteros}, {Basu}, {Bautista}, {Beaton}, {Beers}, {Benavides},
  {Bender}, {Bernardi}, {Bershady}, {Beutler}, {Bidin}, {Bird}, {Bizyaev},
  {Blanc}, {Blanton}, {Boquien}, {Borissova}, {Bovy}, {Brand t}, {Brinkmann},
  {Brownstein}, {Bundy}, {Bureau}, {Burgasser}, {Burtin}, {Cano-D{\'\i}az},
  {Capasso}, {Cappellari}, {Carrera}, {Chabanier}, {Chaplin}, {Chapman},
  {Cherinka}, {Chiappini}, {Doohyun Choi}, {Chojnowski}, {Chung}, {Clerc},
  {Coffey}, {Comerford}, {Comparat}, {da Costa}, {Cousinou}, {Covey}, {Crane},
  {Cunha}, {da Silva Ilha}, {Dai}, {Damsted}, {Darling}, {Davidson}, {Davies},
  {Dawson}, {De}, {de la Macorra}, {De Lee}, {de Andrade Queiroz}, {Deconto
  Machado}, {de la Torre}, {Dell'Agli}, {du Mas des Bourboux},
  {Diamond-Stanic}, {Dillon}, {Donor}, {Drory}, {Duckworth}, {Dwelly},
  {Ebelke}, {Eftekharzadeh}, {Eigenbrot}, {Elsworth}, {Eracleous},
  {Erfanianfar}, {Escoffier}, {Fan}, {Farr}, {Fern{\'a}ndez-Trincado},
  {Feuillet}, {Finoguenov}, {Fofie}, {Fraser-McKelvie}, {Frinchaboy},
  {Fromenteau}, {Fu}, {Galbany}, {Garcia}, {Garc{\'\i}a-Hern{\'a}ndez}, {Garma
  Oehmichen}, {Ge}, {Geimba Maia}, {Geisler}, {Gelfand }, {Goddy},
  {Gonzalez-Perez}, {Grabowski}, {Green}, {Grier}, {Guo}, {Guy}, {Harding},
  {Hasselquist}, {Hawken}, {Hayes}, {Hearty}, {Hekker}, {Hogg}, {Holtzman},
  {Horta}, {Hou}, {Hsieh}, {Huber}, {Hunt}, {Ider Chitham}, {Imig}, {Jaber},
  {Jimenez Angel}, {Johnson}, {Jones}, {J{\"o}nsson}, {Jullo}, {Kim},
  {Kinemuchi}, {Kirkpatrick}, {Kite}, {Klaene}, {Kneib}, {Kollmeier}, {Kong},
  {Kounkel}, {Krishnarao}, {Lacerna}, {Lan}, {Lane}, {Law}, {Le Goff}, {Leung},
  {Lewis}, {Li}, {Lian}, {Lin}, {Long}, {Longa-Pe{\~n}a}, {Lundgren}, {Lyke},
  {Ted Mackereth}, {MacLeod}, {Majewski}, {Manchado}, {Maraston}, {Martini},
  {Masseron}, {Masters}, {Mathur}, {McDermid}, {Merloni}, {Merrifield},
  {M{\'e}sz{\'a}ros}, {Miglio}, {Minniti}, {Minsley}, {Miyaji}, {Mohammad},
  {Mosser}, {Mueller}, {Muna}, {Mu{\~n}oz-Guti{\'e}rrez}, {Myers}, {Nadathur},
  {Nair}, {Nandra}, {do Nascimento}, {Nevin}, {Newman}, {Nidever}, {Nitschelm},
  {Noterdaeme}, {O'Connell}, {Olmstead}, {Oravetz}, {Oravetz}, {Osorio},
  {Pace}, {Padilla}, {Palanque-Delabrouille}, {Palicio}, {Pan}, {Pan},
  {Parker}, {Paviot}, {Peirani}, {Pe{\~n}a Ram{\'r}ez}, {Penny}, {Percival},
  {Perez-Fournon}, {P{\'e}rez-R{\`a}fols}, {Petitjean}, {Pieri},
  {Pinsonneault}, {Poovelil}, {Povick}, {Prakash}, {Price-Whelan}, {Raddick},
  {Raichoor}, {Ray}, {Rembold}, {Rezaie}, {Riffel}, {Riffel}, {Rix}, {Robin},
  {Roman-Lopes}, {Rom{\'a}n-Z{\'u}{\~n}iga}, {Rose}, {Ross}, {Rossi}, {Rowland
  s}, {Rubin}, {Salvato}, {S{\'a}nchez}, {S{\'a}nchez-Menguiano},
  {S{\'a}nchez-Gallego}, {Sayres}, {Schaefer}, {Schiavon}, {Schimoia},
  {Schlafly}, {Schlegel}, {Schneider}, {Schultheis}, {Schwope}, {Seo},
  {Serenelli}, {Shafieloo}, {Shamsi}, {Shao}, {Shen}, {Shetrone}, {Shirley},
  {Silva Aguirre}, {Simon}, {Skrutskie}, {Slosar}, {Smethurst}, {Sobeck},
  {Sodi}, {Souto}, {Stark}, {Stassun}, {Steinmetz}, {Stello}, {Stermer},
  {Storchi-Bergmann}, {Streblyanska}, {Stringfellow}, {Stutz}, {Su{\'a}rez},
  {Sun}, {Taghizadeh-Popp}, {Talbot}, {Tayar}, {Thakar}, {Theriault}, {Thomas},
  {Thomas}, {Tinker}, {Tojeiro}, {Toledo}, {Tremonti}, {Troup}, {Tuttle},
  {Unda-Sanzana}, {Valentini}, {Vargas-Gonz{\'a}lez}, {Vargas-Maga{\~n}a},
  {V{\'a}zquez-Mata}, {Vivek}, {Wake}, {Wang}, {Weaver}, {Weijmans}, {Wild},
  {Wilson}, {Wilson}, {Wolthuis}, {Wood-Vasey}, {Yan}, {Yang}, {Y{\`e}che},
  {Zamora}, {Zarrouk}, {Zasowski}, {Zhang}, {Zhao}, {Zhao}, {Zheng}, {Zheng},
  {Zhu}, \& {Zou}}]{DR16}
{Ahumada}, R., {Allende Prieto}, C., {Almeida}, A., {et~al.} 2020, \apjs, 249,
  3, \dodoi{10.3847/1538-4365/ab929e}

\bibitem[{{Antoja} {et~al.}(2018){Antoja}, {Helmi}, {Romero-G{\'o}mez}, {Katz},
  {Babusiaux}, {Drimmel}, {Evans}, {Figueras}, {Poggio}, {Reyl{\'e}}, {Robin},
  {Seabroke}, \& {Soubiran}}]{Antoja:2018}
{Antoja}, T., {Helmi}, A., {Romero-G{\'o}mez}, M., {et~al.} 2018, \nat, 561,
  360, \dodoi{10.1038/s41586-018-0510-7}

\bibitem[{{Astropy Collaboration} {et~al.}(2013){Astropy Collaboration},
  {Robitaille}, {Tollerud}, {Greenfield}, {Droettboom}, {Bray}, {Aldcroft},
  {Davis}, {Ginsburg}, {Price-Whelan}, {Kerzendorf}, {Conley}, {Crighton},
  {Barbary}, {Muna}, {Ferguson}, {Grollier}, {Parikh}, {Nair}, {Unther},
  {Deil}, {Woillez}, {Conseil}, {Kramer}, {Turner}, {Singer}, {Fox}, {Weaver},
  {Zabalza}, {Edwards}, {Azalee Bostroem}, {Burke}, {Casey}, {Crawford},
  {Dencheva}, {Ely}, {Jenness}, {Labrie}, {Lim}, {Pierfederici}, {Pontzen},
  {Ptak}, {Refsdal}, {Servillat}, \& {Streicher}}]{astropy}
{Astropy Collaboration}, {Robitaille}, T.~P., {Tollerud}, E.~J., {et~al.} 2013,
  \aap, 558, A33, \dodoi{10.1051/0004-6361/201322068}

\bibitem[{{Astropy Collaboration} {et~al.}(2018){Astropy Collaboration},
  {Price-Whelan}, {Sip{\'{o}}cz}, {G{\"u}nther}, {Lim}, {Crawford}, {Conseil},
  {Shupe}, {Craig}, {Dencheva}, {Ginsburg}, {VanderPlas}, {Bradley},
  {P{\'e}rez-Su{\'a}rez}, {de Val- Borro}, {Aldcroft}, {Cruz}, {Robitaille},
  {Tollerud}, {Ardelean}, {Babej}, {Bach}, {Bachetti}, {Bakanov}, {Bamford},
  {Barentsen}, {Barmby}, {Baumbach}, {Berry}, {Biscani}, {Boquien}, {Bostroem},
  {Bouma}, {Brammer}, {Bray}, {Breytenbach}, {Buddelmeijer}, {Burke},
  {Calderone}, {Cano Rodr{\'\i}guez}, {Cara}, {Cardoso}, {Cheedella}, {Copin},
  {Corrales}, {Crichton}, {D'Avella}, {Deil}, {Depagne}, {Dietrich}, {Donath},
  {Droettboom}, {Earl}, {Erben}, {Fabbro}, {Ferreira}, {Finethy}, {Fox},
  {Garrison}, {Gibbons}, {Goldstein}, {Gommers}, {Greco}, {Greenfield},
  {Groener}, {Grollier}, {Hagen}, {Hirst}, {Homeier}, {Horton}, {Hosseinzadeh},
  {Hu}, {Hunkeler}, {Ivezi{\'c}}, {Jain}, {Jenness}, {Kanarek}, {Kendrew},
  {Kern}, {Kerzendorf}, {Khvalko}, {King}, {Kirkby}, {Kulkarni}, {Kumar},
  {Lee}, {Lenz}, {Littlefair}, {Ma}, {Macleod}, {Mastropietro}, {McCully},
  {Montagnac}, {Morris}, {Mueller}, {Mumford}, {Muna}, {Murphy}, {Nelson},
  {Nguyen}, {Ninan}, {N{\"o}the}, {Ogaz}, {Oh}, {Parejko}, {Parley}, {Pascual},
  {Patil}, {Patil}, {Plunkett}, {Prochaska}, {Rastogi}, {Reddy Janga},
  {Sabater}, {Sakurikar}, {Seifert}, {Sherbert}, {Sherwood-Taylor}, {Shih},
  {Sick}, {Silbiger}, {Singanamalla}, {Singer}, {Sladen}, {Sooley},
  {Sornarajah}, {Streicher}, {Teuben}, {Thomas}, {Tremblay}, {Turner},
  {Terr{\'o}n}, {van Kerkwijk}, {de la Vega}, {Watkins}, {Weaver}, {Whitmore},
  {Woillez}, {Zabalza}, \& {Astropy Contributors}}]{astropy:2018}
{Astropy Collaboration}, {Price-Whelan}, A.~M., {Sip{\'{o}}cz}, B.~M., {et~al.}
  2018, \aj, 156, 123, \dodoi{10.3847/1538-3881/aabc4f}

\bibitem[{{Bahcall}(1984)}]{Bahcall:1984}
{Bahcall}, J.~N. 1984, \apj, 276, 169, \dodoi{10.1086/161601}

\bibitem[{{Bailer-Jones}(2015)}]{Bailer-Jones:2015}
{Bailer-Jones}, C. A.~L. 2015, \pasp, 127, 994, \dodoi{10.1086/683116}

\bibitem[{{Beloborodov} \& {Levin}(2004)}]{Beloborodov:2004}
{Beloborodov}, A.~M., \& {Levin}, Y. 2004, \apj, 613, 224,
  \dodoi{10.1086/422908}

\bibitem[{{Bennett} \& {Bovy}(2019)}]{Bennett:2019}
{Bennett}, M., \& {Bovy}, J. 2019, \mnras, 482, 1417,
  \dodoi{10.1093/mnras/sty2813}

\bibitem[{{Binney}(2012)}]{Binney:2012}
{Binney}, J. 2012, \mnras, 426, 1324, \dodoi{10.1111/j.1365-2966.2012.21757.x}

\bibitem[{{Binney} \& {Sanders}(2016)}]{Binney:2016}
{Binney}, J., \& {Sanders}, J.~L. 2016, Astronomische Nachrichten, 337, 939,
  \dodoi{10.1002/asna.201612403}

\bibitem[{{Binney} \& {Tremaine}(2008)}]{Binney:2008}
{Binney}, J., \& {Tremaine}, S. 2008, {Galactic Dynamics: Second Edition}

\bibitem[{{Binney} {et~al.}(2014){Binney}, {Burnett}, {Kordopatis},
  {Steinmetz}, {Gilmore}, {Bienayme}, {Bland-Hawthorn}, {Famaey}, {Grebel},
  {Helmi}, {Navarro}, {Parker}, {Reid}, {Seabroke}, {Siebert}, {Watson},
  {Williams}, {Wyse}, \& {Zwitter}}]{Binney:2014}
{Binney}, J., {Burnett}, B., {Kordopatis}, G., {et~al.} 2014, \mnras, 439,
  1231, \dodoi{10.1093/mnras/stt2367}

\bibitem[{{Bland-Hawthorn} \& {Gerhard}(2016)}]{Bland-Hawthorn:2016}
{Bland-Hawthorn}, J., \& {Gerhard}, O. 2016, \araa, 54, 529,
  \dodoi{10.1146/annurev-astro-081915-023441}

\bibitem[{{Blanton} {et~al.}(2017){Blanton}, {Bershady}, {Abolfathi},
  {Albareti}, {Allende Prieto}, {Almeida}, {Alonso-Garc{\'{\i}}a}, {Anders},
  {Anderson}, {Andrews}, \& et~al.}]{Blanton:2017}
{Blanton}, M.~R., {Bershady}, M.~A., {Abolfathi}, B., {et~al.} 2017, \aj, 154,
  28, \dodoi{10.3847/1538-3881/aa7567}

\bibitem[{{Bonaca} \& {Hogg}(2018)}]{Bonaca:2018}
{Bonaca}, A., \& {Hogg}, D.~W. 2018, \apj, 867, 101,
  \dodoi{10.3847/1538-4357/aae4da}

\bibitem[{{Bovy}(2015)}]{Bovy:2015}
{Bovy}, J. 2015, \apjs, 216, 29, \dodoi{10.1088/0067-0049/216/2/29}

\bibitem[{{Bovy} {et~al.}(2010){Bovy}, {Murray}, \& {Hogg}}]{BoHogg}
{Bovy}, J., {Murray}, I., \& {Hogg}, D.~W. 2010, \apj, 711, 1157,
  \dodoi{10.1088/0004-637X/711/2/1157}

\bibitem[{{Bovy} \& {Rix}(2013)}]{Bovy:2013}
{Bovy}, J., \& {Rix}, H.-W. 2013, \apj, 779, 115,
  \dodoi{10.1088/0004-637X/779/2/115}

\bibitem[{{Bovy} {et~al.}(2012){Bovy}, {Rix}, {Liu}, {Hogg}, {Beers}, \&
  {Lee}}]{Bovy:2012}
{Bovy}, J., {Rix}, H.-W., {Liu}, C., {et~al.} 2012, \apj, 753, 148,
  \dodoi{10.1088/0004-637X/753/2/148}

\bibitem[{{Bovy} {et~al.}(2016){Bovy}, {Rix}, {Schlafly}, {Nidever},
  {Holtzman}, {Shetrone}, \& {Beers}}]{Bovy:2016}
{Bovy}, J., {Rix}, H.-W., {Schlafly}, E.~F., {et~al.} 2016, \apj, 823, 30,
  \dodoi{10.3847/0004-637X/823/1/30}

\bibitem[{{Bovy} {et~al.}(2014){Bovy}, {Nidever}, {Rix}, {Girardi}, {Zasowski},
  {Chojnowski}, {Holtzman}, {Epstein}, {Frinchaboy}, {Hayden}, {Rodrigues},
  {Majewski}, {Johnson}, {Pinsonneault}, {Stello}, {Allende Prieto}, {Andrews},
  {Basu}, {Beers}, {Bizyaev}, {Burton}, {Chaplin}, {Cunha}, {Elsworth},
  {Garc{\'\i}a}, {Garc{\'\i}a-Her{\'n}andez}, {Garc{\'\i}a P{\'e}rez},
  {Hearty}, {Hekker}, {Kallinger}, {Kinemuchi}, {Koesterke},
  {M{\'e}sz{\'a}ros}, {Mosser}, {O'Connell}, {Oravetz}, {Pan}, {Robin},
  {Schiavon}, {Schneider}, {Schultheis}, {Serenelli}, {Shetrone}, {Silva
  Aguirre}, {Simmons}, {Skrutskie}, {Smith}, {Stassun}, {Weinberg}, {Wilson},
  \& {Zamora}}]{Bovy:2014}
{Bovy}, J., {Nidever}, D.~L., {Rix}, H.-W., {et~al.} 2014, \apj, 790, 127,
  \dodoi{10.1088/0004-637X/790/2/127}

\bibitem[{{Bowen} \& {Vaughan}(1973)}]{Bowen:1973}
{Bowen}, I.~S., \& {Vaughan}, A.~H., J. 1973, \ao, 12, 1430,
  \dodoi{10.1364/AO.12.001430}

\bibitem[{{Buch} {et~al.}(2019){Buch}, {Leung}, \& {Fan}}]{Buch:2019}
{Buch}, J., {Leung}, J. S.~C., \& {Fan}, J. 2019, \jcap, 2019, 026,
  \dodoi{10.1088/1475-7516/2019/04/026}

\bibitem[{{Buckley} \& {Peter}(2018)}]{Buckley:2018}
{Buckley}, M.~R., \& {Peter}, A. H.~G. 2018, \physrep, 761, 1,
  \dodoi{10.1016/j.physrep.2018.07.003}

\bibitem[{{Buder} {et~al.}(2018){Buder}, {Asplund}, {Duong}, {Kos}, {Lind},
  {Ness}, {Sharma}, {Bland -Hawthorn}, {Casey}, {de Silva}, {D'Orazi},
  {Freeman}, {Lewis}, {Lin}, {Martell}, {Schlesinger}, {Simpson}, {Zucker},
  {Zwitter}, {Amarsi}, {Anguiano}, {Carollo}, {Casagrande}, {{\v{C}}otar},
  {Cottrell}, {da Costa}, {Gao}, {Hayden}, {Horner}, {Ireland}, {Kafle},
  {Munari}, {Nataf}, {Nordlander}, {Stello}, {Ting}, {Traven}, {Watson},
  {Wittenmyer}, {Wyse}, {Yong}, {Zinn}, {{\v{Z}}erjal}, \& {Galah
  Collaboration}}]{Buder:2018}
{Buder}, S., {Asplund}, M., {Duong}, L., {et~al.} 2018, \mnras, 478, 4513,
  \dodoi{10.1093/mnras/sty1281}

\bibitem[{{Buist} \& {Helmi}(2015)}]{Buist:2015}
{Buist}, H. J.~T., \& {Helmi}, A. 2015, \aap, 584, A120,
  \dodoi{10.1051/0004-6361/201526203}

\bibitem[{{Bullock} \& {Boylan-Kolchin}(2017)}]{Bullock:2017}
{Bullock}, J.~S., \& {Boylan-Kolchin}, M. 2017, \araa, 55, 343,
  \dodoi{10.1146/annurev-astro-091916-055313}

\bibitem[{{Burger} {et~al.}(2020){Burger}, {Pe{\~n}arrubia}, \&
  {Zavala}}]{Burger:2020}
{Burger}, J.~D., {Pe{\~n}arrubia}, J., \& {Zavala}, J. 2020, arXiv e-prints,
  arXiv:2012.00737.
\newblock \doarXiv{2012.00737}

\bibitem[{{Coronado} {et~al.}(2020){Coronado}, {Rix}, {Trick}, {El-Badry},
  {Rybizki}, \& {Xiang}}]{Coronado:2020}
{Coronado}, J., {Rix}, H.-W., {Trick}, W.~H., {et~al.} 2020, \mnras, 495, 4098,
  \dodoi{10.1093/mnras/staa1358}

\bibitem[{{Das} \& {Binney}(2016)}]{Das:2016}
{Das}, P., \& {Binney}, J. 2016, \mnras, 460, 1725,
  \dodoi{10.1093/mnras/stw744}

\bibitem[{{Deng} {et~al.}(2012){Deng}, {Newberg}, {Liu}, {Carlin}, {Beers},
  {Chen}, {Chen}, {Christlieb}, {Grillmair}, {Guhathakurta}, {Han}, {Hou},
  {Lee}, {L{\'e}pine}, {Li}, {Liu}, {Pan}, {Sellwood}, {Wang}, {Wang}, {Yang},
  {Yanny}, {Zhang}, {Zhang}, {Zheng}, \& {Zhu}}]{Deng:2012}
{Deng}, L.-C., {Newberg}, H.~J., {Liu}, C., {et~al.} 2012, Research in
  Astronomy and Astrophysics, 12, 735, \dodoi{10.1088/1674-4527/12/7/003}

\bibitem[{{Drimmel} \& {Poggio}(2018)}]{Drimmel:2018}
{Drimmel}, R., \& {Poggio}, E. 2018, Research Notes of the American
  Astronomical Society, 2, 210, \dodoi{10.3847/2515-5172/aaef8b}

\bibitem[{{Eilers} {et~al.}(2020){Eilers}, {Hogg}, {Rix}, {Frankel}, {Hunt},
  {Fouvry}, \& {Buck}}]{Eilers:2020}
{Eilers}, A.-C., {Hogg}, D.~W., {Rix}, H.-W., {et~al.} 2020, arXiv e-prints,
  arXiv:2003.01132.
\newblock \doarXiv{2003.01132}

\bibitem[{{Eilers} {et~al.}(2019){Eilers}, {Hogg}, {Rix}, \&
  {Ness}}]{Eilers:2019}
{Eilers}, A.-C., {Hogg}, D.~W., {Rix}, H.-W., \& {Ness}, M.~K. 2019, \apj, 871,
  120, \dodoi{10.3847/1538-4357/aaf648}

\bibitem[{{Eisenstein} {et~al.}(2011){Eisenstein}, {Weinberg}, {Agol},
  {Aihara}, {Allende Prieto}, {Anderson}, {Arns}, {Aubourg}, {Bailey},
  {Balbinot}, {Barkhouser}, {Beers}, {Berlind}, {Bickerton}, {Bizyaev},
  {Blanton}, {Bochanski}, {Bolton}, {Bosman}, {Bovy}, {Brandt}, {Breslauer},
  {Brewington}, {Brinkmann}, {Brown}, {Brownstein}, {Burger}, {Busca},
  {Campbell}, {Cargile}, {Carithers}, {Carlberg}, {Carr}, {Chang}, {Chen},
  {Chiappini}, {Comparat}, {Connolly}, {Cortes}, {Croft}, {Cunha}, {da Costa},
  {Davenport}, {Dawson}, {De Lee}, {Porto de Mello}, {de Simoni}, {Dean},
  {Dhital}, {Ealet}, {Ebelke}, {Edmondson}, {Eiting}, {Escoffier}, {Esposito},
  {Evans}, {Fan}, {Femen{\'\i}a Castell{\'a}}, {Dutra Ferreira}, {Fitzgerald},
  {Fleming}, {Font-Ribera}, {Ford}, {Frinchaboy}, {Garc{\'\i}a P{\'e}rez},
  {Gaudi}, {Ge}, {Ghezzi}, {Gillespie}, {Gilmore}, {Girardi}, {Gott}, {Gould},
  {Grebel}, {Gunn}, {Hamilton}, {Harding}, {Harris}, {Hawley}, {Hearty},
  {Hennawi}, {Gonz{\'a}lez Hern{\'a}ndez}, {Ho}, {Hogg}, {Holtzman},
  {Honscheid}, {Inada}, {Ivans}, {Jiang}, {Jiang}, {Johnson}, {Jordan},
  {Jordan}, {Kauffmann}, {Kazin}, {Kirkby}, {Klaene}, {Knapp}, {Kneib},
  {Kochanek}, {Koesterke}, {Kollmeier}, {Kron}, {Lampeitl}, {Lang}, {Lawler},
  {Le Goff}, {Lee}, {Lee}, {Leisenring}, {Lin}, {Liu}, {Long}, {Loomis},
  {Lucatello}, {Lundgren}, {Lupton}, {Ma}, {Ma}, {MacDonald}, {Mack},
  {Mahadevan}, {Maia}, {Majewski}, {Makler}, {Malanushenko}, {Malanushenko},
  {Mand elbaum}, {Maraston}, {Margala}, {Maseman}, {Masters}, {McBride},
  {McDonald}, {McGreer}, {McMahon}, {Mena Requejo}, {M{\'e}nard},
  {Miralda-Escud{\'e}}, {Morrison}, {Mullally}, {Muna}, {Murayama}, {Myers},
  {Naugle}, {Neto}, {Nguyen}, {Nichol}, {Nidever}, {O'Connell}, {Ogando},
  {Olmstead}, {Oravetz}, {Padmanabhan}, {Paegert}, {Palanque-Delabrouille},
  {Pan}, {Pandey}, {Parejko}, {P{\^a}ris}, {Pellegrini}, {Pepper}, {Percival},
  {Petitjean}, {Pfaffenberger}, {Pforr}, {Phleps}, {Pichon}, {Pieri}, {Prada},
  {Price-Whelan}, {Raddick}, {Ramos}, {Reid}, {Reyle}, {Rich}, {Richards},
  {Rieke}, {Rieke}, {Rix}, {Robin}, {Rocha-Pinto}, {Rockosi}, {Roe},
  {Rollinde}, {Ross}, {Ross}, {Rossetto}, {S{\'a}nchez}, {Santiago}, {Sayres},
  {Schiavon}, {Schlegel}, {Schlesinger}, {Schmidt}, {Schneider}, {Sellgren},
  {Shelden}, {Sheldon}, {Shetrone}, {Shu}, {Silverman}, {Simmerer}, {Simmons},
  {Sivarani}, {Skrutskie}, {Slosar}, {Smee}, {Smith}, {Snedden}, {Stassun},
  {Steele}, {Steinmetz}, {Stockett}, {Stollberg}, {Strauss}, {Szalay},
  {Tanaka}, {Thakar}, {Thomas}, {Tinker}, {Tofflemire}, {Tojeiro}, {Tremonti},
  {Vargas Maga{\~n}a}, {Verde}, {Vogt}, {Wake}, {Wan}, {Wang}, {Weaver},
  {White}, {White}, {Wilson}, {Wisniewski}, {Wood-Vasey}, {Yanny}, {Yasuda},
  {Y{\`e}che}, {York}, {Young}, {Zasowski}, {Zehavi}, \&
  {Zhao}}]{Eisenstein:2011}
{Eisenstein}, D.~J., {Weinberg}, D.~H., {Agol}, E., {et~al.} 2011, \aj, 142,
  72, \dodoi{10.1088/0004-6256/142/3/72}

\bibitem[{{Evans} {et~al.}(2009){Evans}, {An}, \& {Walker}}]{Evans:2009}
{Evans}, N.~W., {An}, J., \& {Walker}, M.~G. 2009, \mnras, 393, L50,
  \dodoi{10.1111/j.1745-3933.2008.00596.x}

\bibitem[{{Eyre} \& {Binney}(2011)}]{Eyre:2011}
{Eyre}, A., \& {Binney}, J. 2011, \mnras, 413, 1852,
  \dodoi{10.1111/j.1365-2966.2011.18270.x}

\bibitem[{{Freeman} \& {Bland-Hawthorn}(2002)}]{Freeman:2002}
{Freeman}, K., \& {Bland-Hawthorn}, J. 2002, \araa, 40, 487,
  \dodoi{10.1146/annurev.astro.40.060401.093840}

\bibitem[{{Gaia Collaboration} {et~al.}(2016){Gaia Collaboration}, {Prusti},
  {de Bruijne}, {Brown}, {Vallenari}, {Babusiaux}, {Bailer-Jones}, {Bastian},
  {Biermann}, {Evans}, {Eyer}, {Jansen}, {Jordi}, {Klioner}, {Lammers},
  {Lindegren}, {Luri}, {Mignard}, {Milligan}, {Panem}, {Poinsignon},
  {Pourbaix}, {Randich}, {Sarri}, {Sartoretti}, {Siddiqui}, {Soubiran},
  {Valette}, {van Leeuwen}, {Walton}, {Aerts}, {Arenou}, {Cropper}, {Drimmel},
  {H{\o}g}, {Katz}, {Lattanzi}, {O'Mullane}, {Grebel}, {Holland}, {Huc},
  {Passot}, {Bramante}, {Cacciari}, {Casta{\~n}eda}, {Chaoul}, {Cheek}, {De
  Angeli}, {Fabricius}, {Guerra}, {Hern{\'a}ndez}, {Jean-Antoine-Piccolo},
  {Masana}, {Messineo}, {Mowlavi}, {Nienartowicz}, {Ord{\'o}{\~n}ez-Blanco},
  {Panuzzo}, {Portell}, {Richards}, {Riello}, {Seabroke}, {Tanga},
  {Th{\'e}venin}, {Torra}, {Els}, {Gracia-Abril}, {Comoretto},
  {Garcia-Reinaldos}, {Lock}, {Mercier}, {Altmann}, {Andrae}, {Astraatmadja},
  {Bellas-Velidis}, {Benson}, {Berthier}, {Blomme}, {Busso}, {Carry},
  {Cellino}, {Clementini}, {Cowell}, {Creevey}, {Cuypers}, {Davidson}, {De
  Ridder}, {de Torres}, {Delchambre}, {Dell'Oro}, {Ducourant}, {Fr{\'e}mat},
  {Garc{\'\i}a-Torres}, {Gosset}, {Halbwachs}, {Hambly}, {Harrison}, {Hauser},
  {Hestroffer}, {Hodgkin}, {Huckle}, {Hutton}, {Jasniewicz}, {Jordan},
  {Kontizas}, {Korn}, {Lanzafame}, {Manteiga}, {Moitinho}, {Muinonen},
  {Osinde}, {Pancino}, {Pauwels}, {Petit}, {Recio-Blanco}, {Robin}, {Sarro},
  {Siopis}, {Smith}, {Smith}, {Sozzetti}, {Thuillot}, {van Reeven}, {Viala},
  {Abbas}, {Abreu Aramburu}, {Accart}, {Aguado}, {Allan}, {Allasia},
  {Altavilla}, {{\'A}lvarez}, {Alves}, {Anderson}, {Andrei}, {Anglada Varela},
  {Antiche}, {Antoja}, {Ant{\'o}n}, {Arcay}, {Atzei}, {Ayache}, {Bach},
  {Baker}, {Balaguer-N{\'u}{\~n}ez}, {Barache}, {Barata}, {Barbier}, {Barblan},
  {Baroni}, {Barrado y Navascu{\'e}s}, {Barros}, {Barstow}, {Becciani},
  {Bellazzini}, {Bellei}, {Bello Garc{\'\i}a}, {Belokurov}, {Bendjoya},
  {Berihuete}, {Bianchi}, {Bienaym{\'e}}, {Billebaud}, {Blagorodnova},
  {Blanco-Cuaresma}, {Boch}, {Bombrun}, {Borrachero}, {Bouquillon}, {Bourda},
  {Bouy}, {Bragaglia}, {Breddels}, {Brouillet}, {Br{\"u}semeister},
  {Bucciarelli}, {Budnik}, {Burgess}, {Burgon}, {Burlacu}, {Busonero}, {Buzzi},
  {Caffau}, {Cambras}, {Campbell}, {Cancelliere}, {Cantat-Gaudin}, {Carlucci},
  {Carrasco}, {Castellani}, {Charlot}, {Charnas}, {Charvet}, {Chassat},
  {Chiavassa}, {Clotet}, {Cocozza}, {Collins}, {Collins}, {Costigan}, {Crifo},
  {Cross}, {Crosta}, {Crowley}, {Dafonte}, {Damerdji}, {Dapergolas}, {David},
  {David}, {De Cat}, {de Felice}, {de Laverny}, {De Luise}, {De March}, {de
  Martino}, {de Souza}, {Debosscher}, {del Pozo}, {Delbo}, {Delgado},
  {Delgado}, {di Marco}, {Di Matteo}, {Diakite}, {Distefano}, {Dolding}, {Dos
  Anjos}, {Drazinos}, {Dur{\'a}n}, {Dzigan}, {Ecale}, {Edvardsson}, {Enke},
  {Erdmann}, {Escolar}, {Espina}, {Evans}, {Eynard Bontemps}, {Fabre},
  {Fabrizio}, {Faigler}, {Falc{\~a}o}, {Farr{\`a}s Casas}, {Faye}, {Federici},
  {Fedorets}, {Fern{\'a}ndez-Hern{\'a}ndez}, {Fernique}, {Fienga}, {Figueras},
  {Filippi}, {Findeisen}, {Fonti}, {Fouesneau}, {Fraile}, {Fraser}, {Fuchs},
  {Furnell}, {Gai}, {Galleti}, {Galluccio}, {Garabato}, {Garc{\'\i}a-Sedano},
  {Gar{\'e}}, {Garofalo}, {Garralda}, {Gavras}, {Gerssen}, {Geyer}, {Gilmore},
  {Girona}, {Giuffrida}, {Gomes}, {Gonz{\'a}lez-Marcos},
  {Gonz{\'a}lez-N{\'u}{\~n}ez}, {Gonz{\'a}lez-Vidal}, {Granvik}, {Guerrier},
  {Guillout}, {Guiraud}, {G{\'u}rpide}, {Guti{\'e}rrez-S{\'a}nchez}, {Guy},
  {Haigron}, {Hatzidimitriou}, {Haywood}, {Heiter}, {Helmi}, {Hobbs},
  {Hofmann}, {Holl}, {Holland }, {Hunt}, {Hypki}, {Icardi}, {Irwin}, {Jevardat
  de Fombelle}, {Jofr{\'e}}, {Jonker}, {Jorissen}, {Julbe}, {Karampelas},
  {Kochoska}, {Kohley}, {Kolenberg}, {Kontizas}, {Koposov}, {Kordopatis},
  {Koubsky}, {Kowalczyk}, {Krone-Martins}, {Kudryashova}, {Kull}, {Bachchan},
  {Lacoste-Seris}, {Lanza}, {Lavigne}, {Le Poncin-Lafitte}, {Lebreton},
  {Lebzelter}, {Leccia}, {Leclerc}, {Lecoeur-Taibi}, {Lemaitre}, {Lenhardt},
  {Leroux}, {Liao}, {Licata}, {Lindstr{\o}m}, {Lister}, {Livanou}, {Lobel},
  {L{\"o}ffler}, {L{\'o}pez}, {Lopez-Lozano}, {Lorenz}, {Loureiro},
  {MacDonald}, {Magalh{\~a}es Fernandes}, {Managau}, {Mann}, {Mantelet},
  {Marchal}, {Marchant}, {Marconi}, {Marie}, {Marinoni}, {Marrese},
  {Marschalk{\'o}}, {Marshall}, {Mart{\'\i}n-Fleitas}, {Martino}, {Mary},
  {Matijevi{\v{c}}}, {Mazeh}, {McMillan}, {Messina}, {Mestre}, {Michalik},
  {Millar}, {Miranda}, {Molina}, {Molinaro}, {Molinaro}, {Moln{\'a}r},
  {Moniez}, {Montegriffo}, {Monteiro}, {Mor}, {Mora}, {Morbidelli}, {Morel},
  {Morgenthaler}, {Morley}, {Morris}, {Mulone}, {Muraveva}, {Musella},
  {Narbonne}, {Nelemans}, {Nicastro}, {Noval}, {Ord{\'e}novic},
  {Ordieres-Mer{\'e}}, {Osborne}, {Pagani}, {Pagano}, {Pailler}, {Palacin},
  {Palaversa}, {Parsons}, {Paulsen}, {Pecoraro}, {Pedrosa}, {Pentik{\"a}inen},
  {Pereira}, {Pichon}, {Piersimoni}, {Pineau}, {Plachy}, {Plum}, {Poujoulet},
  {Pr{\v{s}}a}, {Pulone}, {Ragaini}, {Rago}, {Rambaux}, {Ramos-Lerate},
  {Ranalli}, {Rauw}, {Read}, {Regibo}, {Renk}, {Reyl{\'e}}, {Ribeiro},
  {Rimoldini}, {Ripepi}, {Riva}, {Rixon}, {Roelens}, {Romero-G{\'o}mez},
  {Rowell}, {Royer}, {Rudolph}, {Ruiz-Dern}, {Sadowski}, {Sagrist{\`a}
  Sell{\'e}s}, {Sahlmann}, {Salgado}, {Salguero}, {Sarasso}, {Savietto},
  {Schnorhk}, {Schultheis}, {Sciacca}, {Segol}, {Segovia}, {Segransan},
  {Serpell}, {Shih}, {Smareglia}, {Smart}, {Smith}, {Solano}, {Solitro},
  {Sordo}, {Soria Nieto}, {Souchay}, {Spagna}, {Spoto}, {Stampa}, {Steele},
  {Steidelm{\"u}ller}, {Stephenson}, {Stoev}, {Suess}, {S{\"u}veges}, {Surdej},
  {Szabados}, {Szegedi-Elek}, {Tapiador}, {Taris}, {Tauran}, {Taylor},
  {Teixeira}, {Terrett}, {Tingley}, {Trager}, {Turon}, {Ulla}, {Utrilla},
  {Valentini}, {van Elteren}, {Van Hemelryck}, {van Leeuwen}, {Varadi},
  {Vecchiato}, {Veljanoski}, {Via}, {Vicente}, {Vogt}, {Voss}, {Votruba},
  {Voutsinas}, {Walmsley}, {Weiler}, {Weingrill}, {Werner}, {Wevers},
  {Whitehead}, {Wyrzykowski}, {Yoldas}, {{\v{Z}}erjal}, {Zucker}, {Zurbach},
  {Zwitter}, {Alecu}, {Allen}, {Allende Prieto}, {Amorim},
  {Anglada-Escud{\'e}}, {Arsenijevic}, {Azaz}, {Balm}, {Beck}, {Bernstein},
  {Bigot}, {Bijaoui}, {Blasco}, {Bonfigli}, {Bono}, {Boudreault}, {Bressan},
  {Brown}, {Brunet}, {Bunclark}, {Buonanno}, {Butkevich}, {Carret}, {Carrion},
  {Chemin}, {Ch{\'e}reau}, {Corcione}, {Darmigny}, {de Boer}, {de Teodoro}, {de
  Zeeuw}, {Delle Luche}, {Domingues}, {Dubath}, {Fodor}, {Fr{\'e}zouls},
  {Fries}, {Fustes}, {Fyfe}, {Gallardo}, {Gallegos}, {Gardiol}, {Gebran},
  {Gomboc}, {G{\'o}mez}, {Grux}, {Gueguen}, {Heyrovsky}, {Hoar}, {Iannicola},
  {Isasi Parache}, {Janotto}, {Joliet}, {Jonckheere}, {Keil}, {Kim},
  {Klagyivik}, {Klar}, {Knude}, {Kochukhov}, {Kolka}, {Kos}, {Kutka}, {Lainey},
  {LeBouquin}, {Liu}, {Loreggia}, {Makarov}, {Marseille}, {Martayan},
  {Martinez-Rubi}, {Massart}, {Meynadier}, {Mignot}, {Munari}, {Nguyen},
  {Nordlander}, {Ocvirk}, {O'Flaherty}, {Olias Sanz}, {Ortiz}, {Osorio},
  {Oszkiewicz}, {Ouzounis}, {Palmer}, {Park}, {Pasquato}, {Peltzer}, {Peralta},
  {P{\'e}turaud}, {Pieniluoma}, {Pigozzi}, {Poels}, {Prat}, {Prod'homme},
  {Raison}, {Rebordao}, {Risquez}, {Rocca-Volmerange}, {Rosen}, {Ruiz-Fuertes},
  {Russo}, {Sembay}, {Serraller Vizcaino}, {Short}, {Siebert}, {Silva},
  {Sinachopoulos}, {Slezak}, {Soffel}, {Sosnowska}, {Strai{\v{z}}ys}, {ter
  Linden}, {Terrell}, {Theil}, {Tiede}, {Troisi}, {Tsalmantza}, {Tur},
  {Vaccari}, {Vachier}, {Valles}, {Van Hamme}, {Veltz}, {Virtanen}, {Wallut},
  {Wichmann}, {Wilkinson}, {Ziaeepour}, \&
  {Zschocke}}]{Gaia-Collaboration:2016}
{Gaia Collaboration}, {Prusti}, T., {de Bruijne}, J.~H.~J., {et~al.} 2016,
  \aap, 595, A1, \dodoi{10.1051/0004-6361/201629272}

\bibitem[{{Gaia Collaboration} {et~al.}(2018{\natexlab{a}}){Gaia
  Collaboration}, {Brown}, {Vallenari}, {Prusti}, {de Bruijne}, {Babusiaux},
  {Bailer-Jones}, {Biermann}, {Evans}, {Eyer}, {Jansen}, {Jordi}, {Klioner},
  {Lammers}, {Lindegren}, {Luri}, {Mignard}, {Panem}, {Pourbaix}, {Randich},
  {Sartoretti}, {Siddiqui}, {Soubiran}, {van Leeuwen}, {Walton}, {Arenou},
  {Bastian}, {Cropper}, {Drimmel}, {Katz}, {Lattanzi}, {Bakker}, {Cacciari},
  {Casta{\~n}eda}, {Chaoul}, {Cheek}, {De Angeli}, {Fabricius}, {Guerra},
  {Holl}, {Masana}, {Messineo}, {Mowlavi}, {Nienartowicz}, {Panuzzo},
  {Portell}, {Riello}, {Seabroke}, {Tanga}, {Th{\'e}venin}, {Gracia-Abril},
  {Comoretto}, {Garcia-Reinaldos}, {Teyssier}, {Altmann}, {Andrae}, {Audard},
  {Bellas-Velidis}, {Benson}, {Berthier}, {Blomme}, {Burgess}, {Busso},
  {Carry}, {Cellino}, {Clementini}, {Clotet}, {Creevey}, {Davidson}, {De
  Ridder}, {Delchambre}, {Dell'Oro}, {Ducourant},
  {Fern{\'a}ndez-Hern{\'a}ndez}, {Fouesneau}, {Fr{\'e}mat}, {Galluccio},
  {Garc{\'\i}a-Torres}, {Gonz{\'a}lez-N{\'u}{\~n}ez}, {Gonz{\'a}lez-Vidal},
  {Gosset}, {Guy}, {Halbwachs}, {Hambly}, {Harrison}, {Hern{\'a}ndez},
  {Hestroffer}, {Hodgkin}, {Hutton}, {Jasniewicz}, {Jean-Antoine-Piccolo},
  {Jordan}, {Korn}, {Krone-Martins}, {Lanzafame}, {Lebzelter}, {L{\"o}ffler},
  {Manteiga}, {Marrese}, {Mart{\'\i}n-Fleitas}, {Moitinho}, {Mora}, {Muinonen},
  {Osinde}, {Pancino}, {Pauwels}, {Petit}, {Recio-Blanco}, {Richards},
  {Rimoldini}, {Robin}, {Sarro}, {Siopis}, {Smith}, {Sozzetti}, {S{\"u}veges},
  {Torra}, {van Reeven}, {Abbas}, {Abreu Aramburu}, {Accart}, {Aerts},
  {Altavilla}, {{\'A}lvarez}, {Alvarez}, {Alves}, {Anderson}, {Andrei},
  {Anglada Varela}, {Antiche}, {Antoja}, {Arcay}, {Astraatmadja}, {Bach},
  {Baker}, {Balaguer-N{\'u}{\~n}ez}, {Balm}, {Barache}, {Barata}, {Barbato},
  {Barblan}, {Barklem}, {Barrado}, {Barros}, {Barstow}, {Bartholom{\'e}
  Mu{\~n}oz}, {Bassilana}, {Becciani}, {Bellazzini}, {Berihuete}, {Bertone},
  {Bianchi}, {Bienaym{\'e}}, {Blanco-Cuaresma}, {Boch}, {Boeche}, {Bombrun},
  {Borrachero}, {Bossini}, {Bouquillon}, {Bourda}, {Bragaglia}, {Bramante},
  {Breddels}, {Bressan}, {Brouillet}, {Br{\"u}semeister}, {Brugaletta},
  {Bucciarelli}, {Burlacu}, {Busonero}, {Butkevich}, {Buzzi}, {Caffau},
  {Cancelliere}, {Cannizzaro}, {Cantat-Gaudin}, {Carballo}, {Carlucci},
  {Carrasco}, {Casamiquela}, {Castellani}, {Castro-Ginard}, {Charlot},
  {Chemin}, {Chiavassa}, {Cocozza}, {Costigan}, {Cowell}, {Crifo}, {Crosta},
  {Crowley}, {Cuypers}, {Dafonte}, {Damerdji}, {Dapergolas}, {David}, {David},
  {de Laverny}, {De Luise}, {De March}, {de Martino}, {de Souza}, {de Torres},
  {Debosscher}, {del Pozo}, {Delbo}, {Delgado}, {Delgado}, {Di Matteo},
  {Diakite}, {Diener}, {Distefano}, {Dolding}, {Drazinos}, {Dur{\'a}n},
  {Edvardsson}, {Enke}, {Eriksson}, {Esquej}, {Eynard Bontemps}, {Fabre},
  {Fabrizio}, {Faigler}, {Falc{\~a}o}, {Farr{\`a}s Casas}, {Federici},
  {Fedorets}, {Fernique}, {Figueras}, {Filippi}, {Findeisen}, {Fonti},
  {Fraile}, {Fraser}, {Fr{\'e}zouls}, {Gai}, {Galleti}, {Garabato},
  {Garc{\'\i}a-Sedano}, {Garofalo}, {Garralda}, {Gavel}, {Gavras}, {Gerssen},
  {Geyer}, {Giacobbe}, {Gilmore}, {Girona}, {Giuffrida}, {Glass}, {Gomes},
  {Granvik}, {Gueguen}, {Guerrier}, {Guiraud}, {Guti{\'e}rrez-S{\'a}nchez},
  {Haigron}, {Hatzidimitriou}, {Hauser}, {Haywood}, {Heiter}, {Helmi}, {Heu},
  {Hilger}, {Hobbs}, {Hofmann}, {Holland}, {Huckle}, {Hypki}, {Icardi},
  {Jan{\ss}en}, {Jevardat de Fombelle}, {Jonker}, {Juh{\'a}sz}, {Julbe},
  {Karampelas}, {Kewley}, {Klar}, {Kochoska}, {Kohley}, {Kolenberg},
  {Kontizas}, {Kontizas}, {Koposov}, {Kordopatis}, {Kostrzewa-Rutkowska},
  {Koubsky}, {Lambert}, {Lanza}, {Lasne}, {Lavigne}, {Le Fustec}, {Le
  Poncin-Lafitte}, {Lebreton}, {Leccia}, {Leclerc}, {Lecoeur-Taibi},
  {Lenhardt}, {Leroux}, {Liao}, {Licata}, {Lindstr{\o}m}, {Lister}, {Livanou},
  {Lobel}, {L{\'o}pez}, {Managau}, {Mann}, {Mantelet}, {Marchal}, {Marchant},
  {Marconi}, {Marinoni}, {Marschalk{\'o}}, {Marshall}, {Martino}, {Marton},
  {Mary}, {Massari}, {Matijevi{\v{c}}}, {Mazeh}, {McMillan}, {Messina},
  {Michalik}, {Millar}, {Molina}, {Molinaro}, {Moln{\'a}r}, {Montegriffo},
  {Mor}, {Morbidelli}, {Morel}, {Morris}, {Mulone}, {Muraveva}, {Musella},
  {Nelemans}, {Nicastro}, {Noval}, {O'Mullane}, {Ord{\'e}novic},
  {Ord{\'o}{\~n}ez-Blanco}, {Osborne}, {Pagani}, {Pagano}, {Pailler},
  {Palacin}, {Palaversa}, {Panahi}, {Pawlak}, {Piersimoni}, {Pineau}, {Plachy},
  {Plum}, {Poggio}, {Poujoulet}, {Pr{\v{s}}a}, {Pulone}, {Racero}, {Ragaini},
  {Rambaux}, {Ramos-Lerate}, {Regibo}, {Reyl{\'e}}, {Riclet}, {Ripepi}, {Riva},
  {Rivard}, {Rixon}, {Roegiers}, {Roelens}, {Romero-G{\'o}mez}, {Rowell},
  {Royer}, {Ruiz-Dern}, {Sadowski}, {Sagrist{\`a} Sell{\'e}s}, {Sahlmann},
  {Salgado}, {Salguero}, {Sanna}, {Santana-Ros}, {Sarasso}, {Savietto},
  {Schultheis}, {Sciacca}, {Segol}, {Segovia}, {S{\'e}gransan}, {Shih},
  {Siltala}, {Silva}, {Smart}, {Smith}, {Solano}, {Solitro}, {Sordo}, {Soria
  Nieto}, {Souchay}, {Spagna}, {Spoto}, {Stampa}, {Steele},
  {Steidelm{\"u}ller}, {Stephenson}, {Stoev}, {Suess}, {Surdej}, {Szabados},
  {Szegedi-Elek}, {Tapiador}, {Taris}, {Tauran}, {Taylor}, {Teixeira},
  {Terrett}, {Teyssand ier}, {Thuillot}, {Titarenko}, {Torra Clotet}, {Turon},
  {Ulla}, {Utrilla}, {Uzzi}, {Vaillant}, {Valentini}, {Valette}, {van Elteren},
  {Van Hemelryck}, {van Leeuwen}, {Vaschetto}, {Vecchiato}, {Veljanoski},
  {Viala}, {Vicente}, {Vogt}, {von Essen}, {Voss}, {Votruba}, {Voutsinas},
  {Walmsley}, {Weiler}, {Wertz}, {Wevers}, {Wyrzykowski}, {Yoldas},
  {{\v{Z}}erjal}, {Ziaeepour}, {Zorec}, {Zschocke}, {Zucker}, {Zurbach}, \&
  {Zwitter}}]{Gaia-Collaboration:2018}
{Gaia Collaboration}, {Brown}, A.~G.~A., {Vallenari}, A., {et~al.}
  2018{\natexlab{a}}, \aap, 616, A1, \dodoi{10.1051/0004-6361/201833051}

\bibitem[{{Gaia Collaboration} {et~al.}(2018{\natexlab{b}}){Gaia
  Collaboration}, {Katz}, {Antoja}, {Romero-G{\'o}mez}, {Drimmel}, {Reyl{\'e}},
  {Seabroke}, {Soubiran}, {Babusiaux}, {Di Matteo}, {Figueras}, {Poggio},
  {Robin}, {Evans}, {Brown}, {Vallenari}, {Prusti}, {de Bruijne},
  {Bailer-Jones}, {Biermann}, {Eyer}, {Jansen}, {Jordi}, {Klioner}, {Lammers},
  {Lindegren}, {Luri}, {Mignard}, {Panem}, {Pourbaix}, {Randich}, {Sartoretti},
  {Siddiqui}, {van Leeuwen}, {Walton}, {Arenou}, {Bastian}, {Cropper},
  {Lattanzi}, {Bakker}, {Cacciari}, {Casta n}, {Chaoul}, {Cheek}, {De Angeli},
  {Fabricius}, {Guerra}, {Holl}, {Masana}, {Messineo}, {Mowlavi},
  {Nienartowicz}, {Panuzzo}, {Portell}, {Riello}, {Tanga}, {Th{\'e}venin},
  {Gracia-Abril}, {Comoretto}, {Garcia-Reinaldos}, {Teyssier}, {Altmann},
  {Andrae}, {Audard}, {Bellas-Velidis}, {Benson}, {Berthier}, {Blomme},
  {Burgess}, {Busso}, {Carry}, {Cellino}, {Clementini}, {Clotet}, {Creevey},
  {Davidson}, {De Ridder}, {Delchambre}, {Dell'Oro}, {Ducourant},
  {Fern{\'a}ndez-Hern{\'a}ndez}, {Fouesneau}, {Fr{\'e}mat}, {Galluccio},
  {Garc{\'\i}a-Torres}, {Gonz{\'a}lez-N{\'u}{\~n}ez}, {Gonz{\'a}lez-Vidal},
  {Gosset}, {Guy}, {Halbwachs}, {Hambly}, {Harrison}, {Hern{\'a}ndez},
  {Hestroffer}, {Hodgkin}, {Hutton}, {Jasniewicz}, {Jean-Antoine-Piccolo},
  {Jordan}, {Korn}, {Krone-Martins}, {Lanzafame}, {Lebzelter}, {L{\"o}ffler},
  {Manteiga}, {Marrese}, {Mart{\'\i}n-Fleitas}, {Moitinho}, {Mora}, {Muinonen},
  {Osinde}, {Pancino}, {Pauwels}, {Petit}, {Recio-Blanco}, {Richards},
  {Rimoldini}, {Sarro}, {Siopis}, {Smith}, {Sozzetti}, {S{\"u}veges}, {Torra},
  {van Reeven}, {Abbas}, {Abreu Aramburu}, {Accart}, {Aerts}, {Altavilla},
  {{\'A}lvarez}, {Alvarez}, {Alves}, {Anderson}, {Andrei}, {Anglada Varela},
  {Antiche}, {Arcay}, {Astraatmadja}, {Bach}, {Baker},
  {Balaguer-N{\'u}{\~n}ez}, {Balm}, {Barache}, {Barata}, {Barbato}, {Barblan},
  {Barklem}, {Barrado}, {Barros}, {Barstow}, {Bartholom{\'e} Mu{\~n}oz},
  {Bassilana}, {Becciani}, {Bellazzini}, {Berihuete}, {Bertone}, {Bianchi},
  {Bienaym{\'e}}, {Blanco-Cuaresma}, {Boch}, {Boeche}, {Bombrun}, {Borrachero},
  {Bossini}, {Bouquillon}, {Bourda}, {Bragaglia}, {Bramante}, {Breddels},
  {Bressan}, {Brouillet}, {Br{\"u}semeister}, {Brugaletta}, {Bucciarelli},
  {Burlacu}, {Busonero}, {Butkevich}, {Buzzi}, {Caffau}, {Cancelliere},
  {Cannizzaro}, {Cantat-Gaudin}, {Carballo}, {Carlucci}, {Carrasco},
  {Casamiquela}, {Castellani}, {Castro-Ginard}, {Charlot}, {Chemin},
  {Chiavassa}, {Cocozza}, {Costigan}, {Cowell}, {Crifo}, {Crosta}, {Crowley},
  {Cuypers}, {Dafonte}, {Damerdji}, {Dapergolas}, {David}, {David}, {de
  Laverny}, {De Luise}, {De March}, {de Souza}, {de Torres}, {Debosscher}, {del
  Pozo}, {Delbo}, {Delgado}, {Delgado}, {Diakite}, {Diener}, {Distefano},
  {Dolding}, {Drazinos}, {Dur{\'a}n}, {Edvardsson}, {Enke}, {Eriksson},
  {Esquej}, {Eynard Bontemps}, {Fabre}, {Fabrizio}, {Faigler}, {Falc a},
  {Farr{\`a}s Casas}, {Federici}, {Fedorets}, {Fernique}, {Filippi},
  {Findeisen}, {Fonti}, {Fraile}, {Fraser}, {Fr{\'e}zouls}, {Gai}, {Galleti},
  {Garabato}, {Garc{\'\i}a-Sedano}, {Garofalo}, {Garralda}, {Gavel}, {Gavras},
  {Gerssen}, {Geyer}, {Giacobbe}, {Gilmore}, {Girona}, {Giuffrida}, {Glass},
  {Gomes}, {Granvik}, {Gueguen}, {Guerrier}, {Guiraud}, {Guti{\'e}}, {Haigron},
  {Hatzidimitriou}, {Hauser}, {Haywood}, {Heiter}, {Helmi}, {Heu}, {Hilger},
  {Hobbs}, {Hofmann}, {Holland }, {Huckle}, {Hypki}, {Icardi}, {Jan{\ss}en},
  {Jevardat de Fombelle}, {Jonker}, {Juh{\'a}sz}, {Julbe}, {Karampelas},
  {Kewley}, {Klar}, {Kochoska}, {Kohley}, {Kolenberg}, {Kontizas}, {Kontizas},
  {Koposov}, {Kordopatis}, {Kostrzewa-Rutkowska}, {Koubsky}, {Lambert},
  {Lanza}, {Lasne}, {Lavigne}, {Le Fustec}, {Le Poncin-Lafitte}, {Lebreton},
  {Leccia}, {Leclerc}, {Lecoeur-Taibi}, {Lenhardt}, {Leroux}, {Liao}, {Licata},
  {Lindstr{\o}m}, {Lister}, {Livanou}, {Lobel}, {L{\'o}pez}, {Managau}, {Mann},
  {Mantelet}, {Marchal}, {Marchant}, {Marconi}, {Marinoni}, {Marschalk{\'o}},
  {Marshall}, {Martino}, {Marton}, {Mary}, {Massari}, {Matijevi{\v{c}}},
  {Mazeh}, {McMillan}, {Messina}, {Michalik}, {Millar}, {Molina}, {Molinaro},
  {Moln{\'a}r}, {Montegriffo}, {Mor}, {Morbidelli}, {Morel}, {Morris},
  {Mulone}, {Muraveva}, {Musella}, {Nelemans}, {Nicastro}, {Noval},
  {O'Mullane}, {Ord{\'e}novic}, {Ord{\'o}{\~n}ez-Blanco}, {Osborne}, {Pagani},
  {Pagano}, {Pailler}, {Palacin}, {Palaversa}, {Panahi}, {Pawlak},
  {Piersimoni}, {Pineau}, {Plachy}, {Plum}, {Poujoulet}, {Pr{\v{s}}a},
  {Pulone}, {Racero}, {Ragaini}, {Rambaux}, {Ramos-Lerate}, {Regibo}, {Riclet},
  {Ripepi}, {Riva}, {Rivard}, {Rixon}, {Roegiers}, {Roelens}, {Rowell},
  {Royer}, {Ruiz-Dern}, {Sadowski}, {Sagrist{\`a} Sell{\'e}s}, {Sahlmann},
  {Salgado}, {Salguero}, {Sanna}, {Santana-Ros}, {Sarasso}, {Savietto},
  {Schultheis}, {Sciacca}, {Segol}, {Segovia}, {S{\'e}gransan}, {Shih},
  {Siltala}, {Silva}, {Smart}, {Smith}, {Solano}, {Solitro}, {Sordo}, {Soria
  Nieto}, {Souchay}, {Spagna}, {Spoto}, {Stampa}, {Steele},
  {Steidelm{\"u}ller}, {Stephenson}, {Stoev}, {Suess}, {Surdej}, {Szabados},
  {Szegedi-Elek}, {Tapiador}, {Taris}, {Tauran}, {Taylor}, {Teixeira},
  {Terrett}, {Teyssand ier}, {Thuillot}, {Titarenko}, {Torra Clotet}, {Turon},
  {Ulla}, {Utrilla}, {Uzzi}, {Vaillant}, {Valentini}, {Valette}, {van Elteren},
  {Van Hemelryck}, {van Leeuwen}, {Vaschetto}, {Vecchiato}, {Veljanoski},
  {Viala}, {Vicente}, {Vogt}, {von Essen}, {Voss}, {Votruba}, {Voutsinas},
  {Walmsley}, {Weiler}, {Wertz}, {Wevers}, {Wyrzykowski}, {Yoldas},
  {{\v{Z}}erjal}, {Ziaeepour}, {Zorec}, {Zschocke}, {Zucker}, {Zurbach}, \&
  {Zwitter}}]{Gaia-Collaboration:2018b}
{Gaia Collaboration}, {Katz}, D., {Antoja}, T., {et~al.} 2018{\natexlab{b}},
  \aap, 616, A11, \dodoi{10.1051/0004-6361/201832865}

\bibitem[{Gao \& Han(2012)}]{Gao:2012}
Gao, F., \& Han, L. 2012, Computational Optimization and Applications, 51, 259,
  \dodoi{10.1007/s10589-010-9329-3}

\bibitem[{{Garc{\'{\i}}a P{\'e}rez} {et~al.}(2016){Garc{\'{\i}}a P{\'e}rez},
  {Allende Prieto}, {Holtzman}, {Shetrone}, {M{\'e}sz{\'a}ros}, {Bizyaev},
  {Carrera}, {Cunha}, {Garc{\'{\i}}a-Hern{\'a}ndez}, {Johnson}, {Majewski},
  {Nidever}, {Schiavon}, {Shane}, {Smith}, {Sobeck}, {Troup}, {Zamora},
  {Weinberg}, {Bovy}, {Eisenstein}, {Feuillet}, {Frinchaboy}, {Hayden},
  {Hearty}, {Nguyen}, {O'Connell}, {Pinsonneault}, {Wilson}, \&
  {Zasowski}}]{ASPCAP}
{Garc{\'{\i}}a P{\'e}rez}, A.~E., {Allende Prieto}, C., {Holtzman}, J.~A.,
  {et~al.} 2016, \aj, 151, 144, \dodoi{10.3847/0004-6256/151/6/144}

\bibitem[{{Gravity Collaboration} {et~al.}(2018){Gravity Collaboration},
  {Abuter}, {Amorim}, {Anugu}, {Baub{\"o}ck}, {Benisty}, {Berger}, {Blind},
  {Bonnet}, {Brandner}, {Buron}, {Collin}, {Chapron}, {Cl{\'e}net}, {Coud{\'e}
  Du Foresto}, {de Zeeuw}, {Deen}, {Delplancke-Str{\"o}bele}, {Dembet},
  {Dexter}, {Duvert}, {Eckart}, {Eisenhauer}, {Finger}, {F{\"o}rster
  Schreiber}, {F{\'e}dou}, {Garcia}, {Garcia Lopez}, {Gao}, {Gendron},
  {Genzel}, {Gillessen}, {Gordo}, {Habibi}, {Haubois}, {Haug}, {Hau{\ss}mann},
  {Henning}, {Hippler}, {Horrobin}, {Hubert}, {Hubin}, {Jimenez Rosales},
  {Jochum}, {Jocou}, {Kaufer}, {Kellner}, {Kendrew}, {Kervella}, {Kok},
  {Kulas}, {Lacour}, {Lapeyr{\`e}re}, {Lazareff}, {Le Bouquin}, {L{\'e}na},
  {Lippa}, {Lenzen}, {M{\'e}rand}, {M{\"u}ler}, {Neumann}, {Ott}, {Palanca},
  {Paumard}, {Pasquini}, {Perraut}, {Perrin}, {Pfuhl}, {Plewa}, {Rabien},
  {Ram{\'\i}rez}, {Ramos}, {Rau}, {Rodr{\'\i}guez-Coira}, {Rohloff}, {Rousset},
  {Sanchez-Bermudez}, {Scheithauer}, {Sch{\"o}ller}, {Schuler}, {Spyromilio},
  {Straub}, {Straubmeier}, {Sturm}, {Tacconi}, {Tristram}, {Vincent}, {von
  Fellenberg}, {Wank}, {Waisberg}, {Widmann}, {Wieprecht}, {Wiest},
  {Wiezorrek}, {Woillez}, {Yazici}, {Ziegler}, \& {Zins}}]{Gravity:2018}
{Gravity Collaboration}, {Abuter}, R., {Amorim}, A., {et~al.} 2018, \aap, 615,
  L15, \dodoi{10.1051/0004-6361/201833718}

\bibitem[{{Grillmair} \& {Carlin}(2016)}]{Grillmair:2016}
{Grillmair}, C.~J., \& {Carlin}, J.~L. 2016, {Stellar Streams and Clouds in the
  Galactic Halo}, ed. H.~J. {Newberg} \& J.~L. {Carlin}, Vol. 420, 87,
  \dodoi{10.1007/978-3-319-19336-6_4}

\bibitem[{{Gunn} {et~al.}(2006){Gunn}, {Siegmund}, {Mannery}, {Owen}, {Hull},
  {Leger}, {Carey}, {Knapp}, {York}, {Boroski}, {Kent}, {Lupton}, {Rockosi},
  {Evans}, {Waddell}, {Anderson}, {Annis}, {Barentine}, {Bartoszek}, {Bastian},
  {Bracker}, {Brewington}, {Briegel}, {Brinkmann}, {Brown}, {Carr},
  {Czarapata}, {Drennan}, {Dombeck}, {Federwitz}, {Gillespie}, {Gonzales},
  {Hansen}, {Harvanek}, {Hayes}, {Jordan}, {Kinney}, {Klaene}, {Kleinman},
  {Kron}, {Kresinski}, {Lee}, {Limmongkol}, {Lindenmeyer}, {Long}, {Loomis},
  {McGehee}, {Mantsch}, {Neilsen}, {Neswold}, {Newman}, {Nitta}, {Peoples},
  {Pier}, {Prieto}, {Prosapio}, {Rivetta}, {Schneider}, {Snedden}, \&
  {Wang}}]{Gunn:2006}
{Gunn}, J.~E., {Siegmund}, W.~A., {Mannery}, E.~J., {et~al.} 2006, \aj, 131,
  2332, \dodoi{10.1086/500975}

\bibitem[{Harris {et~al.}(2020)Harris, Millman, van~der Walt, Gommers,
  Virtanen, Cournapeau, Wieser, Taylor, Berg, Smith, Kern, Picus, Hoyer, van
  Kerkwijk, Brett, Haldane, del R{\'\i}o, Wiebe, Peterson, G{\'e}rard-Marchant,
  Sheppard, Reddy, Weckesser, Abbasi, Gohlke, \& Oliphant}]{numpy:2020}
Harris, C.~R., Millman, K.~J., van~der Walt, S.~J., {et~al.} 2020, Nature, 585,
  357, \dodoi{10.1038/s41586-020-2649-2}

\bibitem[{{Hayden} {et~al.}(2015){Hayden}, {Bovy}, {Holtzman}, {Nidever},
  {Bird}, {Weinberg}, {Andrews}, {Majewski}, {Allende Prieto}, {Anders},
  {Beers}, {Bizyaev}, {Chiappini}, {Cunha}, {Frinchaboy},
  {Garc{\'\i}a-Her{\'n}and ez}, {Garc{\'\i}a P{\'e}rez}, {Girardi}, {Harding},
  {Hearty}, {Johnson}, {M{\'e}sz{\'a}ros}, {Minchev}, {O'Connell}, {Pan},
  {Robin}, {Schiavon}, {Schneider}, {Schultheis}, {Shetrone}, {Skrutskie},
  {Steinmetz}, {Smith}, {Wilson}, {Zamora}, \& {Zasowski}}]{Hayden:2015}
{Hayden}, M.~R., {Bovy}, J., {Holtzman}, J.~A., {et~al.} 2015, \apj, 808, 132,
  \dodoi{10.1088/0004-637X/808/2/132}

\bibitem[{{Helmi} \& {White}(1999)}]{Helmi:1999}
{Helmi}, A., \& {White}, S. D.~M. 1999, \mnras, 307, 495,
  \dodoi{10.1046/j.1365-8711.1999.02616.x}

\bibitem[{{Hernquist}(1990)}]{Hernquist:1990}
{Hernquist}, L. 1990, \apj, 356, 359, \dodoi{10.1086/168845}

\bibitem[{{Holtzman} {et~al.}(2018){Holtzman}, {Hasselquist}, {Shetrone},
  {Cunha}, {Allende Prieto}, {Anguiano}, {Bizyaev}, {Bovy}, {Casey},
  {Edvardsson}, {Johnson}, {J{\"o}nsson}, {Meszaros}, {Smith}, {Sobeck},
  {Zamora}, {Chojnowski}, {Fernandez-Trincado}, {Garcia-Hernandez}, {Majewski},
  {Pinsonneault}, {Souto}, {Stringfellow}, {Tayar}, {Troup}, \&
  {Zasowski}}]{Holtzman:2018}
{Holtzman}, J.~A., {Hasselquist}, S., {Shetrone}, M., {et~al.} 2018, \aj, 156,
  125, \dodoi{10.3847/1538-3881/aad4f9}

\bibitem[{{Hunt} {et~al.}(2018){Hunt}, {Hong}, {Bovy}, {Kawata}, \&
  {Grand}}]{Hunt:2018}
{Hunt}, J. A.~S., {Hong}, J., {Bovy}, J., {Kawata}, D., \& {Grand}, R. J.~J.
  2018, \mnras, 481, 3794, \dodoi{10.1093/mnras/sty2532}

\bibitem[{Hunter(2007)}]{Hunter:2007}
Hunter, J.~D. 2007, Computing in Science \& Engineering, 9, 90,
  \dodoi{10.1109/MCSE.2007.55}

\bibitem[{{Iben}(1965)}]{Iben:1965}
{Iben}, Icko, J. 1965, \apj, 142, 1447, \dodoi{10.1086/148429}

\bibitem[{{Iorio} \& {Belokurov}(2020)}]{Iorio:2020}
{Iorio}, G., \& {Belokurov}, V. 2020, arXiv e-prints, arXiv:2008.02280.
\newblock \doarXiv{2008.02280}

\bibitem[{{Jeans}(1919)}]{Jeans:1919b}
{Jeans}, J.~H. 1919, {Problems of cosmogony and stellar dynamics}

\bibitem[{{Jeans}(1922)}]{Jeans:1922}
---. 1922, \mnras, 82, 122, \dodoi{10.1093/mnras/82.3.122}

\bibitem[{{Johnston} {et~al.}(1999){Johnston}, {Zhao}, {Spergel}, \&
  {Hernquist}}]{Johnston:1999}
{Johnston}, K.~V., {Zhao}, H., {Spergel}, D.~N., \& {Hernquist}, L. 1999,
  \apjl, 512, L109, \dodoi{10.1086/311876}

\bibitem[{{J{\"o}nsson} {et~al.}(2020){J{\"o}nsson}, {Holtzman}, {Prieto},
  {Cunha}, {Garc{\'\i}a-Hern{\'a}ndez}, {Hasselquist}, {Masseron}, {Osorio},
  {Shetrone}, {Smith}, {Stringfellow}, {Bizyaev}, {Edvardsson}, {Majewski},
  {M{\'e}sz{\'a}ros}, {Souto}, {Zamora}, {Beaton}, {Bovy}, {Donor},
  {Pinsonneault}, {Poovelil}, \& {Sobeck}}]{Jonsson:2020}
{J{\"o}nsson}, H., {Holtzman}, J.~A., {Prieto}, C.~A., {et~al.} 2020, \aj, 160,
  120, \dodoi{10.3847/1538-3881/aba592}

\bibitem[{{Kamdar} {et~al.}(2019){Kamdar}, {Conroy}, {Ting}, {Bonaca}, {Smith},
  \& {Brown}}]{Kamdar:2019}
{Kamdar}, H., {Conroy}, C., {Ting}, Y.-S., {et~al.} 2019, \apjl, 884, L42,
  \dodoi{10.3847/2041-8213/ab4997}

\bibitem[{{Kepler}(1609)}]{Kepler:1609}
{Kepler}, J. 1609, {Astronomia Nova}, 1st edn. (J. Kepler)

\bibitem[{{Khanna} {et~al.}(2019){Khanna}, {Sharma}, {Tepper-Garcia}, {Bland
  -Hawthorn}, {Hayden}, {Asplund}, {Buder}, {Chen}, {De Silva}, {Freeman},
  {Kos}, {Lewis}, {Lin}, {Martell}, {Simpson}, {Nordlander}, {Stello}, {Ting},
  {Zucker}, \& {Zwitter}}]{Khanna:2019}
{Khanna}, S., {Sharma}, S., {Tepper-Garcia}, T., {et~al.} 2019, \mnras, 489,
  4962, \dodoi{10.1093/mnras/stz2462}

\bibitem[{{Koppelman} {et~al.}(2018){Koppelman}, {Helmi}, \&
  {Veljanoski}}]{Koppelman:2018}
{Koppelman}, H., {Helmi}, A., \& {Veljanoski}, J. 2018, \apjl, 860, L11,
  \dodoi{10.3847/2041-8213/aac882}

\bibitem[{{Laporte} {et~al.}(2020){Laporte}, {Belokurov}, {Koposov}, {Smith},
  \& {Hill}}]{Laporte:2020}
{Laporte}, C. F.~P., {Belokurov}, V., {Koposov}, S.~E., {Smith}, M.~C., \&
  {Hill}, V. 2020, \mnras, 492, L61, \dodoi{10.1093/mnrasl/slz167}

\bibitem[{{Laporte} {et~al.}(2019){Laporte}, {Minchev}, {Johnston}, \&
  {G{\'o}mez}}]{Laporte:2019}
{Laporte}, C. F.~P., {Minchev}, I., {Johnston}, K.~V., \& {G{\'o}mez}, F.~A.
  2019, \mnras, 485, 3134, \dodoi{10.1093/mnras/stz583}

\bibitem[{{Lindegren} {et~al.}(2018){Lindegren}, {Hern{\'a}ndez}, {Bombrun},
  {Klioner}, {Bastian}, {Ramos-Lerate}, {de Torres}, {Steidelm{\"u}ller},
  {Stephenson}, {Hobbs}, {Lammers}, {Biermann}, {Geyer}, {Hilger}, {Michalik},
  {Stampa}, {McMillan}, {Casta{\~n}eda}, {Clotet}, {Comoretto}, {Davidson},
  {Fabricius}, {Gracia}, {Hambly}, {Hutton}, {Mora}, {Portell}, {van Leeuwen},
  {Abbas}, {Abreu}, {Altmann}, {Andrei}, {Anglada}, {Balaguer-N{\'u}{\~n}ez},
  {Barache}, {Becciani}, {Bertone}, {Bianchi}, {Bouquillon}, {Bourda},
  {Br{\"u}semeister}, {Bucciarelli}, {Busonero}, {Buzzi}, {Cancelliere},
  {Carlucci}, {Charlot}, {Cheek}, {Crosta}, {Crowley}, {de Bruijne}, {de
  Felice}, {Drimmel}, {Esquej}, {Fienga}, {Fraile}, {Gai}, {Garralda},
  {Gonz{\'a}lez-Vidal}, {Guerra}, {Hauser}, {Hofmann}, {Holl}, {Jordan},
  {Lattanzi}, {Lenhardt}, {Liao}, {Licata}, {Lister}, {L{\"o}ffler},
  {Marchant}, {Martin-Fleitas}, {Messineo}, {Mignard}, {Morbidelli}, {Poggio},
  {Riva}, {Rowell}, {Salguero}, {Sarasso}, {Sciacca}, {Siddiqui}, {Smart},
  {Spagna}, {Steele}, {Taris}, {Torra}, {van Elteren}, {van Reeven}, \&
  {Vecchiato}}]{Gaia-astrometric:2018}
{Lindegren}, L., {Hern{\'a}ndez}, J., {Bombrun}, A., {et~al.} 2018, \aap, 616,
  A2, \dodoi{10.1051/0004-6361/201832727}

\bibitem[{{Mackereth} \& {Bovy}(2020)}]{Mackereth:2020}
{Mackereth}, J.~T., \& {Bovy}, J. 2020, \mnras, 492, 3631,
  \dodoi{10.1093/mnras/staa047}

\bibitem[{{Magorrian}(2014)}]{Magorrian:2014}
{Magorrian}, J. 2014, \mnras, 437, 2230, \dodoi{10.1093/mnras/stt2031}

\bibitem[{{Magorrian}(2019)}]{Magorrian:2019}
---. 2019, \mnras, 484, 1166, \dodoi{10.1093/mnras/stz037}

\bibitem[{{Majewski} {et~al.}(2017){Majewski}, {Schiavon}, {Frinchaboy},
  {Allende Prieto}, {Barkhouser}, {Bizyaev}, {Blank}, {Brunner}, {Burton},
  {Carrera}, {Chojnowski}, {Cunha}, {Epstein}, {Fitzgerald}, {Garc{\'{\i}}a
  P{\'e}rez}, {Hearty}, {Henderson}, {Holtzman}, {Johnson}, {Lam}, {Lawler},
  {Maseman}, {M{\'e}sz{\'a}ros}, {Nelson}, {Nguyen}, {Nidever}, {Pinsonneault},
  {Shetrone}, {Smee}, {Smith}, {Stolberg}, {Skrutskie}, {Walker}, {Wilson},
  {Zasowski}, {Anders}, {Basu}, {Beland}, {Blanton}, {Bovy}, {Brownstein},
  {Carlberg}, {Chaplin}, {Chiappini}, {Eisenstein}, {Elsworth}, {Feuillet},
  {Fleming}, {Galbraith-Frew}, {Garc{\'{\i}}a}, {Garc{\'{\i}}a-Hern{\'a}ndez},
  {Gillespie}, {Girardi}, {Gunn}, {Hasselquist}, {Hayden}, {Hekker}, {Ivans},
  {Kinemuchi}, {Klaene}, {Mahadevan}, {Mathur}, {Mosser}, {Muna}, {Munn},
  {Nichol}, {O'Connell}, {Parejko}, {Robin}, {Rocha-Pinto}, {Schultheis},
  {Serenelli}, {Shane}, {Silva Aguirre}, {Sobeck}, {Thompson}, {Troup},
  {Weinberg}, \& {Zamora}}]{Majewski:2017}
{Majewski}, S.~R., {Schiavon}, R.~P., {Frinchaboy}, P.~M., {et~al.} 2017, \aj,
  154, 94, \dodoi{10.3847/1538-3881/aa784d}

\bibitem[{{Marrese} {et~al.}(2019){Marrese}, {Marinoni}, {Fabrizio}, \&
  {Altavilla}}]{Gaia-crossmatch:2019}
{Marrese}, P.~M., {Marinoni}, S., {Fabrizio}, M., \& {Altavilla}, G. 2019,
  \aap, 621, A144, \dodoi{10.1051/0004-6361/201834142}

\bibitem[{{Martell} {et~al.}(2017){Martell}, {Sharma}, {Buder}, {Duong},
  {Schlesinger}, {Simpson}, {Lind}, {Ness}, {Marshall}, {Asplund},
  {Bland-Hawthorn}, {Casey}, {De Silva}, {Freeman}, {Kos}, {Lin}, {Zucker},
  {Zwitter}, {Anguiano}, {Bacigalupo}, {Carollo}, {Casagrande}, {Da Costa},
  {Horner}, {Huber}, {Hyde}, {Kafle}, {Lewis}, {Nataf}, {Navin}, {Stello},
  {Tinney}, {Watson}, \& {Wittenmyer}}]{Martell:2017}
{Martell}, S.~L., {Sharma}, S., {Buder}, S., {et~al.} 2017, \mnras, 465, 3203,
  \dodoi{10.1093/mnras/stw2835}

\bibitem[{{Martig} {et~al.}(2016){Martig}, {Fouesneau}, {Rix}, {Ness},
  {M{\'e}sz{\'a}ros}, {Garc{\'\i}a-Hern{\'a}ndez}, {Pinsonneault}, {Serenelli},
  {Silva Aguirre}, \& {Zamora}}]{Martig:2016}
{Martig}, M., {Fouesneau}, M., {Rix}, H.-W., {et~al.} 2016, \mnras, 456, 3655,
  \dodoi{10.1093/mnras/stv2830}

\bibitem[{{McMillan} \& {Binney}(2013)}]{McMillan:2013}
{McMillan}, P.~J., \& {Binney}, J.~J. 2013, \mnras, 433, 1411,
  \dodoi{10.1093/mnras/stt814}

\bibitem[{{Miyamoto} \& {Nagai}(1975)}]{Miyamoto:1975}
{Miyamoto}, M., \& {Nagai}, R. 1975, \pasj, 27, 533

\bibitem[{{Monari} {et~al.}(2019){Monari}, {Famaey}, {Siebert}, {Wegg}, \&
  {Gerhard}}]{Monari:2019}
{Monari}, G., {Famaey}, B., {Siebert}, A., {Wegg}, C., \& {Gerhard}, O. 2019,
  \aap, 626, A41, \dodoi{10.1051/0004-6361/201834820}

\bibitem[{{Myeong} {et~al.}(2018){Myeong}, {Evans}, {Belokurov}, {Sand ers}, \&
  {Koposov}}]{Myeong:2018}
{Myeong}, G.~C., {Evans}, N.~W., {Belokurov}, V., {Sand ers}, J.~L., \&
  {Koposov}, S.~E. 2018, \apjl, 856, L26, \dodoi{10.3847/2041-8213/aab613}

\bibitem[{{Navarro} {et~al.}(1996){Navarro}, {Frenk}, \&
  {White}}]{Navarro:1996}
{Navarro}, J.~F., {Frenk}, C.~S., \& {White}, S.~D.~M. 1996, \apj, 462, 563,
  \dodoi{10.1086/177173}

\bibitem[{{Newton}(1687)}]{Newton:1687}
{Newton}, I. 1687, {Philosophi\ae Naturalis Principia Mathematica}, 1st edn.
  (E. Halley)

\bibitem[{{Nidever} {et~al.}(2015){Nidever}, {Holtzman}, {Allende Prieto},
  {Beland}, {Bender}, {Bizyaev}, {Burton}, {Desphande}, {Fleming},
  {Garc{\'{\i}}a P{\'e}rez}, {Hearty}, {Majewski}, {M{\'e}sz{\'a}ros}, {Muna},
  {Nguyen}, {Schiavon}, {Shetrone}, {Skrutskie}, {Sobeck}, \&
  {Wilson}}]{Nidever:2015}
{Nidever}, D.~L., {Holtzman}, J.~A., {Allende Prieto}, C., {et~al.} 2015, \aj,
  150, 173, \dodoi{10.1088/0004-6256/150/6/173}

\bibitem[{{Oort}(1932)}]{Oort:1932}
{Oort}, J.~H. 1932, \bain, 6, 249

\bibitem[{{Pe{\~n}arrubia}(2013)}]{Penarrubia:2013}
{Pe{\~n}arrubia}, J. 2013, \mnras, 433, 2576, \dodoi{10.1093/mnras/stt935}

\bibitem[{P\'erez \& Granger(2007)}]{ipython}
P\'erez, F., \& Granger, B.~E. 2007, Computing in Science and Engineering, 9,
  21, \dodoi{10.1109/MCSE.2007.53}

\bibitem[{{Poggio} {et~al.}(2020){Poggio}, {Drimmel}, {Andrae}, {Bailer-Jones},
  {Fouesneau}, {Lattanzi}, {Smart}, \& {Spagna}}]{Poggio:2020}
{Poggio}, E., {Drimmel}, R., {Andrae}, R., {et~al.} 2020, Nature Astronomy, 4,
  590, \dodoi{10.1038/s41550-020-1017-3}

\bibitem[{{Price-Whelan}(2017)}]{gala}
{Price-Whelan}, A.~M. 2017, The Journal of Open Source Software, 2, 388,
  \dodoi{10.21105/joss.00388}

\bibitem[{Price-Whelan \& Foreman-Mackey(2017)}]{schwimmbad}
Price-Whelan, A.~M., \& Foreman-Mackey, D. 2017, The Journal of Open Source
  Software, 2, \dodoi{10.21105/joss.00357}

\bibitem[{{Price-Whelan} {et~al.}(2014){Price-Whelan}, {Hogg}, {Johnston}, \&
  {Hendel}}]{Price-Whelan:2014}
{Price-Whelan}, A.~M., {Hogg}, D.~W., {Johnston}, K.~V., \& {Hendel}, D. 2014,
  \apj, 794, 4, \dodoi{10.1088/0004-637X/794/1/4}

\bibitem[{{Price-Whelan} {et~al.}(2015){Price-Whelan}, {Johnston}, {Sheffield},
  {Laporte}, \& {Sesar}}]{Price-Whelan:2015}
{Price-Whelan}, A.~M., {Johnston}, K.~V., {Sheffield}, A.~A., {Laporte}, C.
  F.~P., \& {Sesar}, B. 2015, \mnras, 452, 676, \dodoi{10.1093/mnras/stv1324}

\bibitem[{{Queiroz} {et~al.}(2020){Queiroz}, {Anders}, {Chiappini},
  {Khalatyan}, {Santiago}, {Steinmetz}, {Valentini}, {Miglio}, {Bossini},
  {Barbuy}, {Minchev}, {Minniti}, {Garc{\'\i}a Hern{\'a}ndez}, {Schultheis},
  {Beaton}, {Beers}, {Bizyaev}, {Brownstein}, {Cunha},
  {Fern{\'a}ndez-Trincado}, {Frinchaboy}, {Lane}, {Majewski}, {Nataf},
  {Nitschelm}, {Pan}, {Roman-Lopes}, {Sobeck}, {Stringfellow}, \&
  {Zamora}}]{Queiroz:2020}
{Queiroz}, A.~B.~A., {Anders}, F., {Chiappini}, C., {et~al.} 2020, \aap, 638,
  A76, \dodoi{10.1051/0004-6361/201937364}

\bibitem[{{Read}(2014)}]{Read:2014}
{Read}, J.~I. 2014, Journal of Physics G Nuclear Physics, 41, 063101,
  \dodoi{10.1088/0954-3899/41/6/063101}

\bibitem[{{Reid} \& {Brunthaler}(2004)}]{Reid:2004}
{Reid}, M.~J., \& {Brunthaler}, A. 2004, \apj, 616, 872, \dodoi{10.1086/424960}

\bibitem[{{Romanowsky} {et~al.}(2003){Romanowsky}, {Douglas}, {Arnaboldi},
  {Kuijken}, {Merrifield}, {Napolitano}, {Capaccioli}, \&
  {Freeman}}]{Romanowsky:2003}
{Romanowsky}, A.~J., {Douglas}, N.~G., {Arnaboldi}, M., {et~al.} 2003, Science,
  301, 1696, \dodoi{10.1126/science.1087441}

\bibitem[{{Sanders}(2012)}]{Sanders:2012}
{Sanders}, J. 2012, \mnras, 426, 128, \dodoi{10.1111/j.1365-2966.2012.21698.x}

\bibitem[{{Sanders} \& {Binney}(2013)}]{Sanders:2013}
{Sanders}, J.~L., \& {Binney}, J. 2013, \mnras, 433, 1813,
  \dodoi{10.1093/mnras/stt806}

\bibitem[{{Sanders} \& {Binney}(2014)}]{Sanders:2014}
---. 2014, \mnras, 441, 3284, \dodoi{10.1093/mnras/stu796}

\bibitem[{{Sanders} \& {Binney}(2015)}]{Sanders:2015}
---. 2015, \mnras, 449, 3479, \dodoi{10.1093/mnras/stv578}

\bibitem[{{Sanders} \& {Binney}(2016)}]{Sanders:2016}
---. 2016, \mnras, 457, 2107, \dodoi{10.1093/mnras/stw106}

\bibitem[{{Sch{\"o}nrich} \& {Dehnen}(2018)}]{Schonrich:2018}
{Sch{\"o}nrich}, R., \& {Dehnen}, W. 2018, \mnras, 478, 3809,
  \dodoi{10.1093/mnras/sty1256}

\bibitem[{{Shipp} {et~al.}(2018){Shipp}, {Drlica-Wagner}, {Balbinot},
  {Ferguson}, {Erkal}, {Li}, {Bechtol}, {Belokurov}, {Buncher}, {Carollo},
  {Carrasco Kind}, {Kuehn}, {Marshall}, {Pace}, {Rykoff}, {Sevilla-Noarbe},
  {Sheldon}, {Strigari}, {Vivas}, {Yanny}, {Zenteno}, {Abbott}, {Abdalla},
  {Allam}, {Avila}, {Bertin}, {Brooks}, {Burke}, {Carretero}, {Castander},
  {Cawthon}, {Crocce}, {Cunha}, {D'Andrea}, {da Costa}, {Davis}, {De Vicente},
  {Desai}, {Diehl}, {Doel}, {Evrard}, {Flaugher}, {Fosalba}, {Frieman},
  {Garc{\'\i}a-Bellido}, {Gaztanaga}, {Gerdes}, {Gruen}, {Gruendl}, {Gschwend},
  {Gutierrez}, {Hartley}, {Honscheid}, {Hoyle}, {James}, {Johnson}, {Krause},
  {Kuropatkin}, {Lahav}, {Lin}, {Maia}, {March}, {Martini}, {Menanteau},
  {Miller}, {Miquel}, {Nichol}, {Plazas}, {Romer}, {Sako}, {Sanchez},
  {Santiago}, {Scarpine}, {Schindler}, {Schubnell}, {Smith}, {Smith},
  {Sobreira}, {Suchyta}, {Swanson}, {Tarle}, {Thomas}, {Tucker}, {Walker},
  {Wechsler}, \& {DES Collaboration}}]{Shipp:2018}
{Shipp}, N., {Drlica-Wagner}, A., {Balbinot}, E., {et~al.} 2018, \apj, 862,
  114, \dodoi{10.3847/1538-4357/aacdab}

\bibitem[{{Skrutskie} {et~al.}(2006){Skrutskie}, {Cutri}, {Stiening},
  {Weinberg}, {Schneider}, {Carpenter}, {Beichman}, {Capps}, {Chester},
  {Elias}, {Huchra}, {Liebert}, {Lonsdale}, {Monet}, {Price}, {Seitzer},
  {Jarrett}, {Kirkpatrick}, {Gizis}, {Howard}, {Evans}, {Fowler}, {Fullmer},
  {Hurt}, {Light}, {Kopan}, {Marsh}, {McCallon}, {Tam}, {Van Dyk}, \&
  {Wheelock}}]{Skrutskie:2006}
{Skrutskie}, M.~F., {Cutri}, R.~M., {Stiening}, R., {et~al.} 2006, \aj, 131,
  1163, \dodoi{10.1086/498708}

\bibitem[{{Virtanen} {et~al.}(2020){Virtanen}, {Gommers}, {Oliphant},
  {Haberland}, {Reddy}, {Cournapeau}, {Burovski}, {Peterson}, {Weckesser},
  {Bright}, {van der Walt}, {Brett}, {Wilson}, {Millman}, {Mayorov}, {Nelson},
  {Jones}, {Kern}, {Larson}, {Carey}, {Polat}, {Feng}, {Moore}, {Vand erPlas},
  {Laxalde}, {Perktold}, {Cimrman}, {Henriksen}, {Quintero}, {Harris},
  {Archibald}, {Ribeiro}, {Pedregosa}, {van Mulbregt}, \& {SciPy 1. 0
  Contributors}}]{Virtanen:2020}
{Virtanen}, P., {Gommers}, R., {Oliphant}, T.~E., {et~al.} 2020, Nature
  Methods, 17, 261, \dodoi{10.1038/s41592-019-0686-2}

\bibitem[{{Walker} \& {Pe{\~n}arrubia}(2011)}]{Walker:2011}
{Walker}, M.~G., \& {Pe{\~n}arrubia}, J. 2011, \apj, 742, 20,
  \dodoi{10.1088/0004-637X/742/1/20}

\bibitem[{{Watkins} {et~al.}(2010){Watkins}, {Evans}, \& {An}}]{Watkins:2010}
{Watkins}, L.~L., {Evans}, N.~W., \& {An}, J.~H. 2010, \mnras, 406, 264,
  \dodoi{10.1111/j.1365-2966.2010.16708.x}

\bibitem[{{Wheeler} {et~al.}(2020){Wheeler}, {Ness}, {Buder}, {Bland
  -Hawthorn}, {De Silva}, {Hayden}, {Kos}, {Lewis}, {Martell}, {Sharma},
  {Simpson}, {Zucker}, \& {Zwitter}}]{Wheeler:2020}
{Wheeler}, A., {Ness}, M., {Buder}, S., {et~al.} 2020, \apj, 898, 58,
  \dodoi{10.3847/1538-4357/ab9a46}

\bibitem[{{Wilson} {et~al.}(2019){Wilson}, {Hearty}, {Skrutskie}, {Majewski},
  {Holtzman}, {Eisenstein}, {Gunn}, {Blank}, {Henderson}, {Smee}, {Nelson},
  {Nidever}, {Arns}, {Barkhouser}, {Barr}, {Beland}, {Bershady}, {Blanton},
  {Brunner}, {Burton}, {Carey}, {Carr}, {Colque}, {Crane}, {Damke}, {Davidson},
  {Dean}, {Di Mille}, {Don}, {Ebelke}, {Evans}, {Fitzgerald}, {Gillespie},
  {Hall}, {Harding}, {Harding}, {Hammond}, {Hancock}, {Harrison}, {Hope},
  {Horne}, {Karakla}, {Lam}, {Leger}, {MacDonald}, {Maseman}, {Matsunari},
  {Melton}, {Mitcheltree}, {O'Brien}, {O'Connell}, {Patten}, {Richardson},
  {Rieke}, {Rieke}, {Roman-Lopes}, {Schiavon}, {Sobeck}, {Stolberg}, {Stoll},
  {Tembe}, {Trujillo}, {Uomoto}, {Vernieri}, {Walker}, {Weinberg}, {Young},
  {Anthony-Brumfield}, {Bizyaev}, {Breslauer}, {De Lee}, {Downey}, {Halverson},
  {Huehnerhoff}, {Klaene}, {Leon}, {Long}, {Mahadevan}, {Malanushenko},
  {Nguyen}, {Owen}, {S{\'a}nchez-Gallego}, {Sayres}, {Shane}, {Shectman},
  {Shetrone}, {Skinner}, {Stauffer}, \& {Zhao}}]{Wilson:2019}
{Wilson}, J.~C., {Hearty}, F.~R., {Skrutskie}, M.~F., {et~al.} 2019, \pasp,
  131, 055001, \dodoi{10.1088/1538-3873/ab0075}

\bibitem[{{Xu} {et~al.}(2020){Xu}, {Liu}, {Tian}, {Newberg}, {Laporte},
  {Zhang}, {Wang}, {Fu}, {Li}, \& {Deng}}]{Xu:2020}
{Xu}, Y., {Liu}, C., {Tian}, H., {et~al.} 2020, \apj, 905, 6,
  \dodoi{10.3847/1538-4357/abc2cb}

\bibitem[{{Zasowski} {et~al.}(2013){Zasowski}, {Johnson}, {Frinchaboy},
  {Majewski}, {Nidever}, {Rocha Pinto}, {Girardi}, {Andrews}, {Chojnowski},
  {Cudworth}, {Jackson}, {Munn}, {Skrutskie}, {Beaton}, {Blake}, {Covey},
  {Deshpande}, {Epstein}, {Fabbian}, {Fleming}, {Garcia Hernandez}, {Herrero},
  {Mahadevan}, {M{\'e}sz{\'a}ros}, {Schultheis}, {Sellgren}, {Terrien}, {van
  Saders}, {Allende Prieto}, {Bizyaev}, {Burton}, {Cunha}, {da Costa},
  {Hasselquist}, {Hearty}, {Holtzman}, {Garc{\'\i}a P{\'e}rez}, {Maia},
  {O'Connell}, {O'Donnell}, {Pinsonneault}, {Santiago}, {Schiavon}, {Shetrone},
  {Smith}, \& {Wilson}}]{Zasowski:2013}
{Zasowski}, G., {Johnson}, J.~A., {Frinchaboy}, P.~M., {et~al.} 2013, \aj, 146,
  81, \dodoi{10.1088/0004-6256/146/4/81}

\bibitem[{{Zasowski} {et~al.}(2017){Zasowski}, {Cohen}, {Chojnowski},
  {Santana}, {Oelkers}, {Andrews}, {Beaton}, {Bender}, {Bird}, {Bovy},
  {Carlberg}, {Covey}, {Cunha}, {Dell'Agli}, {Fleming}, {Frinchaboy},
  {Garc{\'\i}a-Hern{\'a}ndez}, {Harding}, {Holtzman}, {Johnson}, {Kollmeier},
  {Majewski}, {M{\'e}sz{\'a}ros}, {Munn}, {Mu{\~n}oz}, {Ness}, {Nidever},
  {Poleski}, {Rom{\'a}n-Z{\'u}{\~n}iga}, {Shetrone}, {Simon}, {Smith},
  {Sobeck}, {Stringfellow}, {Szigeti{\'a}ros}, {Tayar}, \&
  {Troup}}]{Zasowski:2017}
{Zasowski}, G., {Cohen}, R.~E., {Chojnowski}, S.~D., {et~al.} 2017, \aj, 154,
  198, \dodoi{10.3847/1538-3881/aa8df9}

\bibitem[{{Zhai} {et~al.}(2018){Zhai}, {Xue}, {Zhang}, {Li}, {Zhao}, \&
  {Yang}}]{Zhai:2018}
{Zhai}, M., {Xue}, X.-X., {Zhang}, L., {et~al.} 2018, Research in Astronomy and
  Astrophysics, 18, 113, \dodoi{10.1088/1674-4527/18/9/113}

\end{thebibliography}

\end{document}